%% file: ms.tex
\documentclass[abstract=on,notitlepage]{scrreprt}

\usepackage[english]{babel}
\usepackage[T1]{fontenc} 
\usepackage[utf8]{inputenc} 
\usepackage{lmodern} 

\usepackage{graphicx} 
\usepackage{caption}
\usepackage{pgf}
\usepackage{tikz}
\usetikzlibrary{fit}
\usetikzlibrary{shapes}
\usetikzlibrary{arrows,automata,decorations.markings}
\usetikzlibrary{positioning}

\tikzstyle{randomVariable}=[circle,fill=white,draw=black,text=black,minimum size=0.8cm]
\tikzset{randomVector/.style={draw=black,inner sep=5pt}}
\tikzstyle{observation}=[circle,fill=gray!40,draw=black,text=black,minimum size=0.8cm]
\tikzstyle{param}=[draw=none,fill=none,text=black,minimum size=0.2cm]

\pgfdeclarelayer{bg}    
\pgfsetlayers{bg,main}  

\usepackage[ruled]{algorithm2e}
\usepackage{amsmath}
\usepackage{amsthm}
\usepackage{amsfonts}
\usepackage{dsfont}
\usepackage{bbm}
\newtheorem*{definition}{Definition}
\newtheorem{prop}{Proposition}[section]
\newtheorem*{properties}{Properties}

\newtheorem{Thm}{Theorem}[section]

\newtheorem*{example}{Example}

\usepackage{a4wide} 
\usepackage{standalone} 
\usepackage[toc,page]{appendix}
\usepackage{enumerate}
\usepackage{color}   
\usepackage{hyperref}
\hypersetup{
    colorlinks=true, 
    linktoc=all,     
    linkcolor=blue,  
}

\newcommand{\indep}{\rotatebox[origin=c]{90}{$\models$}}
\newcommand{\pluseq}{\mathrel{{+}{=}}}
\newcommand{\di}{i}

\begin{document}

\title{The power disaggregation algorithms and their applications to demand dispatch}
\subtitle{Master dissertation}

\author{Arnaud Cadas
\thanks{Inria and the Computer Science Dept. of \'{E}cole Normale Supr\'{e}rieure, CNRS, PSL Research University, Paris, France. Email: \href{mailto:arnaud.cadas@inria.fr}{arnaud.cadas@inria.fr}}
\and Ana Bu\v{s}i\'{c} 
\thanks{Inria and the Computer Science Dept. of \'{E}cole Normale Supr\'{e}rieure, CNRS, PSL Research University, Paris, France. Email: \href{mailto:ana.busic@inria.fr}{ana.busic@inria.fr}}}
\date{October 15, 2017}

  \maketitle
  
  \begin{abstract}
We were interested in solving a power disaggregation problem which comes down to estimating the power consumption of each device given the total power consumption of the whole house. We started by looking at the Factorial Hierarchical Dirichlet Process - Hidden Semi-Markov Model. However, the inference method had a complexity which scales withthe number of observations. Thus, we developed an online algorithm based on particle filters. We applied the method to data from Pecan Street \href{https://dataport.cloud/}{https://dataport.cloud/} using Python. We applied the disaggregation algorithm to the control techniques used in Demand Dispatch.
\bigbreak
\end{abstract}


{\small\tableofcontents} 
\cleardoublepage 

\pagestyle{plain} 


\section*{Introduction}

The increase of renewable energy has made the supply-demand balance of power more complex to handle. In \cite{Busic}, the authors designed randomized controllers to obtain ancillary services to the power grid by harnessing inherent flexibility in many loads. 
\smallbreak
However these controllers suppose that we know the consumption of each device
that we want to control. This introduce the cost and the social constraint of putting sensors on each device of each house. Therefore, our approach was to use Nonintrusive Appliance Load Monitoring (NALM) methods \cite{NALM} to solve a disaggregation problem. The latter comes down to estimating the power consumption of each device given the total power consumption of the whole house.
\smallbreak
We started by looking at the Factorial Hierarchical Dirichlet Process - Hidden Semi-Markov Model (Factorial HDP-HSMM) introduced in \cite{HDPHMM}. In our application, the total power consumption is considered as the observations of this state-space model and the consumption of each device as the state variables. Each of the latter is modeled by an HDP-HSMM which is an extension of a Hidden Markov Model. All the models are presented in chapter~\ref{sec:Model}.
\smallbreak
The inference method used in \cite{HDPHMM} will be developed in chapter~\ref{sec:Inference}. It is based on Gibbs sampling and some of its variations. Our contributions here was to give a detailed proof on how to sample from the posterior distribution of each parameter for each model.
\smallbreak 
However, the inference algorithm has a complexity of $O(T^2 N +T N^2)$ where T is the number of observations and N is the number of hidden states. As our goal is to use the randomized controllers with our estimations, we wanted a method that does not scale with T. Therefore, we developed an online algorithm based on particle filters in chapter~\ref{OnlineInference}. Because we worked in a Bayesian setting, we had to infer the parameters of our model. To do so, we used a method called Particle Learning which is presented in \cite{PL}. The idea is to include the parameters in the state space so that they are tied to the particles. Then, for each (re)sampling step, the parameters are sampled from their posterior distribution with the help of Bayesian sufficient statistics. Smoothing was also introduced as a possible improvement for future work.
\smallbreak
In chapter~\ref{chap:control}, we present the control theory for "demand dispatch" which was developed in \cite{Busic} and which motivated the search for power disaggregation algorithms. We will see how the control architecture work, how to evaluate its performance and how to combine it with the online learning algorithm from the previous chapter.
\smallbreak
We applied the disaggregation method to data from Pecan Street in chapter~\ref{sec:Applications}. Using their Dataport, we collected the power consumption of each device from about a hundred houses. We selected the few devices that consume the most and that are present in most houses. We separated the houses in a training set and a test set. For each device of each house from the training set, we estimated the operating modes with a HDP-HSMM and used these estimations to compute estimators of the priors hyper-parameters. Finally we applied the particle filters method to the test houses using the computed priors. 
\smallbreak
The algorithm performs well for the devices with the highest power consumption, which is the air compressor (of the air conditioning system) in the case of Pecan Street
data. The report ends by an overview of the ongoing work on applying the disaggregation algorithm to the control techniques in \cite{Busic} for thermostatically controlled loads.

\include{Model}

\include{Inference}

\include{Online_inference}

\include{Control}

\include{Applications}

\section*{Conclusion}

To conclude, motivated by the work on "demand dispatch" by Ana Busic, we tackled a power disaggregation problem. We started by looking at the factorial HDP-HSMM to solve it. Thus, we presented the model and how to infer its parameters for a finite number of observations. Because we wanted to apply the disaggregation algorithm to control in real time, we developed an online inference algorithm based on particle filters and the idea of particle learning. Finally, we applied this method to data from Pecan Street. Only the device which consumes the most power was properly retrieved but combining this device estimation with control seems promising.
\smallbreak
We believe that the algorithm could be improved with smoothing by computing a backward filter at regular intervals (once a day for example) and merging it with the forward filter. Our algorithm also depends heavily on the priors. Therefore, future work could be done on computing more precise priors. One idea is to have several classes of priors for each device (the classes could be created by unsupervised learning on the mean consumption for each operating mode) and a learning phase before starting disaggregation. During this phase, we would only look at the aggregated power consumption (in order for the method to stay non intrusive) and we would use supervised learning (using the Pecan Street data) to find the best class for each device.

\addtocontents{toc}{\protect\vspace*{\baselineskip}}




\include{report_appendix}


\end{document}

%% file: Model.tex
\chapter{Models}\label{sec:Model}

During this chapter, we will present the different models introduced in \cite{HDPHMM}. Starting from the Hidden Markov Model, the next models will build upon it, adding new variables and changing the structure so that it is more flexible and could fit more complex data. The final model that we will use for disaggregation is the factorial HDP-HSMM. For this chapter, some knowledge about graphical models and Bayesian statistics is assumed, see appendix~\ref{graphmodel} and appendix~\ref{BayesStats} for more details. Section~\ref{sec:HDPHSMM} also assumes some understanding about the Dirichlet Process and the Hierarchical Dirichlet Process, see appendix~\ref{sec:NonParamBayes} for more details.

\section{Bayesian Hidden Markov Model}\label{sec:BayesianHMM}

The Hidden Markov Model (HMM) is a well known model used for time series analysis. Its structure is more adapted for time series than the usual i.i.d hypothesis as the observations depend on each other through a phenomena that we cannot observe but which evolve in a specific way. Thus, it appears as a good candidate to modelize a signal of the power consumption of a device. 
\smallbreak
The model supposes that we observe a realization of random variables $y_1,\cdots,y_n \in \mathcal{Y}^n$ which we call \textbf{observations} and that depends on latent random variables $x_1,\cdots,x_n \in \mathcal{X}^n$ that we call \textbf{hidden states}. The observations are supposed to be independently distributed given the hidden states: $p(y_1,\cdots,y_n | x_1,\cdots,x_n)=\prod_{i=1}^{n}p(y_i | x_i)$. The hidden states are supposed to evolve like a Markov chain: $p(x_1,\cdots,x_n)=p(x_1)\prod_{i=2}^{n}p(x_i | x_{i-1})$. Here $p$ is a density with respect to a reference measure which will essentially be the Lesbegue measure (if the random variable is continuous) or the counting measure (if the random variable is discrete). We will use this notation $p$ through out the paper.
\smallbreak
The sets $\mathcal{Y}$ and $\mathcal{X}$ can be very general but we intend to use this model for our power disaggregation problem. Therefore, we will focus on the specific sets $\mathcal{Y}=\mathbb{R}^{+}$ and $\mathcal{X}=\{1,\cdots,J\}$. The reason behind this choice is that the hidden states modelize the different operating modes of a device and the observations represent the power consumption of the device in a particular mode. When $| \mathcal{X} |<\infty$ we call this a finite state-space HMM.
\smallbreak
Next, to fully define the model, we need to specify the transition kernel of the hidden Markov chain and the distribution of the observations given the hidden states. Here again, we could use several different distributions but we will focus on the ones adapted to our application. Because we choose $\mathcal{X}=\{1,\cdots,J\}$, it is straightforward that we have a transition matrix as the transition kernel. We will call $(\pi_j)_{1\leq j\leq J}$ the rows of this transition matrix. For the observations, as the power consumption of a device is often very concentrated around a specific value for each operating mode, we will use a normal distribution with a different mean $\theta_j$ and the same variance $\sigma^2$ for each mode. The support of this distribution is not $\mathbb{R}^{+}$ but by using a normal distribution we will greatly ease the inference part later on. If we estimate a negative value for an observation we will set it to zero.
\smallbreak
For the first model, we consider the inferential statistics framework. Thus we suppose that we know $\pi_j$ and $\theta_j$ for $j=1,\cdots,J$ and $\sigma^2$. We can sum up the model this way:
\begin{align*}
x_t | x_{t-1} &\:\sim Cat(\pi_{x_{t-1}})\\
y_t | x_t &\overset{i.i.d}{\sim} \mathcal{N}(\theta_{x_t},\sigma^2)\quad  \text{for } t=1,2,\cdots,n
\end{align*}
where the categorical distribution (noted $Cat(\pi_j)$) is a discrete distribution on the set of $\{1,\cdots,J\}$ (because $dim(\pi_j)=J$) where the probability of each outcome is specified by $\pi_j$. For example, if $x\sim Cat(\pi_j)$, then $x=i$ with probability $\pi_{ji}$. We can also represent this finite state-space HMM with the following graphical model:
\smallbreak
\begin{tikzpicture}[->,>=stealth',shorten >=1pt,auto,node distance=2cm,
                    semithick]
	
	\node[randomVariable]        (X1)                      {$x_{1}$};
	\node[randomVariable]        (X2) [right=0.5cm of X1]  {$x_{2}$};
	\node[randomVariable]         (X3) [right=0.5cm of X2]  {$x_{3}$};
    \node[observation]           (Y1) [below=0.7cm of X1]  {$y_{1}$};
	\node[observation]           (Y2) [right=0.5cm of Y1]  {$y_{2}$};
	\node[observation]          (Y3) [right=0.5cm of Y2]  {$y_{3}$};
	\node[param]         (Xwhite) [right=0.5cm of X3]  {$\cdots$};
	\node[param]        (Ywhite) [right=0.5cm of Y3]  {$\cdots$};
	\node[randomVariable]        (X4) [right=0.5cm of Xwhite]  {$x_{T}$};
	\node[observation]           (Y4) [right=0.5cm of Ywhite]  {$y_{T}$};
	\node[param]        (theta) [above left=0.3cm and 0.6cm of Y1]  {$\theta_j$};
	\node[param]         (pi) [above left=0.3cm and 0.6cm of X1]  {$\pi_j$};

  \path 
	      (X1) edge             node    {}         (X2)
				(X2) edge             node    {}         (X3)
				(X3) edge             node    {}         (Xwhite)
				(Xwhite) edge         node    {}         (X4)
				(X1) edge             node    {}         (Y1)
				(X2) edge             node    {}         (Y2)
				(X3) edge             node    {}         (Y3)
				(X4) edge             node    {}         (Y4)
				(theta.east) edge             node    {}         (Y1.north)
				(theta.east) edge             node    {}         (Y2.north)
				(theta.east) edge             node    {}         (Y3.north)
				(theta.east) edge             node    {}         (Y4.north)
				(pi.east) edge             node    {}         (X1.north)
				(pi.east) edge             node    {}         (X2.north)
				(pi.east) edge             node    {}         (X3.north)
				(pi.east) edge             node    {}         (X4.north);
\end{tikzpicture}
\smallbreak
In this representation, nodes without borders are fixed parameters whereas nodes with borders are random variables. Random variables with a white background are hidden and the ones with a grey background are the ones that we observe.

\smallbreak
Our goal is to solve the power disaggregation problem in a non intrusive way. This means that we suppose having no information about the devices that generated the aggregate signal. Therefore, we suppose that we do not know $\pi_j$ and $\theta_j$ for $j=1,\cdots,J$. To modelize this uncertainty, we will use the Bayesian framework instead of the Inferential one. 
\smallbreak
To do so, we now treat $(\pi_j)_{1\leq j\leq J}$ and $(\theta_j)_{1\leq j\leq J}$ as random variables with given priors. In the context of our application, the prior on $(\theta_j)_{1\leq j\leq J}$ modelize the uncertainty on the version of device. For example, the mean power consumption of an operating mode of a device can vary on the version of this device. However, it is often concentrated around a specific value, so we will use a normal distribution (as the prior) with hyperparameters $\mu_j$ and $\tau_j^2$. Because the $(\pi_j)_{1\leq j\leq J}$ are the rows of a transition matrix, we have the following constraints: $0\leq \pi_{ji} \leq 1$  $\forall j\in\{1,\cdots,J\},i\in\{1,\cdots,J\}$ and $\sum_{i=1}^J \pi_{ji} = 1$ $\forall j\in\{1,\cdots,J\}$ . Thus, we have to choose a prior that has a support which satisfies these constraints. The most natural distribution that comes to our mind is the Dirichlet distribution:

\begin{definition}{Dirichlet distribution}

Let $X=(X_1,\cdots,X_K)$ be a random vector with $K\in\mathbb{N}^{*}$. We say that $X$ is distributed as a Dirichlet of parameter $\alpha=(\alpha_1,\cdots,\alpha_K)\in\mathbb{R}_{+}^K$  (noted $X\sim Dir(\alpha_1,\cdots,\alpha_K)$) if for every $x\in\Delta_{K-1}=\{(t_1,\cdots,t_K): t_i \geq 0, \sum_{i=1}^K t_i =1\}$, its density (with respect to the Lebesgue measure on $\mathbb{R}^{K-1}$) is:
\[ f(x)=\frac{\Gamma(\sum_{k=1}^K \alpha_k)}{\prod_{k=1}^K \Gamma(\alpha_k)}\prod_{k=1}^K x_k^{\alpha_k -1} \text{ with } \Gamma(y)=\int_0^{+\infty} t^{y-1}e^{-t}dt \]
If we have $\alpha_i = 0$ ($i\in \{1,\cdots,K\}$), we say that $X_i$ is degenerate and we put $X_i=0$.
\end{definition}

See appendix~\ref{sec:Dirichlet} for more results on the Dirichlet distribution. The prior parameters will be computed only on specific trainning houses to stay general and to preserve our non intrusive objective, see chapter~\ref{sec:Applications} for more informations on the method. We call this model the (finite state-space) Bayesian HMM and we can summarize this way:

\smallbreak
\begingroup
\begin{center}
\renewcommand{\arraystretch}{1.5}
\begin{tabular}{c|c|c}

 & distribution & priors\\
\hline
hidden states & $x_t | x_{t-1}, (\pi_j)_j \sim Cat(\pi_{x_{t-1}})$ & $\pi_j \overset{i.i.d}{\sim} Dir(\alpha)$\\

observations & $y_t| x_t, (\theta_j)_j \overset{i.i.d}{\sim} \mathcal{N}(\theta_{x_t},\sigma^2)$ & $\theta_j\sim \mathcal{N}(\mu_j,\tau_j^2)$\\
\hline
subscripts & $t=1,\cdots,n$ & $j=1,\cdots,J$ \\

\end{tabular}
\end{center}
\endgroup
\smallbreak

We can also represent this Bayesian HMM with the following graphical model (where we put $\iota=(\mu_j,\tau_j^2)_{1\leq j\leq J}$):
\smallbreak
\begin{tikzpicture}[->,>=stealth',shorten >=1pt,auto,node distance=2cm,
                    semithick]
	
	\node[randomVariable]         (X1)                      {$x_{1}$};
	\node[randomVariable]         (X2) [right=0.5cm of X1]  {$x_{2}$};
	\node[randomVariable]         (X3) [right=0.5cm of X2]  {$x_{3}$};
	\node[observation]         (Y1) [below=0.7cm of X1]  {$y_{1}$};
	\node[observation]         (Y2) [right=0.5cm of Y1]  {$y_{2}$};
	\node[observation]         (Y3) [right=0.5cm of Y2]  {$y_{3}$};
	\node[param]         (Xwhite) [right=0.5cm of X3]  {$\cdots$};
	\node[param]         (Ywhite) [right=0.5cm of Y3]  {$\cdots$};
	\node[randomVariable]          (X4) [right=0.5cm of Xwhite]  {$x_{T}$};
	\node[observation]         (Y4) [right=0.5cm of Ywhite]  {$y_{T}$};
	\node[randomVariable]     (theta2) [above left=0.3cm and 0.6cm of Y1] {$\theta_j$};
	\node[randomVector, label={[shift={(-2.2ex,2.7ex)}]south east:J}, fit=(theta2)]  (theta)   {};
	\node[randomVariable]     (pi2) [above left=0.3cm and 0.6cm of X1] {$\pi_j$};
	\node[randomVector, label={[shift={(-2.2ex,2.7ex)}]south east:J}, fit=(pi2)]     (pi)  {};

	\node[param]         (alpha) [left=0.5cm of pi]  {$\alpha$};
	\node[param]         (lambda) [left=0.5cm of theta]  {$\iota$};

  \path 
	      (X1) edge             node    {}         (X2)
				(X2) edge             node    {}         (X3)
				(X3) edge             node    {}         (Xwhite)
				(Xwhite) edge             node    {}         (X4)
				(X1) edge             node    {}         (Y1)
				(X2) edge             node    {}         (Y2)
				(X3) edge             node    {}         (Y3)
				(X4) edge             node    {}         (Y4)
				(alpha) edge             node    {}         (pi)
				(lambda) edge             node    {}         (theta)
				(theta.east) edge             node    {}         (Y1.north)
				(theta.east) edge             node    {}         (Y2.north)
				(theta.east) edge             node    {}         (Y3.north)
				(theta.east) edge             node    {}         (Y4.north)
				(pi.east) edge             node    {}         (X1.north)
				(pi.east) edge             node    {}         (X2.north)
				(pi.east) edge             node    {}         (X3.north)
				(pi.east) edge             node    {}         (X4.north);
\end{tikzpicture}
\smallbreak
In this representation, a node with a square frame means that there are multiple random variables and the number of them is shown in the bottom-right corner of the square.

\section{Hidden Semi-Markov Model}\label{sec:HSMM}

The HMM is a powerful model which has been proven useful in many situations but it is also limited in certain ways. The fact that the hidden states are a Markov chain makes the time that we stay in a state distributed as a geometric. However, in some cases we would want to have this time following another distribution. For example, here is the histogram of the time that an air compressor stays ON:

\centerline{\includegraphics[width=\paperwidth ,height=.3\textheight]{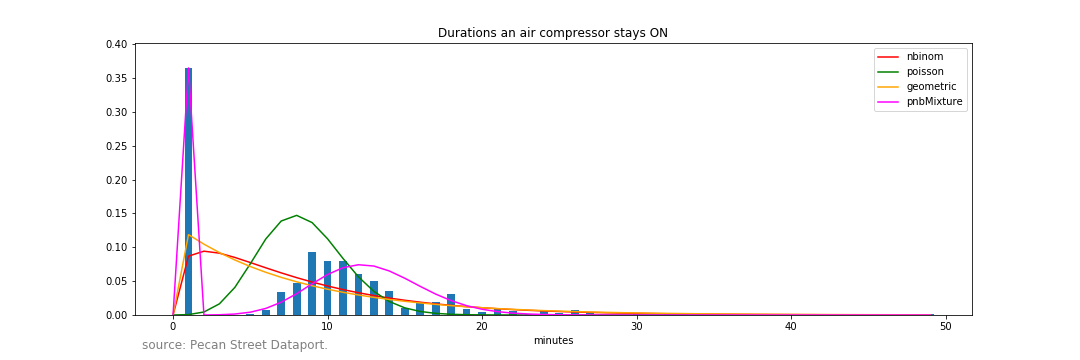}}

We can see that a mixture of a Poisson distribution and a negative binomial distribution is a better fit to these durations than a geometric one. This example motivates us to introduce the Hidden Semi-Markov Model (HSMM) which is presented in \cite{HDPHMM}. This model allows us to choose the distribution we want for the length that a device stays in an operating mode.
\smallbreak
The idea is that the hidden states will behave as a jump process where we choose the arrival time distribution. We now have \textbf{super-states} $z_1,\cdots,z_S$ that represent the jumps, they are like the previous hidden states except that now we do not allow self-transitions. The \textbf{durations} $D_1,\cdots,D_S$ are the time we stay in each state after a jump. $S$ represents the last jump of the process before the end of our observations $T$. We have to specify if our observations ends exactly when the process is about to jump ($T=\sum_{s=1}^S D_s$) or if they are censored. As in our control application, we will have continuously arriving observations, it seems natural to suppose that they are right-censored ($\sum_{s=1}^{S-1} D_s<T\leq\sum_{s=1}^S D_s$) as in \cite{HDPHMM}. The hidden states are entirely determined by the super-states and the durations. The first hidden state $x_1$ is equal to $z_1$ and remains the same until $x_{D_1}$, then $x_{D_1 +1}$ is equal to $z_2$ and remains the same until $x_{D_1 +D_2}$ and so on... 
\smallbreak
The durations are distributed according to a parameter $w_j$ which depends on the current super-state. This represents the fact that a device does not have the same behavior for different modes. As with the observations in the previous section, we can put any (discrete) distribution for the durations but we have chosen one which is well suited to our application: a mixture of a Poisson distribution and a negative binomial distribution. 
\smallbreak
This mixture was most of the time the best fit to our Pecan Street data. The intuition behind this distribution is that a device often stays in a mode around the same time (which is modelized by the Poisson) and sometimes it stays very long compared to the usual habit (for example the owner of the device forget that it is ON or go on vacation and does not use it for a long time, which is modelized by the negative binomial). In this case, we have $w_j=(\phi_j,\lambda_j,r_j,\varphi_j)$ where $\phi_j$ is the proportion (of Poisson) in the mixture, $\lambda_j$ is the parameter of the Poisson distribution, $r_j$ and $\varphi_j$ are the parameters of the negative binomial distribution. We note a mixture distribution as $Mixture(\varpi_{1:M},f_{1:M})$ where $M$ is the number of components and $\varpi_{1:M}$ are the weights associated to $f_{1:M}$ which are the densities of the mixture. 
\smallbreak
Because we work with the bayesian framework, we need to put a prior on these parameters. We choose to put a Beta prior on $\phi_j$ (with hyperparameters $\dot{\alpha}_j$ and $\dot{\beta}_j$), a Gamma prior on $\lambda_j$ (with hyperparameters $\ddot{\alpha}_j$ and $\ddot{\beta}_j$), a Beta prior on $\varphi_j$ (with hyperparameters $\dddot{\alpha}_j$ and $\dddot{\beta}_j$) and no prior on $r_j$. The density (with respect to the counting measure) of a duration given its parameters and the super-state is the following:
\[ p(d|(\phi_j)_j,(\lambda_j)_j, (\varphi_j)_j,z) = \phi_z\dfrac{\lambda_z^d}{d!}e^{-\lambda_z}+(1-\phi_z)\binom{d+r_z-1}{r_z-1}\varphi_z^{d-1}(1-\varphi_z)^{r_z}\]
For the prior on the transition matrix (of the super-states), we need to account the fact that there are no more self-transitions. We note $\pi_j:=(\pi_{j,-j},\pi_{j,j})$ with $\pi_{j,-j}:=(\pi_{j,1},\cdots,\pi_{j,j-1},\pi_{j,j+1},\cdots,\pi_{j,J})$ and $\pi_{j,j}=0$. The prior is now on $\pi_{j,-j}$ and we use an $J-1$ dimensionnal Dirichlet with hyperparameter $\alpha_{-j}:=(\alpha_1,\cdots,\alpha_{j-1},\alpha_{j+1},\cdots,\alpha_J)$. We can sum up the model this way:
\smallbreak
\begingroup
\renewcommand{\arraystretch}{1.5}
\begin{tabular}{c|c|c}

 & distribution & priors\\
\hline
hidden states & $x_t=z_s \quad z_s | z_{s-1}, (\pi_j)_j \sim Cat(\pi_{z_{s-1}})$ & $\pi_{j,-j} \overset{i.i.d}{\sim} Dir(\alpha_{-j})$ \\[3pt]

durations & $\begin{aligned}[c] D_s | z_s,(w_j)_j &\sim Mixture\left((\phi_{z_s},1-\phi_{z_s}),(f_1,f_2)\right) \\
f_1 &= f_{\mathcal{P}(\lambda_{z_s})}\quad f_2=f_{NegBin(r_{z_s},\varphi_{z_s})}\end{aligned}$ & $\begin{aligned}[c] \phi_{j} &\sim Beta(\dot{\alpha}_j,\dot{\beta}_j) \\
\lambda_{j} &\sim Gamma(\ddot{\alpha}_j,\ddot{\beta}_j) \\
\varphi_{j} &\sim Beta(\dddot{\alpha}_j,\dddot{\beta}_j) \end{aligned}$
\\

observations & $y_t| z_s, (\theta_j)_j \overset{i.i.d}{\sim} \mathcal{N}(\theta_{z_s},\sigma^2)$ & $\theta_j\sim \mathcal{N}(\mu_j,\tau_j^2)$\\

\hline
subscripts & $\begin{aligned}[c]
s&=1,\cdots,S \\
t&=t_s^1,\cdots,t_s^2 \text{ with} \\
t_s^2=\sum_{l\leq s}&D_{l},\: t_s^1=t_{s-1}^2 + 1,\: t_0^2=0
\end{aligned}$ & $j=1,\cdots,J$ \\

\end{tabular}
\endgroup
\medbreak
The notations $t_s^1$ and $t_s^2$ represent the start and the end of the $s^{th}$ block of hidden states. A block of hidden states means the sequence of hidden states that are tied to a super-state. So we could also define $t_s^1$ and $t_s^2$ as $t_s^1=min\{t:x_t =z_s\}$ and $t_s^2=max\{t:x_t =z_s\}$. We can represent this Bayesian HSMM with the following graphical model:
\smallbreak
\begin{tikzpicture}[->,>=stealth',shorten >=1pt,auto,node distance=2cm,
                    semithick]
	
	\node[randomVariable]         (Z1)                      {$z_{1}$};
	\node[randomVariable]        (Z2) [right=2cm of Z1]  {$z_{2}$};
	\node[randomVariable]    (D1)[below left=0.5cm and 0.6cm of Z1]{$D_{1}$};
	\node[randomVariable]        (D2) [right=2cm of D1]  {$D_{2}$};
	\node[param]         (Zwhite) [right=0.8cm of Z2]  {$\cdots$};
	\node[param]         (X2) [below=1.7cm of Z1]  {$\cdots$};
	\node[param]         (X5) [below=1.7cm of Z2]  {$\cdots$};
	\node[param]         (Y2) [below=1.2cm of X2]  {$\cdots$};
	\node[param]         (Y5) [below=1.2cm of X5]  {$\cdots$};
	\node[randomVariable]         (Z4) [right=0.5cm of Zwhite]  {$z_{S}$};
	\node[randomVariable] (D4) [below left=0.5cm and 0.6cm of Z4]  {$D_{S}$};
	\node[param]         (X8) [below=1.7cm of Z4]  {$\cdots$};
	\node[param]         (Y8) [below=1.2cm of X8]  {$\cdots$};
	\node[observation]         (Y1) [left=0.1cm of Y2]  {$y_{t_1^1}$};
	\node[observation]         (Y3) [right=0.1cm of Y2]  {$y_{t_1^2}$};
	\node[observation]         (Y4) [left=0.1cm of Y5]  {$y_{t_2^1}$};
	\node[observation]         (Y6) [right=0.1cm of Y5]  {$y_{t_2^2}$};
	\node[observation]         (Y7) [left=0.1cm of Y8]  {$y_{t_S^1}$};
	\node[observation]         (Y9) [right=0.1cm of Y8]  {$y_{t_S^2}$};
	\node[randomVariable]         (X1) [left=0.1cm of X2]  {$x_{t_1^1}$};
	\node[randomVariable]         (X3) [right=0.1cm of X2]  {$x_{t_1^2}$};
	\node[randomVariable]         (X4) [left=0.1cm of X5]  {$x_{t_2^1}$};
	\node[randomVariable]         (X6) [right=0.1cm of X5]  {$x_{t_2^2}$};
	\node[randomVariable]         (X7) [left=0.1cm of X8]  {$x_{t_S^1}$};
	\node[randomVariable]         (X9) [right=0.1cm of X8]  {$x_{t_S^2}$};
	\node[randomVariable]         (theta2) [above left=0.5cm and 1cm of Y1]  {$\theta_j$};
	\node[randomVector, label={[shift={(-2.2ex,2.7ex)}]south east:J}, fit=(theta2)]  (theta)   {};
	\node[randomVariable]         (pi2) [above=3cm of theta]  {$\pi_j$};
	\node[randomVector, label={[shift={(-2.2ex,2.7ex)}]south east:J}, fit=(pi2)]  (pi)   {};
	\node[randomVariable]         (w2) [above=1.45cm of theta]  {$w_j$};
	\node[randomVector, label={[shift={(-2.2ex,2.7ex)}]south east:J}, fit=(w2)]  (w)   {};
	\node[param]         (alpha) [left=0.5cm of pi]  {$\alpha$};
	\node[param]         (kappa) [left=0.5cm of w]  {$\kappa$};
	\node[param]         (lambda) [left=0.5cm of theta]  {$\iota$};

  \path 
	      (Z1) edge             node    {}         (Z2)
	      (Z1) edge             node    {}         (D1)
	      (Z2) edge             node    {}         (D2)
	      (Z4) edge             node    {}         (D4)
				(Z2) edge             node    {}         (Zwhite)
				(Zwhite) edge             node    {}         (Z4)
				(Z1) edge             node{}         (X1.north)
				(Z1) edge             node    {}         (X2.north)
				(Z1) edge             node    {}         (X3.north)
				(D1) edge             node{}         (X1.north)
				(D1) edge             node    {}         (X2.north)
				(D1) edge             node    {}         (X3.north)
				(Z2) edge             node {}         (X4.north)
				(Z2) edge             node    {}         (X5.north)
				(Z2) edge             node    {}         (X6.north)
				(D2) edge             node {}         (X4.north)
				(D2) edge             node    {}         (X5.north)
				(D2) edge             node    {}         (X6.north)
				(Z4) edge             node{}         (X7.north)
				(Z4) edge             node    {}         (X8.north)
				(Z4) edge             node    {}         (X9.north)
				(D4) edge             node{}         (X7.north)
				(D4) edge             node    {}         (X8.north)
				(D4) edge             node    {}         (X9.north)
				(X1) edge             node    {}         (Y1)
				(X3) edge             node    {}         (Y3)
				(X4) edge             node    {}         (Y4)
				(X6) edge             node    {}         (Y6)
				(X7) edge             node    {}         (Y7)
				(X9) edge             node    {}         (Y9)
				(alpha) edge             node    {}         (pi)
				(lambda) edge             node    {}         (theta)
				(kappa) edge             node    {}         (w)
				(theta.east) edge             node    {}         (Y1.north)
				(theta.east) edge             node    {}         (Y3.north)
				(theta.east) edge             node    {}         (Y4.north)
				(theta.east) edge             node    {}         (Y6.north)
				(theta.east) edge             node    {}         (Y7.north)
				(theta.east) edge             node    {}         (Y9.north)
				(pi.east) edge             node    {}         (Z1.north)
				(pi.east) edge             node    {}         (Z2.north)
				(pi.east) edge             node    {}         (Z4.north)
				(w.east) edge             node    {}         (D1.north)
				(w.east) edge             node    {}         (D2.north)
				(w.east) edge             node    {}         (D4.north);
\end{tikzpicture}
\medbreak
with $t_s^2=\sum_{l\leq s}D_{l}$, $t_s^1=t_{s-1}^2 + 1$ and $t_0^2=0$ and with $\kappa=(\dot{\alpha}_j,\dot{\beta}_j,\ddot{\alpha}_j,\ddot{\beta}_j,\dddot{\alpha}_j,\dddot{\beta}_j)_{1\leq j\leq J}$.

\section{Hierarchical Dirichlet Process - HSMM}\label{sec:HDPHSMM}

The Hidden Semi-Markov Model is more flexible than the HMM and had been proved useful in many situations. However, this model imposes us a strong assumption which is the number of operating modes (the number of different hidden states). Because we want to solve our disaggregation problem in a non-intrusive way, we suppose that we do not know the model of a device. The number of operating modes of a device can depend on its model. Therefore, we would want to modelize this uncertainty and be able to use prior knowledge about this number of modes. One solution is to have a model which allows an infinite number of modes and that learns through observations the real number. We call this model a Hierarchical Dirichlet Process - HSMM (HDP-HSMM). see appendix~\ref{sec:NonParamBayes} for more details on the Hierarchical Dirichlet Process.

\smallbreak
The idea behind this model is to have an infinite number of hidden states. Because we now have a transition matrix of infinite dimension, we need a new prior (on each of its rows) which samples an infinite number of values between $0$ and $1$ that all sum up to $1$. A first idea is to use a Dirichlet process as its realizations are discrete probability measure. However, if we choose a non-atomic distribution as the parameter of the DP, the atoms between the rows will be almost surely distincts. This means that with probability $1$ we will always jump from one row to a new one but never to a row previously visited. So, this prior cannot represent a device with a finite number of operating modes which is problematic. One solution is to choose a discrete probability measure as the parameter of the DP prior. We want it to be flexible enough to model various structures in the transitions of operating modes of a device and this leads us to choose again a Dirichlet process. If we put everything together, we have define a hierarchical Dirichlet process as our prior.
\smallbreak
M. J. Johnson proposed in \cite{HDPHMM} to use a HDP prior that can be described (Using the notations of section \ref{sec:HDP}) as:
\begin{align*}
G_0 &\sim DP(\gamma,H)\text{ with }H(\theta,w)=f(\theta)g(w) \\
G_j|G_0 &\sim DP(\alpha,G_0)
\end{align*}
where $f$ is the density of the observations parameters and $g$ is the density of the durations parameters. If we look at the definition of the HDP, the parameter $H$ does not depend on $j$ (which is the row index in our models). This means that we cannot use a different normal distribution for each row $j$ as we did in the HSMM. Therefore, we now choose $f$ (respectively $g$) as a mixture of normal densities (respectively Beta, Gamma and Beta densities). This allows us to have only one $H$ (that does not change with $j$) and still be able to use prior knowledge on the number of operating modes. Indeed, if our prior knowledge tells us that a device has $M$ operating modes, we can choose $f$ as a mixture of $M$ normal densities with each parameters suited to the mode. If we have no prior at all, we can just set $f$ as Gaussian (a mixture of only one component) with randomly chosen parameters.
\smallbreak
In order to better understand how this HDP defines the prior on our transition parameters, we are going to look at its stick-breaking representation:

\begin{align*}
\beta &\sim GEM(\gamma)\quad\quad \theta_k,w_k \overset{i.i.d}{\sim} H \text{ for }k=1,2,\cdots \\
G_0 &=\sum_{k=0}^{\infty}\beta_k \delta_{(\theta_k,w_k)} \\
\pi_j|\beta &\sim DP(\alpha, \beta) \\
G_j &=\sum_{k=0}^{\infty}\pi_{jk} \delta_{(\theta_k,w_k)}\quad \text{ for }j=1,2,\cdots 
\end{align*}

As we have described in section~\ref{sec:stickDP}, each transition row $\pi_j$ is attached to a $(\theta_j,w_j)$. We can see that the transitions are between the observations (and durations) parameters instead of the hidden states. By this, we means that we jump from $(\theta_j,w_j)$ to $(\theta_k,w_k)$ with probability $\pi_{jk}$. To link this HDP prior to our previous models, we can introduce the super-states as labels which represent the rows associated to the observations. We get the following generative process (looking only at the observations parameters):

\begin{align*}
\beta &\:\sim GEM(\gamma) & \\
\pi_j | \beta &\overset{i.i.d}{\sim} DP(\alpha,\beta) \quad (\theta_j,w_j)\sim H \quad &\text{ for } j=1,2,\cdots,\\
\bar{\pi}_j &\: =\frac{\pi_{j,-j}}{1-\pi_{j,j}} & \\
z_s | z_{s-1},(\pi_j)_j &\: \sim \bar{\pi}_{z_{s-1}}  &  \\
y_{t_s^1 : t_s^2}|z_s,(\theta_j)_j&\overset{i.i.d}{\sim} f(\theta_{z_s}) & \text{ for } s=1,\cdots,S
\end{align*}
where $\bar{\pi}_j$ is the transition row without $\pi_{j,j}$ (and renormalized) because we do not allow self-transition to be able to have different distribution on the durations. We can now see more clearly the link with the HSMM and how with this HDP prior, we have defined priors on our new transition matrix of infinite dimension. The HDP-HSMM can be summarized as:

\medbreak
\begingroup
\renewcommand{\arraystretch}{2}
\hspace*{-2.5cm}\begin{tabular}{c|c|c}

 & distribution & priors\\
\hline
hidden states & $x_t=z_s \quad z_s | z_{s-1}, (\bar{\pi}_j)_j \sim Cat(\bar{\pi}_{z_{s-1}})$ & $\bar{\pi}_j=\frac{\pi_{j,-j}}{1-\pi_{j,j}}\quad \pi_{j}| \beta \overset{i.i.d}{\sim} DP(\alpha,\beta)\quad \beta \sim GEM(\gamma)$\\

durations & $\begin{aligned}[c] D_s | z_s,(w_j)_j &\sim Mixture\left((\phi_{z_s},1-\phi_{z_s}),(f_1,f_2)\right) \\
f_1 &= f_{\mathcal{P}(\lambda_{z_s})}\quad f_2=f_{NegBin(r_{z_s},\varphi_{z_s})}\end{aligned}$ & $\begin{aligned}[c] &\phi_{j},\lambda_{j},\varphi_{j}\sim Mixture\left(\varpi_{1:M},f_{1:M}\right) \\
f_m = &f_{Beta(\dot{\alpha}_m,\dot{\beta}_m)}f_{Gamma(\ddot{\alpha}_m,\ddot{\beta}_m)}f_{Beta(\dddot{\alpha}_m,\dddot{\beta}_m)} \end{aligned} $\\

observations & $y_t| z_s, (\theta_j)_j \overset{i.i.d}{\sim} \mathcal{N}(\theta_{z_s},\sigma^2)$ & $\theta_j\sim Mixture\left(p_{1:M},\left(f_{\mathcal{N}(\mu_m,\tau_m^2)}\right)_m\right)$\\
\hline
subscripts & $\begin{aligned}[c]
s&=1,\cdots,S \\
t&=t_s^1,\cdots,t_s^2 \text{ with} \\
t_s^2=\sum_{l\leq s}&D_{l},\: t_s^1=t_{s-1}^2 + 1,\: t_0^2=0
\end{aligned}$ & $\begin{aligned}[c] j&=1,2,\cdots \\
m&=1,\cdots,M \end{aligned}$ \\

\end{tabular}
\endgroup
\medbreak
We can also represent the HDP-HSMM with the following graphical model:
\medbreak
\begin{tikzpicture}[->,>=stealth',shorten >=1pt,auto,node distance=2cm,
                    semithick]
	
	\node[randomVariable]         (Z1)                      {$z_{1}$};
	\node[randomVariable]        (Z2) [right=2cm of Z1]  {$z_{2}$};
	\node[randomVariable]    (D1)[below left=0.5cm and 0.6cm of Z1]{$D_{1}$};
	\node[randomVariable]        (D2) [right=2cm of D1]  {$D_{2}$};
	\node[param]         (Zwhite) [right=0.8cm of Z2]  {$\cdots$};
	\node[param]         (X2) [below=1.7cm of Z1]  {$\cdots$};
	\node[param]         (X5) [below=1.7cm of Z2]  {$\cdots$};
	\node[param]         (Y2) [below=1.2cm of X2]  {$\cdots$};
	\node[param]         (Y5) [below=1.2cm of X5]  {$\cdots$};
	\node[randomVariable]         (Z4) [right=0.5cm of Zwhite]  {$z_{S}$};
	\node[randomVariable] (D4) [below left=0.5cm and 0.6cm of Z4]  {$D_{S}$};
	\node[param]         (X8) [below=1.7cm of Z4]  {$\cdots$};
	\node[param]         (Y8) [below=1.2cm of X8]  {$\cdots$};
	\node[observation]         (Y1) [left=0.1cm of Y2]  {$y_{t_1^1}$};
	\node[observation]         (Y3) [right=0.1cm of Y2]  {$y_{t_1^2}$};
	\node[observation]         (Y4) [left=0.1cm of Y5]  {$y_{t_2^1}$};
	\node[observation]         (Y6) [right=0.1cm of Y5]  {$y_{t_2^2}$};
	\node[observation]         (Y7) [left=0.1cm of Y8]  {$y_{t_S^1}$};
	\node[observation]         (Y9) [right=0.1cm of Y8]  {$y_{t_S^2}$};
	\node[randomVariable]         (X1) [left=0.1cm of X2]  {$x_{t_1^1}$};
	\node[randomVariable]         (X3) [right=0.1cm of X2]  {$x_{t_1^2}$};
	\node[randomVariable]         (X4) [left=0.1cm of X5]  {$x_{t_2^1}$};
	\node[randomVariable]         (X6) [right=0.1cm of X5]  {$x_{t_2^2}$};
	\node[randomVariable]         (X7) [left=0.1cm of X8]  {$x_{t_S^1}$};
	\node[randomVariable]         (X9) [right=0.1cm of X8]  {$x_{t_S^2}$};
	\node[randomVariable]         (theta2) [above left=0.5cm and 1cm of Y1]  {$\theta_j$};
	\node[randomVector, label={[shift={(-3.3ex,2ex)}]south east:{$\infty$}}, fit=(theta2)]  (theta)   {};
	\node[randomVariable]         (pi2) [above=3cm of theta]  {$\pi_j$};
	\node[randomVector, label={[shift={(-3.3ex,2ex)}]south east:{$\infty$}}, fit=(pi2)]  (pi)   {};
	\node[randomVariable]         (w2) [above=1.45cm of theta]  {$w_j$};
	\node[randomVector, label={[shift={(-3.3ex,2ex)}]south east:{$\infty$}}, fit=(w2)]  (w)   {};
	\node[randomVariable]         (beta) [above=0.5cm of pi]  {$\beta$};
	\node[param]         (alpha) [left=0.5cm of pi]  {$\alpha$};
	\node[param]         (kappa) [left=0.5cm of w]  {$\kappa$};
	\node[param]         (lambda) [left=0.5cm of theta]  {$\iota$};
	\node[param]         (gamma) [above=1.05cm of alpha]  {$\gamma$};

  \path 
	      (Z1) edge             node    {}         (Z2)
	      (Z1) edge             node    {}         (D1)
	      (Z2) edge             node    {}         (D2)
	      (Z4) edge             node    {}         (D4)
				(Z2) edge             node    {}         (Zwhite)
				(Zwhite) edge             node    {}         (Z4)
				(Z1) edge             node{}         (X1.north)
				(Z1) edge             node    {}         (X2.north)
				(Z1) edge             node    {}         (X3.north)
				(D1) edge             node{}         (X1.north)
				(D1) edge             node    {}         (X2.north)
				(D1) edge             node    {}         (X3.north)
				(Z2) edge             node {}         (X4.north)
				(Z2) edge             node    {}         (X5.north)
				(Z2) edge             node    {}         (X6.north)
				(D2) edge             node {}         (X4.north)
				(D2) edge             node    {}         (X5.north)
				(D2) edge             node    {}         (X6.north)
				(Z4) edge             node{}         (X7.north)
				(Z4) edge             node    {}         (X8.north)
				(Z4) edge             node    {}         (X9.north)
				(D4) edge             node{}         (X7.north)
				(D4) edge             node    {}         (X8.north)
				(D4) edge             node    {}         (X9.north)
				(X1) edge             node    {}         (Y1)
				(X3) edge             node    {}         (Y3)
				(X4) edge             node    {}         (Y4)
				(X6) edge             node    {}         (Y6)
				(X7) edge             node    {}         (Y7)
				(X9) edge             node    {}         (Y9)
				(alpha) edge             node    {}         (pi)
				(lambda) edge             node    {}         (theta)
				(kappa) edge             node    {}         (w)
				(theta.east) edge             node    {}         (Y1.north)
				(theta.east) edge             node    {}         (Y3.north)
				(theta.east) edge             node    {}         (Y4.north)
				(theta.east) edge             node    {}         (Y6.north)
				(theta.east) edge             node    {}         (Y7.north)
				(theta.east) edge             node    {}         (Y9.north)
				(pi.east) edge             node    {}         (Z1.north)
				(pi.east) edge             node    {}         (Z2.north)
				(pi.east) edge             node    {}         (Z4.north)
				(w.east) edge             node    {}         (D1.north)
				(w.east) edge             node    {}         (D2.north)
				(w.east) edge             node    {}         (D4.north)
				(gamma) edge             node    {}         (beta)
				(beta) edge             node    {}         (pi);
\end{tikzpicture}

\section{Factorial HDP-HSMM}\label{sec:FactorialHDPHSMM}

All the models that we have presented were designed to describe the consumption of a device with multiple operating modes. However, our goal is to solve a disaggregation problem. This means that we only have access to the total consumption of a house which is the sum of the consumption of each device. Therefore, we need a final model that describes this aggregation using our previous model. For this purpose, \cite{HDPHMM} introduced the Factorial HDP-HSMM. The idea is that we want to represent each device by a HDP-HSMM (where the different states are the different modes of the device). So now, each device will have a hidden emission $y_n^{(k)}$ (which represents the power consumption of the device) and the observation will be the aggregate power $\bar{y}_n=\sum_{k=1}^{K}y_n^{(k)}+\epsilon$ (with $K$ the number of devices and $\epsilon$ a white noise). In order to ease the computations for the inference of the model, we suppose that the hidden emissions are independent given the hidden states:
\[ p(y_n^{(1)},\cdots,y_n^{(K)}|x_n^{(1)},\cdots,x_n^{(K)})=\prod_{k=1}^{K}p(y_n^{(k)}|x_n^{(k)})\] 
To illustrate this structure, we can draw the hidden states part with only the $x_{1:n}^{(k)}$ and the $y_{1:n}^{(k)}$ to simplify it (the transitions, durations and emissions parameters just have to be copied from what was done in the previous section for each device):
\medbreak
\begin{tikzpicture}[->,>=stealth',shorten >=1pt,auto,node distance=2cm,
                    semithick]
	
	\node[randomVariable]         (X1)                      {$x_{1}^{(1)}$};
	\node[randomVariable]         (X2) [right=2cm of X1]  {$x_{2}^{(1)}$};
	\node[randomVariable]         (X21) [above right=0.5cm and 0.3cm of X1]  {$x_{1}^{(2)}$};
	\node[randomVariable]         (X22) [right=2cm of X21]  {$x_{2}^{(2)}$};
	\node[param]         (Xwhite) [right=0.5cm of X2]  {$\cdots$};
	\node[param]         (Xwhite2) [right=0.5cm of X22]  {$\cdots$};
	\node[randomVariable]         (X4) [right=0.5cm of Xwhite]  {$x_{n}^{(1)}$};
	\node[randomVariable]         (X24) [right=0.5cm of Xwhite2]  {$x_{n}^{(2)}$};
	\node[randomVariable]         (Y1) [below=1.2cm of X1]  {$y_{1}^{(1)}$};
	\node[randomVariable]         (Y2) [below=1.2cm of X2]  {$y_{2}^{(1)}$};
	\node[randomVariable]         (Y4) [below=1.2cm of X4]  {$y_{n}^{(1)}$};
	\node[randomVariable]         (Y21) [below=1.2cm of X21]  {$y_{1}^{(2)}$};
	\node[randomVariable]         (Y22) [below=1.2cm of X22]  {$y_{2}^{(2)}$};
	\node[randomVariable]         (Y24) [below=1.2cm of X24]  {$y_{n}^{(2)}$};
	\node[observation]         (Ybar1) [below right=0.7cm and 0.5cm of Y1]  {$\bar{y}_{1}$};
	\node[observation]         (Ybar2) [below right=0.7cm and 0.5cm of Y2]  {$\bar{y}_{2}$};
	\node[observation]         (Ybar4) [below right=0.7cm and 0.5cm of Y4]  {$\bar{y}_{n}$};

  \path 
	      (X1) edge             node    {}         (X2)
				(X2) edge             node    {}         (Xwhite)
				(Xwhite) edge             node    {}         (X4)
				(X1) edge             node  {}   (Y1)
				(X2) edge             node    {}         (Y2)
				(X4) edge             node    {}         (Y4)
				(X21) edge             node    {}         (X22)
				(X22) edge             node    {}         (Xwhite2)
				(Xwhite2) edge             node    {}         (X24)
				(X21) edge             node  {}   (Y21)
				(X22) edge             node    {}         (Y22)
				(X24) edge             node    {}         (Y24)
				(Y1) edge             node    {}         (Ybar1)
				(Y2) edge             node    {}         (Ybar2)
				(Y4) edge             node    {}         (Ybar4)
				(Y21) edge             node    {}         (Ybar1)
				(Y22) edge             node    {}         (Ybar2)
				(Y24) edge             node    {}         (Ybar4);
\end{tikzpicture}

%% file: Inference.tex
\chapter{Batch Inference} \label{sec:Inference}

Now that we have defined our models, we want to infer the parameters that will make our models fit the best our data from Pecan Street. We are working in a Bayesian setting, so we want to compute Bayesian estimators of the parameters of our models. To do so, we want to sample from the posterior distribution of the model which means the joint distribution of all the parameters of the model given the observations. Because it is too complex to directly sample from this joint distribution, we are going to use a method called Gibbs sampling. This algorithm can produce approximate samples from the joint distribution if we know how to sample from each parameter given the others and the observations, see appendix~\ref{sec:GibbsSampling} for more details. During this chapter, we will first briefly present Gibbs sampling. Then, we will see how to sample from the posterior distribution of each parameter for each model introduced in the previous chapter.

\section{Bayesian HMM Inference}

In the Bayesian Hidden Markov Model, defined in section~\ref{sec:BayesianHMM}, the parameters of the model are the hidden states $x_{1:T}$, the observations parameters $(\theta_j)_j$ and the transitions parameters $(\pi_j)_j$. Therefore, we will use the following posterior joint density as the target density in a Gibbs sampler:
\[ p( x_{1:T}, (\theta_j)_j, (\pi_j)_j | y_{1:T}) \]
with $x_{1:T}=(x_1,\cdots,x_T)$ and $y_{1:T}=(y_1,\cdots,y_T)$. In order to compute the inference algorithm, we have to be able to sample from the distribution of each of these parameters given the observations and the other parameters. During this section, we will show how to sample each parameter. First we will see how to sample the hidden states which will lead us to another variation of the Gibbs sampler: the blocked Gibbs sampler. Then, we will see how to easily sample the observations and transitions parameters through conjugacy.

\subsection{Posterior of the hidden states}\label{sec:HMMPosteriorStates}

Most of the inference problems associated with the Hidden Markov models comes down to computing the posterior distribution of the hidden variables given the observations (whether it is the joint distribution or its marginals). The algorithms to solve these problems often use "forward and backward messages". Because we are working in a Bayesian setting, we are more interested in sampling from this posterior distribution than computing it. First, we will define these "messages" and then, we will see how we can use them to obtain samples from the posterior distribution. It is also important to remember that we supposed in section~\ref{sec:BayesianHMM} that $|\mathcal{X}|<\infty$. This hypothesis is essential in order to be able to compute these "messages".

\begin{definition}{Forward messages}

Let $(f_t)_{1\leq t\leq T}$ be a sequence of functions with for all $t\in\{1,\cdots,T\},\: f_t :\mathcal{X} \mapsto \mathbb{R}$.
We define the first function as $f_1(x_1):=p(x_1)p(y_1|x_1)$ and then, we define $f_t(x_t)=p(y_t|x_t)\sum_{x_{t-1}}f_{t-1}(x_{t-1})p(x_t |x_{t-1})$ recursively until $t=T$.
\end{definition}
\begin{prop}\label{eqn:forward}
\[ f_t(x_t)=p(y_{1:t},x_t)\quad \forall t=1,\cdots,T  \] 
\end{prop}
\begin{proof}
The property is true for $f_1(x_1)=p(y_1,x_1)$. Suppose that \ref{eqn:forward} is true at time $t-1$, then we have:
\begin{align*}
f_t(x_t)&=p(y_t|x_t)\sum_{x_{t-1}}p(y_{1:t-1},x_{t-1})p(x_t |x_{t-1}) \\
&=\frac{p(y_{1:t},x_t)}{p(y_{1:t-1},x_t)}\sum_{x_{t-1}}p(y_{1:t-1},x_{t-1})\frac{p(y_{1:t-1},x_t,x_{t-1})}{y_{1:t-1},x_{t-1})} \\
&=\frac{p(y_{1:t},x_t)}{p(y_{1:t-1},x_t)}\sum_{x_{t-1}}p(y_{1:t-1},x_{t-1},x_t) \\
&=p(y_{1:t},x_t)
\end{align*}
By induction, it is true for all $t\in\{1,\cdots,T\}$
\end{proof}
\begin{definition}{Backward messages}

Let $(b_t)_{1\leq t\leq T}$ be a sequence of functions with for all $t\in\{1,\cdots,T\},\: b_t :\mathcal{X} \mapsto \mathbb{R}$.
We define the last function as $b_T\equiv 1$ and then, we define $b_t(x_t)=\sum_{x_{t+1}}b_{t+1}(x_{t+1})p(y_{t+1}|x_{t+1})p(x_{t+1} |x_t)$ recursively until $t=1$.
\end{definition}
\begin{prop}\label{eqn:backward}
\[ b_t(x_t)=p(y_{t+1:T} | x_t) \quad \forall t=1,\cdots,T-1 \]\end{prop}
\begin{proof}
The property is true for 
\[ b_{T-1}(x_{T-1})=\sum_{x_T}p(y_T|x_T)p(x_T |x_{T-1})=\sum_{x_T}\frac{p(y_T,x_T,x_{T-1})}{p(x_T,x_{T-1})}\frac{p(x_T ,x_{T-1})}{p(x_{T-1})}=p(y_T | x_{T-1}) \] 
Suppose that \ref{eqn:backward} is true at time $t+1$, then we have:
\begin{align*}
b_t(x_t)&=\sum_{x_{t+1}}p(y_{t+2:T}  | x_{t+1})p(y_{t+1}|x_{t+1})p(x_{t+1} |x_t) \\
&=\sum_{x_{t+1}}\frac{p(y_{t+1:T} ,x_t,x_{t+1})}{p(y_{t+1},x_t,x_{t+1})}\frac{p(y_{t+1},x_t,x_{t+1})}{p(x_t,x_{t+1})}\frac{p(x_t,x_{t+1})}{p(x_t)} \\
&=p(y_{t+1:T}  | x_t)
\end{align*}
By induction, it is true for all $t\in\{1,\cdots,T-1\}$
\end{proof}

One example of using the messages is to compute marginals of the posterior distribution, i.e $p(x_t | y_{1:T}),\:\forall t\in\{1,\cdots,T\},\:\forall x_t\in\mathcal{X}$. To do so, we multiply both messages:
\[ p(x_t | y_{1:T}) \propto f_t(x_t)b_t(x_t) \]
because $f_t(x_t)b_t(x_t)=p(y_{1:t},x_t)p(y_{t+1:T} | x_t)=p(y_{1:t},x_t)\frac{p(y_{1:T},x_t)}{p(y_{1:t},x_t)}=p(y_{1:T},x_t)$ and $p(x_t | y_{1:T})=\frac{p(y_{1:T},x_t)}{p(y_{1:T})}\propto p(y_{1:T},x_t)$.
\smallbreak
However, these marginals do not help us directly to obtain samples from $p(x_t | y_{1:T}, x_{-t})$ which we would need to create a Gibbs sampler. Therefore, we are going to use another variation of the Gibbs sampler which is called the blocked Gibbs sampler. This method consists in sampling from the joint distribution given the observations $y_{1:T}$ instead of sampling each hidden state individually given the observations and the other hidden states $x_{-t}$. This way, we sample the whole hidden states sequence in one go. To achieve this, we are going to decompose the posterior joint distribution with Bayes' rule and graphical models properties and then we are going to see how to simulate each elements with values we know.
\begin{align*}
p(x_{1:T} | y_{1:T})&=p(x_{2:T} | y_{1:T},x_1)p(x_1 | y_{1:T})\quad\text{by Bayes' rule} \\
&=p(x_T | y_{1:T},x_{1:T-1})p(x_{T-1} | y_{1:T},x_{1:T-2})\cdots p(x_2 | y_{1:T}, x_1)p(x_1 | y_{1:T}) \\
&=p(x_T | y_{1:T},x_{T-1})p(x_{T-1} | y_{1:T},x_{T-2})\cdots p(x_2 | y_{1:T}, x_1)p(x_1 | y_{1:T})\:\text{by Markov chain properties} \\
&=p(x_T | y_{T},x_{T-1})p(x_{T-1} | y_{T-1:T},x_{T-2})\cdots p(x_2 | y_{2:T}, x_1)p(x_1 | y_{1:T})
\end{align*}
by graphical models properties. If we look at each individual elements, we can see that we can simulate from them with the transitions probabilities $p(x_t | x_{t-1})$, the likelyhood terms $p(y_t | x_t)$ and the backward messages $b_t(x_t)=p(y_{t+1:T} | x_t)$:
\begin{align*}
p(x_1 | y_{1:T})&\propto p(x_1, y_{1:T})\quad\text{by Bayes' rule} \\
&=p(x_1)p(y_{1:T}|x_1) \\
&=p(x_1)p(y_1 | y_{2:T},x_1)p(y_{2:T}|x_1) \\
&= p(x_1) p(y_1 | x_1) b_1(x_1)\quad\text{by graphical model properties} \\
\forall t\in\{2,\cdots,T\}\quad p(x_t | y_{t:T}, x_{t-1})&\propto \frac{p(x_t, x_{t-1}, y_{t:T})}{p(x_{t-1})}\quad\text{by Bayes' rule} \\
&= p(x_t | x_{t-1})p(y_{t:T}|x_t,x_{t-1}) \\
&= p(x_t | x_{t-1}) p(y_t | y_{t+1:T},x_t,x_{t-1}) p(y_{t+1:T} | x_t, x_{t-1}) \\
&= p(x_t | x_{t-1}) p(y_t | x_t) b_t(x_t)\quad\text{by graphical model properties} 
\end{align*}
In conclusion, we can sample the whole hidden state sequence by first sampling $\tilde{x}_1$ following $p(x_1 | y_{1:T})\propto p(x_1) p(y_1 | x_1) b_1(x_1)$. Then, we iterate this process by sampling $\tilde{x}_t$ following $p(x_t | y_{t:T}, \tilde{x}_{t-1})\propto p(x_t | \tilde{x}_{t-1}) p(y_t | x_t) b_t(x_t)$.

\subsection{Posterior of the observations and transitions parameters}\label{sec:HMMPosteriorObsTrans}

In order to sample the observations and transitions parameters, we will use conjugacy. In Bayesian statistics, if the posterior distribution is in the same family as the prior distribution, we say that the prior is conjugate to the likelihood function. This means that if it is easy to sample from the prior distribution, it will be easy to sample from the posterior distribution provided that you know how to compute the new parameters for the posterior distribution. In section~\ref{sec:BayesianHMM}, we carefully choose the prior distributions so that it is coherent with our application but also because they are conjugate to their likelihood functions. 
\smallbreak
First, let us look at the observations parameters. Each parameter is tied to a hidden state and thus only depends on  the observations which are associated to this hidden state. We want to be able to sample from $p(\theta_j | (y_t)_{t\in \mathcal{T}_j} )$ (with $\mathcal{T}_j=\{t : x_t = j\}$) for all $j=1,\cdots,J$ and we know that $y_t | x_t \overset{i.i.d}{\sim} \mathcal{N}(\theta_{x_t},\sigma^2)$ and that $\theta_j\sim \mathcal{N}(\mu_j,\tau_j^2)$. A Gaussian prior is conjugate to a Gaussian likelihood and we get that
\[ \theta_j | (y_t)_{t\in \mathcal{T}_j} \sim \mathcal{N}(\tilde{\mu}_j,\tilde{\tau}_j^2)\text{ with }\tilde{\mu}_j = \left( \dfrac{\mu_j}{\tau_j^2}+\dfrac{\sum_{t\in \mathcal{T}_j}y_t}{\sigma^2} \right)\times\tilde{\tau}_j^2 \text{ and } \tilde{\tau}_j^2 = \left( \dfrac{1}{\tau_j^2}+\dfrac{|\mathcal{T}_j|}{\sigma^2} \right)^{-1}\]

The $(\pi_j)_j$ are i.i.d and their posterior distribution depends only on the $(x_t)_t$. So, we only need to know how to sample from $p(\pi_j | (x_t)_{t : x_{t-1}=j})$ for all $j$. We have that $x_t\sim \pi_{x_{t-1}}$ so for all $t$ such that $x_{t-1}=j$, we have $x_t\sim Cat(\pi_j)$. Because $\pi_j \sim Dir(\alpha)$ and because the Dirichlet distribution is conjugate to the categorical distribution, we have $\pi_j | (x_t)_{s : x_{t-1}=j} \sim Dir(\alpha+c)$ with $c=(c_1,\cdots,c_N)$ and $c_i$ which is equal to the number of $x_t = i$ with $t$ such that $x_{t-1}=j$. 

\section{HSMM Inference}

The difference between the Hidden Semi-Markov Model (defined in section~\ref{sec:HSMM}) and the Bayesian HMM is that we added hidden super-states $z_{1:S}$, durations $D_{1:S}$ and their parameters $(w_j)_j$. Moreover, the hidden states are now entirely defined by the super-states and the durations. Therefore, we only need to sample these two following their posterior distribution to obtain samples of the hidden states. Thus, we will now use the following posterior joint density as the target density in a Gibbs sampler:
\[ p( z_{1:S}, D_{1:S}, (w_j)_j, (\theta_j)_j, (\pi_j)_j | y_{1:T}) \]
We have already seen how to sample from the posterior distribution of the observation and transitions parameters in section~\ref{sec:HMMPosteriorObsTrans}. So, during this section, we will first see how to sample the hidden super-states and the durations with a blocked Gibbs sampler. Then, we will see how to sample the durations parameters with a Gibbs sampler for a mixture distribution.

\subsection{Posterior of the hidden states}\label{sec:HSMMStates}

\smallbreak
We use a similar method (blocked Gibbs sampling) as in section~\ref{sec:HMMPosteriorStates} to sample from the posterior of the hidden super-states and the posterior of the durations. As previously, we will need to compute backward messages. However, we cannot use the ones defined in section~\ref{sec:HMMPosteriorStates} as the HSMM introduced durations variables. Thus, we will use the backward messages presented in \cite{HDPHMM}:

\begin{align*}
B_t(i)&:=p(y_{t+1:T}|x_t=i,F_t=1) \\
&=\sum_j B_t^{*}(j)p(x_{t+1}=j|x_t=i) \\
B_t^{*}(i)&:=p(y_{t+1:T}|x_{t+1}=i,F_t=1) \\
&=\sum_{d=1}^{T-t}B_{t+d}(i)p(D_{t+1}=d|x_{t+1}=i)p(y_{t+1:t+d}|x_{t+1}=i,D_{t+1}=d) \\
&+p(D_{t+1}>T-t|x_{t+1}=i)p(y_{t+1:T}|x_{t+1}=i,D_{t+1}>T-t) \\
B_T(i)&:=1
\end{align*}
where $F_t$ is a variable which is equal to $1$ if we jump to a new super-state at time $t+1$ (i.e, $\exists s\in\{1,\cdots,S\}$ such that $\sum_{l<=s}D_l = t$). $D_{t+1}$ is the duration variable associated to the super-state we jumped to at time $t+1$ (i.e, $D_{t+1}:=D_{s+1}$ with $s$ such that $\sum_{l<=s}D_l = t$).
\smallbreak
To sample the super-states, we are going to use the Blocked Gibbs sampler as we did in section~\ref{sec:HMMPosteriorStates}. To do so, we are going to again decompose the posterior joint distribution with Bayes' rule and graphical models properties:
\begin{align*}
p(z_{1:S} | y_{1:T})&=p(z_S | y_{1:T},z_{1:S-1})p(z_{S-1} | y_{1:T},z_{1:S-2})\cdots p(z_2 | y_{1:T}, z_1)p(z_1 | y_{1:T}) \\
&=p(z_S | y_{1:T},z_{S-1})p(z_{S-1} | y_{1:T},z_{S-2})\cdots p(z_2 | y_{1:T}, z_1)p(z_1 | y_{1:T}) \\
&=p(z_S | y_{t_{S}^1:T},z_{S-1})p(z_{S-1} | y_{t_{S-1}^1:T},z_{S-2})\cdots p(z_2 | y_{t_{2}^1:T}, z_1)p(z_1 | y_{1:T})
\end{align*}
with $t_{s}^2=\sum_{l\leq s}D_l$, $t_{s}^1=t_{s-1}^2+1$ and $t_0^2=0$. If we look at each individual elements, we can see that we can simulate from them with the transitions probabilities and the backward messages $B_{t}^{*}$:
\begin{align*}
p(z_1 | y_{1:T})&=\frac{p(z_1, y_{1:T})}{p(y_{1:T})}\propto p(z_1, y_{1:T}) = p(z_1)p(y_{1:T}|z_1) = p(z_1) B_{0}^{*}(z_1) \\
p(z_s | y_{t_{s}^1:T}, z_{s-1})&\propto \frac{p(z_s, y_{t_{s}^1:T}, z_{s-1})}{p(z_{s-1})} = p(z_s | z_{s-1})p(y_{t_{s}^1:T}|z_s,z_{s-1}) = p(z_s | z_{s-1}) B_{t_{s-1}^2}^{*}(z_s)\; \forall\: s\in\{2,\cdots,S\}
\end{align*}

To sample the durations, we can use the same method (Blocked Gibbs sampler) and tricks (Bayes' rule and graphical models properties):
\[ p(D_{1:S} | y_{1:T},z_{1:S})=\prod_{s=1}^{S}p(D_s | y_{t_s^1:T},z_s)\:\text{by graphical models properties}\]
If we look at each individual elements, we can see that we can simulate from them with likelihoods on the observations and on the durations and the backward messages $B_{t}$ and $B_{t}^{*}$:
\begin{align}
p(D_s  =d | y_{t_s^1:T},z_s)&=\frac{p(D_s=d, y_{t_s^1:T}|z_s)}{p(y_{t_s^1:T}|z_s)}\nonumber\\
&=\frac{p(D_s=d |z_s)p( y_{t_s^1:T}|D_s=d ,z_s)}{p(y_{t_s^1:T}|z_s)}\nonumber\\
&=\frac{p(D_s=d |z_s)p( y_{t_s^1:t_s^1+d-1}|D_s=d,z_s)p( y_{t_s^1+d:T}|D_s=d,z_s)}{p(y_{t_s^1:T}|z_s)}\nonumber\\
&=\frac{p(D_s=d|z_s)p( y_{t_s^1:t_s^1+d-1}|D_s = d,z_s)B_{t_s^1+d-1}(z_s)}{B_{t_{s-1}^2}^{*}(z_s)} \label{eqn:Ds}
\end{align}
In particular, \begin{equation}\label{eqn:D1}p(D_1 = d | y_{1:T},z_1)=\frac{p(D_1 = d|z_1)p( y_{1:d}|D_1 = d,z_1)B_{d}(z_1)}{B_{0}^{*}(z_1)} \end{equation}
\medbreak
In conclusion, we can sample the whole hidden state sequence by first sampling $\tilde{z}_1$ following $p(z_1 | y_{1:T})\propto p(z_1) B_{0}^{*}(z_1)$, then we sample $\tilde{D}_1$ following \eqref{eqn:D1}. We iterate this process by sampling $\tilde{z}_s$ following $p(z_s | y_{\tilde{t}_{s}^1:T}, \tilde{z}_{s-1})\propto p(z_s | \tilde{z}_{s-1}) B_{\tilde{t}_{s-1}^2}^{*}(z_s)$ and $\tilde{D}_s$ following \eqref{eqn:Ds} with $\tilde{t}_{s}^2=\sum_{l\leq s}\tilde{D}_l$ and $\tilde{t}_{s}^1=\tilde{t}_{s-1}^2+1$. Finally, we set $\tilde{x}_{\tilde{t}_s^1:\tilde{t}_s^2}=\tilde{z}_s$ $\forall s\in\{1,\cdots,S\}$.

\subsection{Posterior of the duration parameters}\label{sec:HSMMDur}

When we defined the HSMM in section~\ref{sec:HSMM}, we choose a mixture distribution as the distribution for the durations. So, to sample from the posterior of the duration parameters, we have to know how to sample from the posterior of a mixture distribution. Because we will need this result in the next sections, we will compute the posterior of a mixture distribution in a general setting.
\smallbreak
Let $\xi_1 ,\cdots, \xi_n$ be independent random variables with $\xi_i$ distributed according to the mixture distribution $Mixture(\varpi_{1:M},f_{1:M})$ for all $i=1,\cdots,n$. We use the same notation as in section~\ref{sec:HSMM} which means that $M$ is the number of components and $\varpi_{1:M}$ are the weights associated to $f_{1:M}$ which are the densities of the mixture. Let $\varpi$ be a $Dirichlet(\alpha)$-distributed random vector and suppose that each density $f_m$ of the mixure has a parameter $\vartheta_m$ which has a conjugate prior (for $f_m$). Moreover, let $Z_1 ,\cdots, Z_n$ be the labels which indicate the component associated to each observation (i.e realisation of $\xi_1 ,\cdots, \xi_n$). This means that $\forall\: i\in\{1,\cdots,n\},\:\forall\: m\in\{1,\cdots,M\}, p(Z_i = m) = \varpi_m$. In order to understand the interaction between all these variables, we can look at the graphical model:
\smallbreak
\begin{tikzpicture}[->,>=stealth',shorten >=1pt,auto,node distance=2cm,
                    semithick]
	
	\node[randomVariable]        (X1)                      {$Z_{1}$};
	\node[randomVariable]        (X2) [right=0.5cm of X1]  {$Z_{2}$};
	\node[randomVariable]         (X3) [right=0.5cm of X2]  {$Z_{3}$};
    \node[observation]           (Y1) [below=0.7cm of X1]  {$\xi_{1}$};
	\node[observation]           (Y2) [right=0.5cm of Y1]  {$\xi_{2}$};
	\node[observation]          (Y3) [right=0.5cm of Y2]  {$\xi_{3}$};
	\node[param]         (Xwhite) [right=0.5cm of X3]  {$\cdots$};
	\node[param]        (Ywhite) [right=0.5cm of Y3]  {$\cdots$};
	\node[randomVariable]        (X4) [right=0.5cm of Xwhite]  {$Z_{n}$};
	\node[observation]           (Y4) [right=0.5cm of Ywhite]  {$\xi_{n}$};
	\node[randomVariable]     (theta2) [above left=0.3cm and 0.6cm of Y1] {$\vartheta_m$};
	\node[randomVector, label={[shift={(-3ex,2.5ex)}]south east:M}, fit=(theta2)]  (theta)   {};
	\node[randomVariable]     (pi) [above left=0.3cm and 0.6cm of X1] {$\varpi$};

  \path 
				(X1) edge             node    {}         (Y1)
				(X2) edge             node    {}         (Y2)
				(X3) edge             node    {}         (Y3)
				(X4) edge             node    {}         (Y4)
				(theta.east) edge             node    {}         (Y1.north)
				(theta.east) edge             node    {}         (Y2.north)
				(theta.east) edge             node    {}         (Y3.north)
				(theta.east) edge             node    {}         (Y4.north)
				(pi.east) edge             node    {}         (X1.north)
				(pi.east) edge             node    {}         (X2.north)
				(pi.east) edge             node    {}         (X3.north)
				(pi.east) edge             node    {}         (X4.north);
\end{tikzpicture}
\smallbreak
To sample from the posterior of this mixture distribution, we will use a Gibbs sampler with the following target density:
\[ p(\varpi,(\vartheta_m)_m,Z_{1:n}|\xi_{1:n})\]
Therefore, we need to be able to sample from those three densities:
\[ p(\varpi|\xi_{1:n},(\vartheta_m)_m,Z_{1:n})=p(\varpi|Z_{1:n})\]
\[ p((\vartheta_m)_m|\xi_{1:n},Z_{1:n},\varpi)=\prod_{m=1}^M p(\vartheta_m|\xi_{1:n},Z_{1:n})=\prod_{m=1}^M p(\vartheta_m|\xi_{l\in\mathcal{Z}_m})\text{ with }\mathcal{Z}_m =\{l:Z_l=m\}\]
\[ p(Z_{1:n}|\xi_{1:n},\varpi,(\vartheta_m)_m)=\prod_{l=1}^n p(Z_{l}|\xi_{l},\varpi,(\vartheta_m)_m)\propto \prod_{l=1}^n p(\xi_{l}|(\vartheta_m)_m,Z_{l})p(Z_{l}|\varpi)\]
The first one is easy to sample from because $\varpi$ is $Dirichlet(\alpha)$-distributed which is a conjugate prior to the categorical likelihood of the labels $Z_{1:n}$. For the same reasons, the second can also be easily sampled from because we choose conjugate priors for the distributions of the parameters $(\vartheta_m)_m$. Finally, for the third density, if $M$ is not too big, we can compute $p(\xi_{l}|\vartheta_m,Z_{l}=m)p(Z_{l}=m|\varpi)=f_m(\xi_l)\varpi_m$ for all $m=1,\cdots,M$. Then, we can sample from these probabilities. Therefore, we can derive the Gibbs sampler: Algorithm~\ref{MixtureGibbs}.

\begin{algorithm}
\caption{Gibbs sampler for Mixture distribution}\label{MixtureGibbs}
Sample $x^1$ with an initial distribution.

\For{$k\in\{1,\cdots,K\}$}{

    Sample the components parameters:
    
    \For{$m\in\{1,\cdots,m\}$}{
    
           Compute $\mathcal{Z}_m^{k}=\{l:Z_l^k=m\}$
    
           Sample $\vartheta_m^{k+1}$ following $p(\vartheta_m|\xi_{l\in\mathcal{Z}_m^{k}})$
    }
    
    Sample the mixture weights $\varpi^{k+1}$ following $p(\varpi|Z_{1:n}^{k})$
    
    Sample the labels:
    
    \For{$l\in\{1,\cdots,n\}$}{
           Sample $Z_{l}^{k+1}$ following $p(\xi_{l}|(\vartheta_m^{k+1})_m,Z_{l})p(Z_{l}|\varpi^{k+1})$
    } 
    
}
\end{algorithm}

In the end, we can use Algorithm~\ref{MixtureGibbs} to sample from the posterior distribution of the duration parameters. All we need is to specify the variables using the notations from section~\ref{sec:HSMM} and the posterior distributions of the components parameters. For the HSMM, we have $D_s | z_s,(w_j)_j \sim Mixture\left((\phi_{z_s},1-\phi_{z_s}),(f_{\mathcal{P}(\lambda_{z_s})},f_{Beta(r_{z_s},\varphi_{z_s})})\right)$ with $D_1,\cdots,D_S$ which are independent given $z_1,\cdots,z_S$.  So, we have $J$ mixtures (one for each possible hidden states). For each mixture, the observations are $(D_s)_{s: z_s=j}$ and the parameters are $M=2$, $\vartheta_1=\lambda_j$, $\vartheta_2=(r_{j},\varphi_{j})$ and $\varpi=(\phi_j,1-\phi_j)$. The labels were not defined in the section, so we have to introduce new variables that we note $U_{1:S}$. For the posterior distributions of the components parameters, we use conjugacy:
\[ \lambda_j | (D_s)_{s\in\mathcal{S}_{j1}} \sim Gamma(\ddot{\alpha}_j +\sum_{s\in\mathcal{S}_{j1}}D_s,\ddot{\beta}_j+|\mathcal{S}_{j1}|) \text{ with }\mathcal{S}_{j1}=\{s: z_s =j,U_s=1 \}\]
\[ \varphi_j | (D_s)_{s\in\mathcal{S}_{j2}} \sim Beta(\dddot{\alpha}_j +\sum_{s\in\mathcal{S}_{j2}}D_s,\dddot{\beta}_j+r_j|\mathcal{S}_{j2}|) \text{ with }\mathcal{S}_{j2}=\{s: z_s =j,U_s=2 \}\]

\section{HDP-HSMM Inference}

The Hierarchical Dirichlet Process - Hidden Semi-Markov Model builds upon the HSMM by having an infinite number of possible hidden states. It also introduces an HDP prior and adds a new transition parameter $\beta$. Therefore, we will now use the following posterior joint density as the target density in a Gibbs sampler:
\[ p( z_{1:S}, D_{1:S}, (w_j)_j, (\theta_j)_j, (\pi_j)_j, \beta | y_{1:T}) \]
We have already seen how to sample from the posterior distribution of the super-sates and durations in section~\ref{sec:HSMMStates}. However, the method supposed that $|\mathcal{X}|<\infty$ to make the computations tractable. We will see during this section how we can sample from the posterior distribution of the transitions parameters using a finite-dimensional approximation of the HDP. Thus, we also get back tractability for the hidden states inference. Finally, the last difference between the HDP-HSMM and the HSMM is the distributions chosen for the observations and durations parameters. For the observations parameters, we chose a Gaussian mixture. So, to sample from its posterior distribution, we can use Algorithm~\ref{MixtureGibbs} presented in section~\ref{sec:HSMMDur}. For the durations parameters, each $w_j$ is now distributed as a mixture distribution. To sample from their posterior distribution, we can use the same method as in section~\ref{sec:HSMMDur}. However, now the components are mixture distributions themselves and thus, to sample them we have to reuse Algorithm~\ref{MixtureGibbs} instead of conjugacy.

\subsection{Posterior of the transitions parameters}

Let us recall that our transitions parameters are defined by a Hierarchical Dirichlet Process:
\begin{align*}
\beta &\:\sim GEM(\gamma) & \\
\pi_j | \beta &\overset{i.i.d}{\sim} DP(\alpha,\beta) \quad (\theta_j,w_j)\sim H \quad &\text{ for } j=1,2,\cdots,\\
\bar{\pi}_j &\: =\frac{\pi_{j,-j}}{1-\pi_{j,j}} & \\
z_s | z_{s-1},(\pi_j)_j &\: \sim \bar{\pi}_{z_{s-1}}  &  \text{ for } s=1,\cdots,S 
\end{align*}
To sample from the posterior of this HDP, we need to sample the posterior of the $(\pi_j)_j$ and the posterior of $\beta$. However, there is an infinity of $\pi_j$ and so we cannot compute all of them. There are a few differents methods to solve this problem. We are going to look at one of them which is presented in \cite{HDPHMM}: the Weak-Limit Gibbs Sampler. This method creates $L$-dimensional Dirichlet distributions which approximate the HDP when $L$ grows, then it uses Gibbs Sampling to sample from the posterior of these distributions (which approximates the sampling from the posterior of the HDP). For our model, the approximation is:
\begin{align*}
\beta &\sim Dir(\frac{\gamma}{L},\cdots,\frac{\gamma}{L}) & \\
\pi_j|\beta &\overset{i.i.d}{\sim} Dir(\alpha\beta_1,\cdots,\alpha\beta_L) \quad (\theta_j,w_j)\sim H \quad  &\text{for } j=1,\cdots,L\\
\bar{\pi}_j&=\frac{\pi_{j,-j}}{1-\pi_{j,j}} & \\
z_s| z_{s-1},(\pi_j)_j &\sim \bar{\pi}_{z_{s-1}}  \quad &\text{for } s=1,\cdots,S 
\end{align*}
The approximation of the HDP by $L$-dimensional Dirichlet distributions is presented in \cite{TehHDP} and is justified by a result of \cite{Ishwaran}. This finite approximation can now be computed, we need to determine the posterior distribution of $(\pi_j)_j$ and $\beta$:
\[p((\pi_j)_j,\beta | \gamma, \alpha, (z_s)_s)=p((\pi_j)_j |\beta, \gamma, \alpha, (z_s)_s)p(\beta | \gamma, \alpha, (z_s)_s)\]
\medbreak
\subsubsection{Sampling from posterior of $(\pi_j)_j$}
\medbreak
First, let us recall that the $\pi_1,\cdots,\pi_L$ are conditionally independent and identically distributed given $ \beta $:
\[p(\pi_1,\cdots,\pi_L |\beta, \gamma, \alpha, (z_s)_s)=\prod_{j=1}^L p(\pi_j |\beta, \alpha, (z_s)_{s\in S_j})\quad with\: S_j=\{s:z_{s-1}=j\}\]
Now, let us look at only one of them:
\begin{align*}
p(\pi_j |\beta, \alpha, (z_s)_{s\in S_j})&\propto p(\pi_j |\beta, \alpha)p((z_s)_{s\in S_j}|\pi_j, \beta, \alpha) \\
&= p(\pi_j |\beta, \alpha)p((z_s)_{s\in S_j}|\pi_j) \\
&\propto \pi_{j1}^{\alpha\beta_1 -1}\cdots \pi_{jL}^{\alpha\beta_L -1}\left( \frac{\pi_{j1}}{1-\pi_{jj}}\right)^{n_{j1}}\cdots\left( \frac{\pi_{jL}}{1-\pi_{jj}}\right)^{n_{jL}}
\end{align*}
with $n_{jk}=|\{s:z_s=k,z_{s-1}=j\}|$ (note that $n_{jj}=0$ so $\left( \frac{\pi_{jj}}{1-\pi_{jj}}\right)^{n_{jj}}=1$ because we do not allow self-transitions). Normally, the $(z_s)_{s\in S_j}$ follow a categorical law with $\pi_j$ as a Dirichlet prior and so the posterior should be also a Dirichlet distribution with new parameters (actualised with our observations) because they are conjugate. However, because we do not allow for self-transitions, a new term $\left( \frac{1}{1-\pi_{jj}}\right)^{n_{jk}}$ appears (it comes from the normalization of $\pi_j$ into $\bar{\pi}_j$) and we lose conjugacy.
\smallbreak
One way to recover conjugacy, is to use the data augmentation technique described in Van Dyk and Meng (2001). This technique can be used here by creating new variables that will compensate the additional terms in our computations. To do so, we introduce the $ (\rho_{ji})_{ji} $:
\[\rho_{ji}|\pi_{jj} \overset{i.i.d}{\sim} Geo(1-\pi_{jj}) \quad \text{for } j=1,\cdots,L \quad \text{and } i=1,\cdots,n_{j\cdot}\]
with $n_{j\cdot}=\sum_{k=1}^{L}n_{jk}$. We can also show how these new variables are placed in our graphical model (with only the HDP part):
\medbreak
\begin{tikzpicture}[->,>=stealth',shorten >=1pt,auto,node distance=2cm,
                    semithick]
	
	\tikzstyle{every state}=[fill=white,draw=black,text=black,minimum size=0.2cm]
	
	\node[state]         (X1)                      {$z_{1}$};
	\node[state]         (X2) [right=2cm of X1]  {$z_{2}$};
	
	\tikzstyle{every state}=[fill=white,draw=none,text=black,minimum size=0.2cm]
	
	\node[state]         (Xwhite) [right=0.5cm of X2]  {$\cdots$};

	\tikzstyle{every state}=[fill=white,draw=black,text=black,minimum size=0.2cm]
	
	\node[state]         (X4) [right=0.5cm of Xwhite]  {$z_{S}$};
	\node[state]         (pi) [above left=0.5cm and 1cm of X1]  {$\pi_j$};
	\node[state]         (beta) [above=0.5cm of pi]  {$\beta$};
	\node[state]         (rho) [above=0.7cm of X1]  {$\rho_{ji}$};
	
	\tikzstyle{every state}=[fill=white,draw=none,text=black,minimum size=0.2cm]
	\node[state]         (alpha) [left=0.5cm of pi]  {$\alpha$};
	\node[state]         (gamma) [left=0.5cm of beta]  {$\gamma$};

  \path 
	      (X1) edge             node    {}         (X2)
				(X2) edge             node    {}         (Xwhite)
				(Xwhite) edge             node    {}         (X4)
				(alpha) edge             node    {}         (pi)
				(gamma) edge             node    {}         (beta)
				(beta) edge             node    {}         (pi)
				(pi.east) edge             node    {}         (X1.north)
				(pi.east) edge             node    {}         (X2.north)
				(pi.east) edge             node    {}         (X4.north)
				(pi.east) edge             node    {}         (rho.west);
\end{tikzpicture}
\medbreak
Now, if we computes again the posterior of the $\pi_1,\cdots,\pi_L$ and $(\rho_{ji})_{ji}$:
\[p(\pi_1,\cdots,\pi_L, (\rho_{ji})_{ji}|\beta, \gamma, \alpha, (z_s)_s)\]
We use Gibbs sampling to sample first the $(\rho_{ji})_{ji}$:
\[p((\rho_{ji})_{ji}|(\pi_j)_j,\beta, \gamma, \alpha, (z_s)_s)=\prod_{j=1}^{L}\prod_{i=1}^{n_{j\cdot}}p(\rho_{ji}|\pi_{jj})\]
Then, we sample from the posterior of the $\pi_1,\cdots,\pi_L$:
\[p(\pi_1,\cdots,\pi_L|\beta, \gamma, \alpha, (z_s)_s, (\rho_{ji})_{ji})=\prod_{j=1}^L p(\pi_j |\beta, \alpha, (z_s)_{s\in S_j},(\rho_{ji})_{1\leq i\leq n_{j\cdot}})\]

\begin{align*}
p(\pi_j |\beta, \alpha, &(z_s)_{s\in S_j},(\rho_{ji})_{1\leq i\leq n_{j\cdot}})\\
&\propto p(\pi_j |\beta, \alpha)p((z_s)_{s\in S_j}|\pi_j, \beta, \alpha)p((\rho_{ji})_{1\leq i\leq n_{j\cdot}}|\pi_j, \beta, \alpha, (z_s)_{s\in S_j}) \\
&= p(\pi_j |\beta, \alpha)p((z_s)_{s\in S_j}|\pi_j)\prod_{i=1}^{n_{j\cdot}}p(\rho_{ji}|\pi_{jj}) \\
&\propto \pi_{j1}^{\alpha\beta_1 -1}\cdots \pi_{jL}^{\alpha\beta_L -1}\left( \frac{\pi_{j1}}{1-\pi_{jj}}\right)^{n_{j1}}\cdots\left( \frac{\pi_{jL}}{1-\pi_{jj}}\right)^{n_{jL}}\left( \prod_{i=1}^{n_{j\cdot}} \pi_{jj}^{\rho_{ji}}(1-\pi_{jj})\right) \\
&=\pi_{j1}^{\alpha\beta_1 + n_{j1} -1}\cdots\pi_{jj-1}^{\alpha\beta_{j-1} + n_{jj-1} -1}\pi_{jj}^{\alpha\beta_j + \sum_{i=1}^{n_{j\cdot}}\rho_{ji} -1}\pi_{jj+1}^{\alpha\beta_{j+1} + n_{jj+1} -1}\cdots \pi_{jL}^{\alpha\beta_L + n_{jL}-1}\\
&\propto Dir(\alpha\beta_1 + n_{j1},\cdots,\alpha\beta_{j-1} + n_{jj-1}, \alpha\beta_j + \sum_{i=1}^{n_{j\cdot}}\rho_{ji},\alpha\beta_{j+1} + n_{jj+1},\cdots,\alpha\beta_L+ n_{jL})
\end{align*}
We can see that we get back conjugacy and so we can easily sample from the posterior of the $\pi_1,\cdots,\pi_L$ and $(\rho_{ji})_{ji}$ with Gibbs sampling and known distributions.
\medbreak
\subsubsection{Sampling from posterior of $\beta$}
\medbreak
To complete the Gibbs sampling from the posterior of our HDP, we need to sample from the posterior of $\beta$ given all other variables. This means that we have to compute:
\[p(\beta | \{\pi_j \}, \gamma, \alpha, (z_s)_s, (\rho_{ji})_{ji})\]
However, this distribution is quite hard to compute because of the $\{\pi_j \}$ so we will compute the distribution without them and we will see later how it will be enough for our sampling method. We can decompose this probability in values we know with Bayes' rule:
\[p(\beta | \gamma, \alpha, (z_s)_s, (\rho_{ji})_{ji})=\dfrac{p((\pi_j)_j,\beta | \gamma, \alpha, (z_s)_s, (\rho_{ji})_{ji})}{p((\pi_j)_j | \beta, \gamma, \alpha, (z_s)_s, (\rho_{ji})_{ji})}\]
Where the denominator is something we already computed in the previous section. From now on, we define $ n_{jj}:= \sum_{i=1}^{n_{j\cdot}}\rho_{ji}$ to simplify the notations.
\begin{align*}
p((\pi_j)_j | \beta, \gamma, \alpha, (z_s)_s, (\rho_{ji})_{ji})&=\prod_{j=1}^{L}\dfrac{\Gamma(\sum_{i=1}^{L}\alpha\beta_{i}+n_{ji})}{\prod_{i=1}^{L}\Gamma(\alpha\beta_{i}+n_{ji})}\prod_{i=1}^{L}\pi_{ji}^{\alpha\beta_{i}+n_{ji}-1}\\
&=\prod_{j=1}^{L}\dfrac{\Gamma(\alpha+n_{j\cdot})}{\prod_{i=1}^{L}\Gamma(\alpha\beta_{i}+n_{ji})}\prod_{i=1}^{L}\pi_{ji}^{\alpha\beta_{i}+n_{ji}-1}
\end{align*}
For the numerator, we can also decompose it in values we know:
\begin{align*}
p((\pi_j)_j,\beta | \gamma, \alpha, (z_s)_s, (\rho_{ji})_{ji})&\propto p((\rho_{ji})_{ji} | (\pi_j)_j,\beta, \gamma, \alpha, (z_s)_s)p((z_s)_s | (\pi_j)_j,\beta, \gamma, \alpha)p((\pi_j)_j| \beta, \gamma, \alpha)p(\beta| \gamma, \alpha) \\
&= p((\rho_{ji})_{ji} | (\pi_j)_j)p((z_s)_s | (\pi_j)_j)p((\pi_j)_j| \beta, \alpha)p(\beta| \gamma)\\
&=\dfrac{\Gamma(\sum_{j=1}^{L}\dfrac{\gamma}{L})}{\prod_{i=1}^{L}\Gamma(\dfrac{\gamma}{L})}\prod_{j=1}^{L}\beta_{j}^{\dfrac{\gamma}{L}}(1-\pi_{jj})^{L}\dfrac{\Gamma(\sum_{i=1}^{L}\alpha\beta_{i})}{\prod_{i=1}^{L}\Gamma(\alpha\beta_{i})}\prod_{i=1}^{L}\pi_{ji}^{\alpha\beta_{i}+n_{ji}-1} \\
&=\dfrac{\Gamma(\gamma)}{\Gamma(\dfrac{\gamma}{L})^{L}}\prod_{j=1}^{L}\beta_{j}^{\dfrac{\gamma}{L}}(1-\pi_{jj})^{L}\dfrac{\Gamma(\alpha)}{\prod_{i=1}^{L}\Gamma(\alpha\beta_{i})}\prod_{i=1}^{L}\pi_{ji}^{\alpha\beta_{i}+n_{ji}-1}
\end{align*}
Finally, we can compute the fraction where the $ \pi_{ji} $ term is simplified and we keep only the terms that have $ \beta $ in it (because we only need to be proportional to):
\begin{align*}
p(\beta | \gamma, \alpha, (z_s)_s, (\rho_{ji})_{ji})&\propto\dfrac{\dfrac{\Gamma(\gamma)}{\Gamma(\dfrac{\gamma}{L})^{L}}\prod_{j=1}^{L}\beta_{j}^{\dfrac{\gamma}{L}}(1-\pi_{jj})^{L}\dfrac{\Gamma(\alpha)}{\prod_{i=1}^{L}\Gamma(\alpha\beta_{i})}\prod_{i=1}^{L}\pi_{ji}^{\alpha\beta_{i}+n_{ji}-1}}{\prod_{j=1}^{L}\dfrac{\Gamma(\alpha+n_{j\cdot})}{\prod_{i=1}^{L}\Gamma(\alpha\beta_{i}+n_{ji})}\prod_{i=1}^{L}\pi_{ji}^{\alpha\beta_{i}+n_{ji}-1}}\\
&=\dfrac{\Gamma(\gamma)}{\Gamma(\dfrac{\gamma}{L})^{L}}\prod_{j=1}^{L}\beta_{j}^{\dfrac{\gamma}{L}}(1-\pi_{jj})^{L}\dfrac{\Gamma(\alpha)}{\prod_{i=1}^{L}\Gamma(\alpha\beta_{i})}\dfrac{\prod_{i=1}^{L}\Gamma(\alpha\beta_{i}+n_{ji})}{\Gamma(\alpha+n_{j\cdot})}\\
&\propto\prod_{j=1}^{L}\beta_{j}^{\dfrac{\gamma}{L}}\prod_{i=1}^{L}\dfrac{\Gamma(\alpha\beta_{i}+n_{ji})}{\Gamma(\alpha\beta_{i})}
\end{align*}
As with the sampling of the $ \pi_{j} $, we can see that we have nearly a Dirichlet distribution (we have $ \prod_{j=1}^{L}\beta_{j}^{\dfrac{\gamma}{L}} $) but the term $\prod_{i=1}^{L}\dfrac{\Gamma(\alpha\beta_{i}+n_{ji})}{\Gamma(\alpha\beta_{i})}$ breaks it. In the same way as the previous section, we will use a data augmentation technique. This means that we are going to introduce new variables $(m_{kj})_{1\leq k,j \leq L}$ that are quick to compute and that give us back the Dirichlet distribution. These new variables follow this distribution:
\[ p(m_{kj} | \beta, \alpha, (z_s)_s, (\rho_{ji})_{ji})=\frac{\Gamma(\alpha\beta_{j})}{\Gamma(\alpha\beta_{j}+n_{kj})}|s(n_{kj},m_{kj})|(\alpha\beta_{j})^{m_{kj}} \]
Where $ n_{kj} $ are the same as $ n_{ji} $ from before (this means that $ n_{kj} $ is the number of transitions from $k$ to $j$ in the $(z_s)_s$ with $ n_{kk}:= \sum_{i=1}^{n_{k\cdot}}\rho_{ki}$) and $|s(n_{kj},m_{kj})|$ are unsigned Stirling numbers of the first kind. This distribution is also called the "Antoniak equation" and is due to Antoniak (1974). We can as with the $(\rho_{ji})_{ji}$ show the place of our new variables in our graphical model (the HDP part):
\medbreak
\begin{tikzpicture}[->,>=stealth',shorten >=1pt,auto,node distance=2cm,
                    semithick]
	
	\tikzstyle{every state}=[fill=white,draw=black,text=black,minimum size=0.2cm]
	
	\node[state]         (X1)                      {$z_{1}$};
	\node[state]         (X2) [right=2cm of X1]  {$z_{2}$};
	
	\tikzstyle{every state}=[fill=white,draw=none,text=black,minimum size=0.2cm]
	
	\node[state]         (Xwhite) [right=0.5cm of X2]  {$\cdots$};

	\tikzstyle{every state}=[fill=white,draw=black,text=black,minimum size=0.2cm]
	
	\node[state]         (X4) [right=0.5cm of Xwhite]  {$z_{S}$};
	\node[state]         (pi) [above left=0.5cm and 1cm of X1]  {$\pi_j$};
	\node[state]         (beta) [above=0.5cm of pi]  {$\beta$};
	\node[state]         (rho) [above=0.7cm of X1]  {$\rho_{ji}$};
	\node[state]         (m) [right=1.7cm of beta]  {$m_{kj}$};
	
	\tikzstyle{every state}=[fill=white,draw=none,text=black,minimum size=0.2cm]
	\node[state]         (alpha) [left=0.5cm of pi]  {$\alpha$};
	\node[state]         (gamma) [left=0.5cm of beta]  {$\gamma$};

  \path 
	      (X1) edge             node    {}         (X2)
	      (X1) edge             node    {}         (m.south)
				(X2) edge             node    {}         (Xwhite)
				(X2) edge             node    {}         (m.south)
				(Xwhite) edge             node    {}         (X4)
				(X4) edge             node    {}         (m.south)
				(alpha) edge             node    {}         (pi)
				(alpha) edge             node    {}         (m.west)
				(gamma) edge             node    {}         (beta)
				(beta) edge             node    {}         (pi)
				(beta) edge             node    {}         (m)
				(pi.east) edge             node    {}         (X1.north)
				(pi.east) edge             node    {}         (X2.north)
				(pi.east) edge             node    {}         (X4.north)
				(pi.east) edge             node    {}         (rho.west)
				(rho) edge             node    {}         (m);
\end{tikzpicture}
\medbreak
\begin{definition}
The unsigned Stirling numbers of the first kind, written $|s(n,m)|$, count the number of permutations of $n$ elements with $m$ disjoint cycles.
\end{definition}
\begin{properties} ~ 

\begin{itemize}
\item $|s(0,0)|=|s(1,1)|=1$
\item $|s(n,0)|=0$ for $n\geq 1$
\item $|s(n,m)|=0$ for $m> n$
\item $|s(n+1,m)|=|s(n,m-1)|+n|s(n,m)|$
\end{itemize}
\end{properties}
However, the unsigned Stirling numbers of the first kind can be quite long to compute if $L$ is big, so we can use another way to compute these variables. To do so, we use Algorithm~\ref{samplem}.

\begin{algorithm}
\caption{Sampling method for the $(m_{kj})_{k,j}$}\label{samplem}
\For{$(k,j)\in\{1,\cdots,L\}^{2}$}{
	\For{$i\in\{0,\cdots,n_{kj}\}$}{
	sample $b_{i}\sim Ber\left(\dfrac{\alpha\beta_{j}}{i+\alpha\beta_{j}}\right)$
	}
	put $m_{kj}=\sum_{i=1}^{n_{kj}}b_{i}$
}
\end{algorithm}
This sampling method gives us samples from the distribution we introduced and the proof follows the explication of the "Antoniak equation" by Tom Stepleton \cite{Antoniak}. Let us compute the probability of a sequence $\textbf{b}=(b_{i})_{0\leq i\leq n-1}$ of length $n$ with $m=\sum_{i=0}^{n-1}b_{i}$, for example:
\[ p(\textbf{b}=1,1,0,1,0)=\left( \dfrac{\alpha\beta_{j}}{\alpha\beta_{j}}\right)\left( \dfrac{\alpha\beta_{j}}{\alpha\beta_{j}+1}\right)\left( \dfrac{2}{\alpha\beta_{j}+2}\right)\left( \dfrac{\alpha\beta_{j}}{\alpha\beta_{j}+3}\right)\left( \dfrac{4}{\alpha\beta_{j}+4}\right)  \]
We can factorize it in:
\[ p(\textbf{b}=1,1,0,1,0)=\underbrace{(\alpha\beta_{j})^{m}\dfrac{\Gamma(\alpha\beta_{j})}{\Gamma(\alpha\beta_{j}+n)}}_{G}\underbrace{1\cdot 1\cdot 2\cdot 1\cdot 4}_{Q}  \]
If we fix $n$ and $m$, the first part $G$ doesn't change, only the part $Q$ change. Suppose $n$ fixed, the probability that $m=k$ is equal to $G$ multiplied by the sum of all possible $Q$ sequences for this $n$ and this $k$. Let us look at this sum of possible $Q$ that we call $SQ(n,k)$: we have $SQ(0,0)=1$ because this means we sample only $b_0 \sim Ber(\dfrac{\alpha\beta_{j}}{\alpha\beta_{j}})=Ber(1)$, $SQ(1,0)=0$ because we cannot have $m=0$ as $b_0 = 1$ necessarily and $SQ(1,1)=1$ because to sample $b_0$, $b_1$ and have $m=1$ there can be only one possible $Q$ which is $Q=1\cdot 1=1$. Now, suppose that $SQ(j,l)=|s(j,l)|$ for $0\leq l \leq j \leq n-1$, then for $ 0< k < n $ we have:
\[ SQ(n,k)=SQ(n-1,k-1)+n\times SQ(n-1,k)\overset{hyp}{=}|s(n-1,k-1)|+n\times |s(n-1,k)|=|s(n,k)| \]
Because you can have a $\textbf{b}$ sequence of length $n$ which sum to $k$ with a $\textbf{b}$ sequence of length $n-1$ which sum to $k-1$ (this means the next draw is a $1$ so we multiply $Q$ by 1) or a $\textbf{b}$ sequence of length $n-1$ which sum to $k$ (this means the next draw is a $0$ so we multiply $Q$ by $n$). Moreover, we have $SQ(n,0)$ because we cannot have $m=0$ as $b_0 = 1$ necessarily and $SQ(n,n)=1$ because there is only one possible $Q=1\cdots 1 = 1$. Therefor, by induction we have proved that $SQ(n,k)=|s(n,k)|$ for all $n\in\mathbb{N}$ and for $0\leq k \leq n$. This prove that if our $(m_{kj})_{1\leq k,j \leq L}$ variables are sampled from Algorithm~\ref{samplem} then we have for $n_{kj}\in\mathbb{N}$ fixed:
\[ p(m_{kj} | \beta, \alpha, (z_s)_s, (\rho_{ji})_{ji})=G\cdot SQ(n_{kj},m_{kj})=(\alpha\beta_{j})^{m_{kj}}\frac{\Gamma(\alpha\beta_{j})}{\Gamma(\alpha\beta_{j}+n_{kj})}|s(n_{kj},m_{kj})|\]
Now that we have a simple way to sample these new variables, let us see how they can give us back the Dirichlet distribution for the posterior of $\beta$. We sample $ \beta $ with the same method as developed before but with the added $(m_{kj})_{kj}$:
\[p(\beta | \gamma, \alpha, (z_s)_s, (m_{kj})_{kj}, (\rho_{ji})_{ji})=\dfrac{p((\pi_j)_j,\beta | \gamma, \alpha, (z_s)_s, (m_{kj})_{kj}, (\rho_{ji})_{ji})}{p((\pi_j)_j | \beta, \gamma, \alpha, (z_s)_s, (m_{kj})_{kj}, (\rho_{ji})_{ji})}\]
The denominator stays the same because of graphical model properties:\begin{align*}
p((\pi_j)_j | \beta, \gamma, \alpha, (z_s)_s, (m_{kj})_{kj}, (\rho_{ji})_{ji})&=p((\pi_j)_j | \beta, \gamma, \alpha, (z_s)_s, (\rho_{ji})_{ji})\\
&=\prod_{j=1}^{L}\dfrac{\Gamma(\alpha+n_{j\cdot})}{\prod_{i=1}^{L}\Gamma(\alpha\beta_{i}+n_{ji})}\prod_{i=1}^{L}\pi_{ji}^{\alpha\beta_{i}+n_{ji}-1}
\end{align*}
For the numerator, the distribution of the $(m_{kj})_{kj}$ is introduced:
\begin{align*}
p(\{&\pi_j\},\beta | \gamma, \alpha, (z_s)_s, (m_{kj})_{kj}, (\rho_{ji})_{ji})\\
&\propto p((m_{kj})_{kj} | (\pi_j)_j,\beta, \gamma, \alpha, (z_s)_s, (\rho_{ji})_{ji} )p((\rho_{ji})_{ji} | (\pi_j)_j,\beta, \gamma, \alpha, (z_s)_s) \\
&\quad\quad \times p((z_s)_s | (\pi_j)_j,\beta, \gamma, \alpha)p((\pi_j)_j| \beta, \gamma, \alpha)p(\beta| \gamma, \alpha) \\
&= p((m_{kj})_{kj} | \beta, \alpha, (z_s)_s,(\rho_{ji})_{ji})p((\rho_{ji})_{ji} | (\pi_j)_j) p((z_s)_s | (\pi_j)_j)p((\pi_j)_j| \beta, \alpha)p(\beta| \gamma)\\
&=\dfrac{\Gamma(\sum_{j=1}^{L}\dfrac{\gamma}{L})}{\prod_{i=1}^{L}\Gamma(\dfrac{\gamma}{L})}\prod_{j=1}^{L}(\alpha\beta_{j})^{m_{\cdot j}}\beta_{j}^{\dfrac{\gamma}{L}}(1-\pi_{jj})^{L} \dfrac{\Gamma(\sum_{i=1}^{L}\alpha\beta_{i})}{\prod_{i=1}^{L}\Gamma(\alpha\beta_{i})}\prod_{i=1}^{L}\pi_{ji}^{\alpha\beta_{i}+n_{ji}-1}\dfrac{\Gamma(\alpha\beta_{j})}{\Gamma(\alpha\beta_{j}+n_{ij})}|s(n_{ij},m_{ij})|    \\
&=\dfrac{\Gamma(\gamma)}{\Gamma(\dfrac{\gamma}{L})^{L}}\prod_{j=1}^{L}\alpha^{m_{\cdot j}}\beta_{j}^{\dfrac{\gamma}{L}+m_{\cdot j}}(1-\pi_{jj})^{L}\dfrac{\Gamma(\alpha)}{\prod_{i=1}^{L}\Gamma(\alpha\beta_{i})}\prod_{i=1}^{L}\pi_{ji}^{\alpha\beta_{i}+n_{ji}-1}\dfrac{\Gamma(\alpha\beta_{j})}{\Gamma(\alpha\beta_{j}+n_{ij})}|s(n_{ij},m_{ij})|
\end{align*}
Finally, we can compute the fraction where the $ \pi_{ji} $ term is simplified and we keep only the terms that have $ \beta $ in it (because we only need to be proportional to):
\begin{align*}
p(\beta& | \gamma, \alpha, (z_s)_s, (m_{kj})_{kj}, (\rho_{ji})_{ji})\\
&\propto\dfrac{\dfrac{\Gamma(\gamma)}{\Gamma(\dfrac{\gamma}{L})^{L}}\prod_{j=1}^{L}\alpha^{m_{\cdot j}}\beta_{j}^{\dfrac{\gamma}{L}+m_{\cdot j}}(1-\pi_{jj})^{L}\dfrac{\Gamma(\alpha)}{\prod_{i=1}^{L}\Gamma(\alpha\beta_{i})}\prod_{i=1}^{L}\pi_{ji}^{\alpha\beta_{i}+n_{ji}-1}\dfrac{\Gamma(\alpha\beta_{j})}{\Gamma(\alpha\beta_{j}+n_{ij})}|s(n_{ij},m_{ij})|}{\prod_{j=1}^{L}\dfrac{\Gamma(\alpha+n_{j\cdot})}{\prod_{i=1}^{L}\Gamma(\alpha\beta_{i}+n_{ji})}\prod_{i=1}^{L}\pi_{ji}^{\alpha\beta_{i}+n_{ji}-1}}\\
&=\dfrac{\Gamma(\gamma)}{\Gamma(\dfrac{\gamma}{L})^{L}}\prod_{j=1}^{L}\alpha^{m_{\cdot j}}\beta_{j}^{\dfrac{\gamma}{L}+m_{\cdot j}}(1-\pi_{jj})^{L}\dfrac{\Gamma(\alpha)}{\Gamma(\alpha+n_{j\cdot})}\prod_{i=1}^{L}\dfrac{\Gamma(\alpha\beta_{i}+n_{ji})}{\Gamma(\alpha\beta_{i})}\dfrac{\Gamma(\alpha\beta_{j})}{\Gamma(\alpha\beta_{j}+n_{ij})}|s(n_{ij},m_{ij})|\\
&\propto\prod_{j=1}^{L}\beta_{j}^{\dfrac{\gamma}{L}+m_{\cdot j}}\prod_{i=1}^{L}\dfrac{\Gamma(\alpha\beta_{i}+n_{ji})}{\Gamma(\alpha\beta_{i})}\dfrac{\Gamma(\alpha\beta_{j})}{\Gamma(\alpha\beta_{j}+n_{ij})}\\
&=\left( \prod_{j=1}^{L}\beta_{j}^{\dfrac{\gamma}{L}+m_{\cdot j}}\right) \left( \dfrac{\Gamma(\beta_{1})^{L}\cdots \Gamma(\beta_{L})^{L}}{(\Gamma(\beta_{1})\cdots \Gamma(\beta_{L}))^{L}}\cdot\dfrac{\Gamma(\alpha\beta_{1}+n_{11})\cdots\Gamma(\alpha\beta_{L}+n_{1L})\cdots\Gamma(\alpha\beta_{1}+n_{L1})\cdots\Gamma(\alpha\beta_{L}+n_{LL})}{\Gamma(\alpha\beta_{1}+n_{11})\cdots\Gamma(\alpha\beta_{1}+n_{L1})\cdots\Gamma(\alpha\beta_{L}+n_{1L})\cdots\Gamma(\alpha\beta_{L}+n_{LL})}\right) \\
&=\prod_{j=1}^{L}\beta_{j}^{\dfrac{\gamma}{L}+m_{\cdot j}}
\end{align*}
with $m_{\cdot j}=\sum_{k=1}^{L}m_{kj}$. In the end, we can see that by introducing these new variables $(m_{kj})_{kj}$, we were able to get back the Dirichlet distribution for the posterior of $ \beta $ with new parameters: $\dfrac{\gamma}{L}+m_{\cdot 1},\cdots ,\dfrac{\gamma}{L}+m_{\cdot L}$. We have computed the posterior distributions of all the variables included in the HDP part of the model but some doesn't match with the distributions you should know for a Gibbs sampling. Indeed, the joint posterior distribution of the HDP part of our model is $p((\pi_j)_j, (\rho_{ji})_{ji}, \beta, (m_{kj})_{kj} | \gamma, \alpha, (z_s)_s)$, so a Gibbs sampling method would be Algorithm~\ref{sampleGibbsHDP}.
\begin{algorithm}
\caption{Gibbs sampling method for the posterior of the HDP part of the model}\label{sampleGibbsHDP}
sample $(\pi_j)_j\sim p((\pi_j)_j | \gamma, \alpha, (z_s)_s, (\rho_{ji})_{ji}, \beta, (m_{kj})_{kj})$

sample $(\rho_{ji})_{ji}\sim p((\rho_{ji})_{ji} | \gamma, \alpha, (z_s)_s, (\pi_j)_j, \beta, (m_{kj})_{kj})$

sample $\beta\sim p(\beta | \gamma, \alpha, (z_s)_s, (\pi_j)_j, (\rho_{ji})_{ji}, (m_{kj})_{kj})$

sample $(m_{kj})_{kj}\sim p((m_{kj})_{kj} | \gamma, \alpha, (z_s)_s, (\pi_j)_j, (\rho_{ji})_{ji}, \beta)$

\end{algorithm}
However, for $ \beta $ we only know the distribution without the $(\pi_j)_j$;  $p(\beta | \gamma, \alpha, (z_s)_s, (\rho_{ji})_{ji}, (m_{kj})_{kj})$ and for $(\rho_{ji})_{ji}$ we only know the distribution without the $(m_{kj})_{kj}$; $p((\rho_{ji})_{ji} | \gamma, \alpha, (z_s)_s, (\pi_j)_j, \beta)$. This means that our sampler is not a Gibbs sampler and looks like Algorithm~\ref{sampleNotGibbsHDP}. 
\begin{algorithm}
\caption{Sampling method for the posterior of the HDP part of the model}\label{sampleNotGibbsHDP}
sample $(\pi_j)_j\sim p((\pi_j)_j | \gamma, \alpha, (z_s)_s, (\rho_{ji})_{ji}, \beta, (m_{kj})_{kj})$

sample $(\rho_{ji})_{ji}\sim p((\rho_{ji})_{ji} | \gamma, \alpha, (z_s)_s, (\pi_j)_j, \beta)$

sample $\beta\sim p(\beta | \gamma, \alpha, (z_s)_s, (\rho_{ji})_{ji}, (m_{kj})_{kj})$

sample $(m_{kj})_{kj}\sim p((m_{kj})_{kj} | \gamma, \alpha, (z_s)_s, (\pi_j)_j, (\rho_{ji})_{ji}, \beta)$

\end{algorithm}

To solve this problem, we are going to use a Partially Collapsed Gibbs Sampler which has been introduced by David A. van Dyk and Taeyoung Park \cite{PartialGibbs}. We will use the $3$ tools (Marginalization, Permutation and Trimming) the authors explained to modify the Gibbs sampler in the way that the joint posterior distribution of our model (that we want to sample from) is still stationnary to the kernel of the Markov chain created by the new algorithm. First, we marginalize Algorithm~\ref{sampleGibbsHDP} which gives us Algorithm~\ref{sampleMargeGibbsHDP}.
\begin{algorithm}
\caption{Marginalized Gibbs sampler}\label{sampleMargeGibbsHDP}
sample $(\pi_j)_j\sim p((\pi_j)_j | \gamma, \alpha, (z_s)_s, (\rho_{ji})_{ji}, \beta, (m_{kj})_{kj})$

sample $(\rho_{ji})_{ji},(m_{kj})_{kj}\sim p((\rho_{ji})_{ji}, (m_{kj})_{kj} | \gamma, \alpha, (z_s)_s, (\pi_j)_j, \beta)$

sample $\beta, (\pi_j)_j\sim p(\beta, (\pi_j)_j | \gamma, \alpha, (z_s)_s, (\rho_{ji})_{ji}, (m_{kj})_{kj})$

sample $(m_{kj})_{kj}\sim p((m_{kj})_{kj} | \gamma, \alpha, (z_s)_s, (\pi_j)_j, (\rho_{ji})_{ji}, \beta)$
\end{algorithm}

Then we use permutation on Algorithm~\ref{sampleMargeGibbsHDP} to get Algorithm~\ref{samplePermGibbsHDP}.

\begin{algorithm}
\caption{Permuted Gibbs sampler}\label{samplePermGibbsHDP}

sample $(\rho_{ji})_{ji},(m_{kj})_{kj}\sim p((\rho_{ji})_{ji}, (m_{kj})_{kj} | \gamma, \alpha, (z_s)_s, (\pi_j)_j, \beta)$

sample $(m_{kj})_{kj}\sim p((m_{kj})_{kj} | \gamma, \alpha, (z_s)_s, (\pi_j)_j, (\rho_{ji})_{ji}, \beta)$

sample $\beta, (\pi_j)_j\sim p(\beta, (\pi_j)_j | \gamma, \alpha, (z_s)_s, (\rho_{ji})_{ji}, (m_{kj})_{kj})$

sample $(\pi_j)_j\sim p((\pi_j)_j | \gamma, \alpha, (z_s)_s, (\rho_{ji})_{ji}, \beta, (m_{kj})_{kj})$

\end{algorithm}

Finally, we use trimming to transform Algorithm~\ref{samplePermGibbsHDP} into Algorithm~\ref{sampleTrimGibbsHDP}.

\begin{algorithm}
\caption{Trimmed Gibbs sampler}\label{sampleTrimGibbsHDP}

sample $(\rho_{ji})_{ji}\sim p((\rho_{ji})_{ji} | \gamma, \alpha, (z_s)_s, (\pi_j)_j, \beta)$

sample $(m_{kj})_{kj}\sim p((m_{kj})_{kj} | \gamma, \alpha, (z_s)_s, (\pi_j)_j, (\rho_{ji})_{ji}, \beta)$

sample $\beta\sim p(\beta | \gamma, \alpha, (z_s)_s, (\rho_{ji})_{ji}, (m_{kj})_{kj})$

sample $(\pi_j)_j\sim p((\pi_j)_j | \gamma, \alpha, (z_s)_s, (\rho_{ji})_{ji}, \beta, (m_{kj})_{kj})$

\end{algorithm}

In the end, we can follow Algorithm~\ref{sampleTrimGibbsHDP} because this method only needs the distributions we computed and David A. van Dyk and Taeyoung Park \cite{PartialGibbs} explain that this new algorithm will alter the kernel of the Markov chain created but not its stationnary distribution compared to the original Gibbs sampler. We note that the proof for this last property is not clearly given by the authors and we should later get back to this point to write it ourselves. Supposing this property is true, our algorithm should gives us samples from the joint posterior distribution we wanted. We can developped Algorithm~\ref{sampleTrimGibbsHDP} with the distributions we computed during last sections to get Algorithm~\ref{sampleHDP}.

\begin{algorithm}
\caption{Sampling method for the posterior of the HDP part of the model}\label{sampleHDP}
\For{$j\in\{1,\cdots,L\}$}{
	\For{$i\in\{1,\cdots,n_{j\cdot}\}$}{
	sample $\rho_{ji}|\pi_{jj}\sim Geo(1-\pi_{jj})$
	}
	put $n_{jj}=\sum_{i=1}^{n_{j\cdot}}\rho_{ji}$
}
\For{$k\in\{1,\cdots,L\}$}{
	\For{$j\in\{1,\cdots,L\}$}{
	sample $m_{kj}$ following Algorithm~\ref{samplem}
	}
}
sample $\beta\sim Dir(\dfrac{\gamma}{L}+m_{\cdot 1}, \cdots ,\dfrac{\gamma}{L}+m_{\cdot L})$

\For{$j\in\{1,\cdots,L\}$}{
	sample $\pi_{j}\sim Dir(\alpha\beta_{1}+n_{j1},\cdots ,\alpha\beta_{L}+n_{jL})$
}

\end{algorithm}

The $ (m_{jk})_{jk} $ variables can seem to have appeared out of nowhere but they are motivated by a different representation of a Hierarchical Dirichlet Process. Indeed, these variables should represents the table counts in a Chinese Restaurant Franchise which is another representation of the HDP that we saw in section~\ref{sec:CRF}.

%% file: Online_inference.tex
\chapter{Online inference}\label{OnlineInference}

In the previous part, we developped a method to infer the parameters of our HDP-HSMM. However, this method is essentially based on a variant of Gibbs sampling and it uses block sampling to sample the whole sequence of hidden states at once. This means that it will not be efficient with data arriving continuously. Suppose we use the method on observations $y_1,\cdots,y_n$, then we get a new observation $y_{n+1}$, we'll have to use the method on observations $y_1,\cdots,y_{n+1}$. Therefore, the time needed to infer the parameters of the model will grow with each new observations to the point where it is no more sustainable. This is a major drawback as we want to combine the desaggregation problem with control, so we are in a context of continuously arriving observations and we want that the computational time needed to desaggregate does not depend on how long the process has been active.
\smallbreak
To solve this problem, we want to use online learning to infer the parameters of our model. Ample literature can be find on particle filters. These methods suppose that data is structered as an HMM (which matches with our model) and are broadly speaking Monte-Carlo estimators of the posterior distribution of the hidden states that update quickly (by quickly we mean that each update does not depend on how long it has been running) as each new observation arrives.
\smallbreak
We will first present the Sequential Monte Carlo (SMC) framework that A. Doucet introduced in \cite{Doucet} and explain some of the algorithms displayed in the paper that we believe are useful to understand Particle filters and the building blocks of our final algorithm. Then, we will use the method of parameters estimation explained by A. Rodriguez \cite{Rodriguez} in the SMC framework to develop a Particle filter for Bayesian HMM under certain conditions. Finally, we will extend the latter algorithm to Factorial Bayesian HMM with the assumption that the chains behave independantly.

\section{Sequential Monte Carlo}

This section is a general presentation of \cite{Doucet}, its purpose is to introduce the Particles filters to the reader, to show some intuition on the object and how it behaves and to explain the building blocks that will help us construct our final algorithm. More detailed explications, theoritical results or other varieties of the particles filters algorithms that fit in the SMC framework can be find in \cite{Doucet}. Before talking about SMC, we will make a brief review of the classic Monte Carlo method and importance sampling. More informations on Monte Carlo methods can be found in \cite{MonteCarlo}.
 
\subsection{Monte Carlo and importance sampling}

The Sequential Monte Carlo framework makes the assumption that data follows a specific structure: we collect observations in sequence (usually it is associated with time) that are independant given hidden states. This mirrors the context of the HMM that we discussed earlier. However, here the states can live in a more general space and so these models are often called state space models. Using similar notations as before, let us define our state space model: let $y_{1:n}\in \mathcal{Y}^n$ be the sequence of observations, $x_{1:n}\in \mathcal{X}^n$ (with $|\mathcal{X}|<\infty$) be the sequence of hidden states. Suppose that we know the initial density $p(x_1)$, the transitions $p(x_i | x_{i-1})$ (for $i=2,\cdots,n$) and the likelihood $p(y_i | x_i)$ (for $i=1,\cdots,n$).
\smallbreak
Our goal is to compute the distribution of the hidden states given the observations: $p(x_{1:n}|y_{1:n})$, at any time $n\in\mathbb{N}^{*}$. For each $n$, we can construct an estimation of the target with a simple Monte Carlo method. To give a quick reminder of this method, it consists in estimating a value of the form $I:=\mathbb{E}[\phi(Z)]=\int \phi(z)f(z)\nu(dz)$ (where $f$ is the density of $Z$ with respect to a reference measure $\nu$) by sampling $z_1,\cdots,z_k$ according to $f$ and constructing the following estimator: $I^{MC}:= \dfrac{1}{k}\sum_{i=1}^{k}\phi(z_i)$. One can easily show that this estimator is consistent (with the law of large numbers) and it's convergence speed (with the central limit theorem). If we suppose that we know how to sample from $p(x_{1:n}|y_{1:n})$ (but we do not know how to compute it), we can use this method to estimate $p(x_{1:n}|y_{1:n})$. Instead of $Z$, we have $X_{1:n}$. $f(x_{1:n})=p(x_{1:n}|y_{1:n})$ and we choose $\phi(X_{1:n}):=\delta_{x_{1:n}}(X_{1:n})$ where $\delta_x$ is the Dirac function in $x$.
\smallbreak
However, we can not easily sample from $p(x_{1:n}|y_{1:n})$ because we can not compute directly this distribution and our Gibbs sampling method introduced before has a complexity that grows with $n$. To solve this problem, we will use a famous tool used in the field of MC methods which is importance sampling. The idea behind it is suppose that we can sample $\xi_1,\cdots,\xi_k$ from an other distribution (that we call the importance distribution) with density $q$  and that we can compute the following quantity $w(\xi):=\dfrac{f(\xi)}{q(\xi)} $ for all $\xi$, then we can create another estimator for $I$ which is $I^{IS}:=\dfrac{1}{k}\sum_{i=1}^{k}w(\xi_i)\phi(\xi_i)$ because we have:
\[\int \phi(z)f(z)\nu(dz) = \int w(\xi)\phi(\xi)q(\xi)\nu(d\xi) \]
and $I^{IS}$ is an MC estimator for the right side of the equation. Again, one can easily show that this estimator is consistent (with the law of large numbers) and it's convergence speed (with the central limit theorem). 
\smallbreak
We now have to choose an importance distribution $q$ and we would like to choose the optimal one. Usually, we define an importance distribution as optimal (noted as $q^{opt}$) if it minimizes the asymptotic variance of the estimator. In chapter 2 section 2.2.1 of \cite{MonteCarlo}, it is shown that $q^{opt}$ is defined as:
\[ q^{opt}(\xi):=\dfrac{|\phi(\xi)|f(\xi)}{\int |\phi(\xi)|f(\xi)\nu(d\xi)}\]
We can see that $q^{opt}$ depends on $\phi$. This is an issue because our goal is to estimate a distribution of the form $I_n:=\int \phi_n(z_{1:n})f_n(z_{1:n})\nu(dz_{1:n})$ for each $n\in\mathbb{N}^{*}$, so if we compute $q_{n-1}^{opt}(\xi_{1:n-1})$ (which is optimal for $\phi_{n-1}(z_{1:n-1})$) there is no guarantee that it will be equal to $q_{n}^{opt}(\xi_{1:n-1})$ (which is the marginal of $q_{n}^{opt}(\xi_{1:n})$ which is optimal for $\phi_{n}(z_{1:n})$). So, we must use another way to get optimality that does not depend on $\phi$. A. Doucet propose to minimize the variance of the importance weights $w$. This is justified by the following result:
\[ \dfrac{Var_f (\phi(z))}{Var_q (\phi(\xi)w(\xi))}\simeq \dfrac{1}{1+Var_q (w(\xi))} \]
which was proved by Liu in \cite{Liu} (page 35-36). As $Var_f (\phi(z))$ is fixed (by what we want to estimate), we can minimize the variance of our estimator by minimizing the variance of the importance weights. From now on, we will define an importance distribution as optimal if it minimizes the variance of the importance weights. It is obvious that we have $q^{opt}(\xi)=f(\xi)$ but we can't use $f$ as we supposed that we can't sample from $f$ (this is why we use Importance Sampling) but this shows that we have to find an importance distribution that is close to the target.
\smallbreak
A good approximation of $q^{opt}$ can be found most of the time but until now, we worked with $n$ fixed. This means that for each $n$, we want to estimate a new distribution $p(x_{1:n}|y_{1:n})$, we have to find a good importance distribution, sample from it and compute the estimator. Although each time step is done independantly, we would want to use the estimator at time $n-1$ to construct the estimator at time $n$ and so leverage our past work. To do so, we will use Sequential Monte Carlo methods.

\subsection{Sequential Importance Sampling}\label{sec:SIS}

From now on, we will use the same notations as in \cite{Doucet}. This means that the distribution we want to estimate is called $\pi_n(x_{1:n}):=p(x_{1:n}|y_{1:n})$, we define $\gamma_n(x_{1:n}):=p(x_{1:n}, y_{1:n})$ and $Z_n:=p(y_{1:n})$. So we have $\pi_n(x_{1:n})=\dfrac{\gamma_n(x_{1:n})}{Z_n}$. Moreover, the target is now $\gamma_n(x_{1:n})$, so we define an importance distribution $q_n(x_{1:n})$ and weights $w_n(x_{1:n}):=\dfrac{\gamma_n(x_{1:n})}{q_n(x_{1:n})}$. It doesn't matter that our target is now $\gamma_n(x_{1:n})$, even if we are interested in $\pi_n(x_{1:n})$ because we can still construct a Monte Carlo estimator for $\pi_n(x_{1:n})$:
\begin{equation} \hat{\pi}_n(x_{1:n}):=\dfrac{1}{N}\sum_{i=1}^{N}W_n^i \delta_{X_{1:n}^{i}}(x_{1:n})\text{ with } W_n^i:=\dfrac{w_n(X_{1:n}^i)}{\sum_{j=1}^{N}w_n(X_{1:n}^j)} \label{eq:pi_estimator}\end{equation}
and where $X_{1:n}^1,\cdots,X_{1:n}^N$ are sampled from $q_n(x_{1:n})$.
\smallbreak
In addition, we want to have a link between $\hat{\pi}_{n-1}(x_{1:n-1})$ and $\hat{\pi}_n(x_{1:n})$. This seems achievable because we have this result:
\[\label{gamman}\gamma_n(x_{1:n})=p(x_{1:n-1}, y_{1:n-1})p(x_n |x_{n-1})p(y_n | x_n)=\gamma_{n-1}(x_{1:n-1})p(x_n |x_{n-1})p(y_n | x_n)\tag{$\ast$}\]
because data is structered like a state-space model and so we have:
\[\pi_n(x_{1:n})=p(x_{1:n}|y_{1:n})=p(x_{1:n-1}|y_{1:n-1})\dfrac{p(x_n |x_{n-1})p(y_n | x_n)}{p(y_n | y_{1:n-1})}=\pi_{n-1}(x_{1:n-1})\dfrac{p(x_n |x_{n-1})p(y_n | x_n)}{p(y_n | y_{1:n-1})}\]
Therefore, to leverage this induction, we suppose that we can decompose $q_n(x_{1:n})$ like this:
\[q_n(x_{1:n})=q_{n-1}(x_{1:n-1})q_n(x_n|x_{1:n-1})=q_1(x_1)\prod_{i=2}^{n}q_i(x_i|x_{1:i-1})\]
where $q_i(x_i|x_{1:i-1}):=\dfrac{q_i(x_{1:i})}{q_{i-1}(x_{1:i-1})}$ and that we can easily sample from $q_1(x_1)$ and $q_i(x_i|x_{1:i-1})$ for all $i=2,\cdots,n$. This creates induction in the weights also:
\[w_n(x_{1:n})=\dfrac{\gamma_n(x_{1:n})}{q_n(x_{1:n})}=\dfrac{\gamma_{n-1}(x_{1:n-1})}{q_{n-1}(x_{1:n-1})}\dfrac{\gamma_{n}(x_{1:n})}{\gamma_{n-1}(x_{1:n-1})q_{n}(x_n|x_{1:n-1})}=w_{n-1}(x_{1:n-1})\alpha_n(x_{1:n})\]
where $\alpha_n(x_{1:n}):=p(x_n |x_{n-1})p(y_n | x_n)q_{n}(x_n|x_{1:n-1})$ because of \eqref{gamman}. All of this means that from an estimator of $\pi_{n-1}(x_{1:n-1})$ (so knowing the samples $X_{1:n-1}^1,\cdots,X_{1:n-1}^N$ and weights $W_{n-1}^1,\cdots,W_{n-1}^N$), we can easily compute an estimator of $\pi_{n}(x_{1:n})$ by sampling $X_{n}^1,\cdots,X_{n}^N$ given $X_{1:n-1}^1,\cdots,X_{1:n-1}^N$ and multiplying the weights of time step $n-1$ by $\alpha_n$. This leads to Algorithm~\ref{SIS}.

{\SetAlgoNoLine\begin{algorithm}
\caption{Sequential Importance Sampling}\label{SIS}
At time $n=1$:

\Indp\For{$i\in\{1,\cdots,N\}$}{
	sample $X_{1}^i\sim q_1(x_1)$.
	
	compute the unormalized weights $w_1(X_{1}^i)$
}
compute the normalized weights $W_1^1,\cdots,W_1^N$ with $W_1^i=\dfrac{w_1(X_{1}^i)}{\sum_{j=1}^N w_1(X_{1}^j)}$

\Indm At time $n\geq2$:

\Indp\For{$i\in\{1,\cdots,N\}$}{
	sample $X_{n}^i\sim q_n(x_n|X_{1:n-1}^i)$.
	
	compute the unormalized weights $w_n(X_{1:n}^i)=w_{n-1}(X_{1:n-1}^i)\alpha_n(X_{1:n}^i)$
}
compute the normalized weights $W_n^1,\cdots,W_n^N$ with $W_n^i=\dfrac{w_n(X_{1:n}^i)}{\sum_{j=1}^N w_n(X_{1:n}^j)}$

\end{algorithm}}

To choose a good importance sampling for this method, we can look at the optimal one. We have seen in the previous section that $q_n^{opt}=\gamma_n$, we also have $q_n^{opt}=\pi_n$ (because we only multiplied by a constant, so the variance is still $0$). This means that we should be able to sample easily from $q_1^{opt}(x_1)=\pi_1(x_1)=p(x_1|y_1)$ and 
\[ q_i^{opt}(x_i|x_{1:i-1})=\dfrac{q_i^{opt}(x_{1:i})}{q_{i-1}^{opt}(x_{1:i-1})}=\dfrac{\pi_i(x_{1:i})}{\pi_{i-1}(x_{1:i-1})}=\dfrac{p(x_{1:i}|y_{1:i})}{p(x_{1:i-1}|y_{1:i-1})}=p(x_i|x_{1:i-1},y_{1:i}) \]
which can be developed as:
\begin{align*}
p(x_i|x_{1:i-1},y_{1:i})&=\dfrac{p(x_i,y_i|x_{1:i-1},y_{1:i-1})}{p(y_i|x_{1:i-1},y_{1:i-1})} \\
&=\dfrac{p(y_i|x_{1:i},y_{1:i-1})p(x_i|x_{1:i-1},y_{1:i-1})}{p(y_i|x_{1:i-1},y_{1:i-1})} \\
&=\dfrac{p(y_i|x_{i})p(x_i|x_{i-1})}{p(y_i|x_{i-1})} 
\end{align*}
for all $i=2,\cdots,n$ (with $p(y_i|x_{i-1})=\int p(y_i|x_{i})p(x_i|x_{i-1})dx_i$). Then, we would have:
\[ \alpha_n(x_{1:n})=\dfrac{\gamma_{n}(x_{1:n})}{\gamma_{n-1}(x_{1:n-1})p(x_n|x_{1:n-1},y_{1:n})}=\dfrac{p(x_n,y_n|x_{1:n-1},y_{1:n-1})}{p(x_n,y_n|x_{1:n-1},y_{1:n-1})}p(y_n|x_{n-1})=p(y_n | x_{n-1})\]
However, most of the time we can not sample from $p(x_n|x_{1:n-1},y_{1:n})$ and so we should choose an importance distribution that estimates this distribution.
\smallbreak
By construction, this algorithm gives at each time step $n$ an estimator $\hat{\pi}_n(x_{1:n})$ (defined as in \eqref{eq:pi_estimator}) of $\pi_n(x_{1:n})$. In the same way as previous MC methods, we can easily show (with the law of large numbers) that:
\[\hat{\pi}_n(x_{1:n})\xrightarrow[N\rightarrow +\infty]{p.s} \pi_n(x_{1:n})\]

\subsection{Sequential Importance Resampling}

This previous algorithm can be improved by what we call "Resampling". At each time step, Algorithm~\ref{SIS} samples new states given particles from the previous step that were sampled from the importance distribution. However, at each time step, we also construct an estimator of the target distribution so we could resample new particles with this estimator (this is called resample because these particles were already sampled from the importance distribution and we sample them again). As the estimator is the empirical distribution associated to the target, the particles with the most weight (so the more likely hidden states given the observations) will be the most often sampled during this resampling step. This way, we only "keep" the "good" particles.
\smallbreak
The resampling step can be done with various techniques. The idea behind all of them is to sample offsprings $N_n^1,\cdots, N_n^N$ from a multinomial distribution with parameters $N$ and $W_n^{1:N}$ and then to keep $N_n^i$ times the particle $X_{1:n}^i$ for $i=1,\cdots,N$. The most famous technique is the Systematic Resampling which goes as follows: Sample $U_1 \sim \mathcal{U}[0,\frac{1}{N}]$ and define $U_i = U_1 + \frac{i-1}{N}$ for $i=2,\cdots,N$, then set $N_n^i=\left| \left\{ U_j:\sum_{k=1}^{i-1}W_n^k \leq U_j \leq \sum_{k=1}^{i}W_n^k \right\} \right|$ with the convention $\sum_{k=1}^0:=0$. Other resampling techniques can be found in \cite{Doucet}.
\smallbreak
There is also a change in the weights because of this resampling step. Indeed, at time step $n>1$, we have particles that are sampled approximately from $\pi_{n-1}(x_{1:n-1})$, then we sample the new states according to $q(x_n | x_{1:n-1})$ which means that weights are approximately equal to :
\[ w_n(x_{1:n})\simeq \dfrac{\gamma_n(x_{1:n})}{\pi_{n-1}(x_{1:n-1})q(x_n | x_{1:n-1})}=Z_{n-1}\dfrac{\gamma_n(x_{1:n})}{\gamma_{n-1}(x_{1:n-1}))q(x_n | x_{1:n-1})}\propto \alpha_n(x_{1:n})\]
All of this leads to Algorithm~\ref{SIR}.

{\SetAlgoNoLine\begin{algorithm}
\caption{Sequential Importance Resampling}\label{SIR}
At time $n=1$:

\Indp\For{$i\in\{1,\cdots,N\}$}{
	sample $X_{1}^i\sim q_1(x_1)$.
	
	compute the unormalized weights $w_1(X_{1}^i)$
}
compute the normalized weights $W_1^1,\cdots,W_1^N$ with $W_1^i=\dfrac{w_1(X_{1}^i)}{\sum_{j=1}^N w_1(X_{1}^j)}$

\For{$i\in\{1,\cdots,N\}$}{
	resample $\overline{X}_{1}^i\sim \hat{\pi}_1(x_1)$.
	
	set uniform weights $W_1^i=\dfrac{1}{N}$
} 

\medbreak

\Indm At time $n\geq2$:

\Indp\For{$i\in\{1,\cdots,N\}$}{
	sample $X_{n}^i\sim q_n(x_n|\overline{X}_{1:n-1}^i)$. Put $X_{1:n}^i = (\overline{X}_{1:n-1}^i , X_{n}^i)$
	
	compute the unormalized weights $\alpha_n(X_{1:n}^i)$
}
compute the normalized weights $W_n^1,\cdots,W_n^N$ with $W_n^i=\dfrac{\alpha_n(X_{1:n}^i)}{\sum_{j=1}^N \alpha_n(X_{1:n}^j)}$

\For{$i\in\{1,\cdots,N\}$}{
	resample $\overline{X}_{1:n}^i\sim \hat{\pi}_n(x_{1:n})$.
	
	set uniform weights $W_n^i=\dfrac{1}{N}$
}

\end{algorithm}}

\subsection{Auxiliary Particle Filtering}

The next improvement to our algorithm is to perform the resampling step before the sampling step. This means that the resampling step will use the next observation, so it will only "keep" the particles that will propagate with high probability towards the most likely hidden states given the observations. However, to be able to switch these two steps, we need to have weights that does not involve the current state in their computation. Therefore, we need to carefully choose our importance distribution. 
\smallbreak
We've shown in previous section that the optimal importance distribution is $q^{opt}_n(x_n |x_{1:n-1})=p(x_n |x_{1:n-1},y_{1:n})$ and that for this distribution, we have weights proportionnal to $\alpha_n(x_{1:n})=p(y_n | x_{n-1})$ which can be computed without knowing $x_n$. From now on, we will use this optimal importance distribution because we are going to apply these SMC algorithms in a context where we can easily sample from $p(x_n |x_{1:n-1},y_{1:n})$ (see the derivations in section~\ref{sec:BPF}), this leads to Algorithm~\ref{APF}. However, if the reader is interested in using this algorithm in a context where he can't easily sample from $p(x_n |x_{1:n-1},y_{1:n})$, Doucet explains in \cite{Doucet} how to get it around by changing the target distribution (adding a predictive likelyhood to it) and making adjusments to the weights.

{\SetAlgoNoLine\begin{algorithm}
\caption{Auxiliary Particle Filter}\label{APF}
At time $n=1$:

\Indp\For{$i\in\{1,\cdots,N\}$}{
	sample $X_{1}^i\sim q_1(x_1)$.
	
	compute the unormalized weights $w_1(X_{1}^i)$
}
compute the normalized weights $W_1^1,\cdots,W_1^N$ with $W_1^i=\dfrac{w_1(X_{1}^i)}{\sum_{j=1}^N w_1(X_{1}^j)}$

\For{$i\in\{1,\cdots,N\}$}{
	resample $\overline{X}_{1}^i\sim \hat{\pi}_1(x_1)$.
	
	set uniform weights $W_1^i=\dfrac{1}{N}$
} 

\medbreak

\Indm At time $n\geq2$:

\Indp compute the unormalized weights $\alpha_{n-1}(X_{1:n-1}^1),\cdots,\alpha_{n-1}(X_{1:n-1}^N)$ 

compute the normalized weights $W_{n-1}^1,\cdots,W_{n-1}^N$ with $W_{n-1}^i=\dfrac{\alpha_{n-1}(X_{1:n-1}^i)}{\sum_{j=1}^N \alpha_{n-1}(X_{1:n-1}^j)}$

\For{$i\in\{1,\cdots,N\}$}{
	resample $\overline{X}_{1:n-1}^i\sim \hat{\pi}_{n-1}(x_{1:n-1})$.
	
	set uniform weights $W_{n-1}^i=\dfrac{1}{N}$
}

\For{$i\in\{1,\cdots,N\}$}{
	sample $X_{n}^i\sim q_n(x_n|\overline{X}_{1:n-1}^i)$. Put $X_{1:n}^i = (\overline{X}_{1:n-1}^i , X_{n}^i)$

}

\end{algorithm}}

\section{Bayesian particle filter}\label{sec:BPF}

The Sequential Monte Carlo framework and all the algorithms that we derived in the previous sections are powerful tools to do online learning with a state-space model, yet it doesn't exactly match our needs. Let's recall that the model we constructed for our disaggregation problem (see chapter ~\ref{sec:Model}) added several layers on top of the HMM structure (durations, bayesian non parametric prior on transitions, etc...). We also want to estimate the power consumption of each device, this means that we are interested not only in the posterior law of the hidden states but also in the posterior law of the parameters. 
\smallbreak
To account for these, we will use what is called Particle Learning which was introduced by C. M. Carvalho in \cite{PL}. This method allows us to estimate the parameters of our model under the condition that there exists a low dimensionnal vector $r$ of bayesian sufficient statistics for the posterior distribution of the parameters and that we can update them recursively.

\begin{definition}
Let $X$ a random variable distributed according to $p(x|\theta)$, $\theta$ a random variable with prior $p(\theta)$ and $T(X)$ a statistic. We define $T(X)$ as a bayesian sufficient statistic if for almost every $x$, we have:
\[p(\theta | X=x) = p(\theta | T(X)=T(x))\]
\end{definition}

A. Rodriguez already used Particle Learning to derive an algorithm for the infinite HMM in \cite{PLHMM}. His model is not quite the same as ours and he integrates out the emission parameter (which we are highly interested in) so although his work has greatly inspired us, the rest of the section will have some differences in the derivations.
\smallbreak
First of all, all our work in chapter~\ref{sec:Inference} proves that there exists bayesian sufficient statistics for the posterior distribution of our parameters (essentially because we put conjugate priors) and we will use it to derive the algorithm. Therefore, we can use Particle Learning in our context and now let's present it. The main idea behind PL is to treat the parameters and their bayesian sufficient statistics as part of the state space. Let's call $r_n$ the vector of bayesian sufficient statistics (computed with observations up to time $n$) and $\zeta_n$ all the parameters (estimated with observations up to time $n$). The target distribution is now the posterior distribution of the hidden states, the statistics and the parameters: 
\[\pi_n(x_{1:n},r_{1:n},\zeta_{1:n})=\dfrac{\gamma_n(x_{1:n},r_{1:n},\zeta_{1:n})}{Z_n} \text{ with }\gamma_n(x_{1:n},r_{1:n},\zeta_{1:n}):=p(x_{1:n},r_{1:n},\zeta_{1:n}, y_{1:n})\] 
For an Auxiliary particle filter, we would then have the following importance distribution:
\[ q^{opt}_n(x_n, r_n, \zeta_n |x_{1:n-1},r_{1:n-1},\zeta_{1:n-1})=p(x_n, r_n, \zeta_n |x_{1:n-1},r_{1:n-1},\zeta_{1:n-1},y_{1:n}) \]
that we can then develop this way:
\begin{align*}
p(x_n, r_n, \zeta_n |x_{1:n-1},r_{1:n-1},\zeta_{1:n-1},y_{1:n})&=p(x_n |x_{1:n-1},r_{1:n-1},\zeta_{1:n-1},y_{1:n})\\
&\quad \times p(r_n|x_{1:n},r_{1:n-1},\zeta_{1:n-1},y_{1:n}) \\
&\quad \times p(\zeta_n |x_{1:n},r_{1:n},\zeta_{1:n-1},y_{1:n}) \\
&=p(x_n |x_{n-1},\zeta_{n-1},y_{n})\\
&\quad \times p(r_n|x_{n},r_{n-1},y_{n}) \\
&\quad \times p(\zeta_n |r_{n})
\end{align*}
because $r_n$ is a vector of bayesian sufficient statistics for $\zeta_n $ and where $p(r_n|x_{n},r_{n-1},y_{n})=\delta_{\mathcal{R}(r_{n-1},x_n,y_n)}(r_n)$ with $\mathcal{R}(r_{n-1},x_n,y_n)$ the function to update sequentially the statistics. We can also develop $p(x_n |x_{n-1},\zeta_{n-1},y_{n})$:
\[p(x_n |x_{n-1},\zeta_{n-1},y_{n})=\dfrac{p(y_n | x_n, \zeta_{n-1})p(x_n | x_{n-1},\zeta_{n-1})}{p(y_n | x_{n-1},\zeta_{n-1})} \]
with $p(y_n | x_{n-1},\zeta_{n-1})=\int p(y_n | x_n, \zeta_{n-1})p(x_n | x_{n-1},\zeta_{n-1}) dx_n$. Moreover, the importance weights are define as $\alpha_n(x_{1:n},r_{1:n},\zeta_{1:n})=p(y_n | x_{n-1},\zeta_{n-1})$. Now, each particle has a vector of statistics and parameters associated with it. During the resampling step, we will keep only the statistics and parameters associated to the particles we keep. During a sampling step, we will first sample the hidden state given the new observation, then we update the statistics with the observation and the estimated hidden state, finally we sample new parameters given the observation and the estimated hidden state (through the updated statistics). This leads to Algorithm~\ref{BayesianPF}

{\SetAlgoNoLine\begin{algorithm}
\caption{Bayesian Particle Filter}\label{BayesianPF}
At time $n=1$:

\Indp\For{$i\in\{1,\cdots,N\}$}{
	sample $X_{1}^i\sim q_1(x_1)$.
	
	compute the bayesian sufficient statistics $r_1^i = \mathcal{R}(X_1^i,y_1)$.
	
	sample the parameters $\zeta_1^i \sim p(\zeta_1 | r_1^i)$.
	
	compute the unormalized weights $w_1(X_{1}^i,r_1^i,\zeta_1^i)$.
}
compute the normalized weights $W_1^1,\cdots,W_1^N$ with $W_1^i=\dfrac{w_1(X_{1}^i,r_1^i,\zeta_1^i)}{\sum_{j=1}^N w_1(X_{1}^j,r_1^j,\zeta_1^j)}$

\For{$i\in\{1,\cdots,N\}$}{
	resample $\overline{X}_{1}^i,\overline{r}_{1}^i,\overline{\zeta}_{1}^i\sim \hat{\pi}_1(x_1,r_1,\zeta_1)$.
	
	set uniform weights $W_1^i=\dfrac{1}{N}$
} 

\medbreak

\Indm At time $n\geq2$:

\Indp compute the unormalized weights $\alpha_{n-1}(X_{1:n-1}^1,r_{1:n-1}^1,\zeta_{1:n-1}^1),\cdots,\alpha_{n-1}(X_{1:n-1}^N,r_{1:n-1}^N,\zeta_{1:n-1}^N)$ 

compute the normalized weights $W_{n-1}^1,\cdots,W_{n-1}^N$ with $W_{n-1}^i=\dfrac{\alpha_{n-1}(X_{1:n-1}^i,r_{1:n-1}^i,\zeta_{1:n-1}^i)}{\sum_{j=1}^N \alpha_{n-1}(X_{1:n-1}^j,r_{1:n-1}^j,\zeta_{1:n-1}^j)}$

\For{$i\in\{1,\cdots,N\}$}{
	resample $\overline{X}_{1:n-1}^i,\overline{r}_{1:n-1}^i,\overline{\zeta}_{1:n-1}^i\sim \hat{\pi}_{n-1}(x_{1:n-1},r_{1:n-1},\zeta_{1:n-1})$.
	
	set uniform weights $W_{n-1}^i=\dfrac{1}{N}$
}

\For{$i\in\{1,\cdots,N\}$}{
	sample $X_{n}^i\sim p(x_n|\overline{X}_{n-1}^i,\overline{\zeta}_{n-1}^i,y_{n})$. 
	
	compute the bayesian sufficient statistics $r_n^i = \mathcal{R}(r_{n-1}^i,X_n^i,y_n)$.
	
	sample the parameters $\zeta_n^i \sim p(\zeta_n | r_n^i)$.
	
	Put $X_{1:n}^i = (\overline{X}_{1:n-1}^i , X_{n}^i)$, $r_{1:n}^i = (\overline{r}_{1:n-1}^i , r_{n}^i)$ and $\zeta_{1:n}^i = (\overline{\zeta}_{1:n-1}^i , \zeta_{n}^i)$.

}
\end{algorithm}}
\smallbreak
Let's compute all the needed distributions with the Bayesian HMM (which we have presented in chapter~\ref{sec:Model} section~\ref{sec:BayesianHMM}) to show an example of this algorithm and to prove that we can easily sample from $p(x_n |x_{n-1},\zeta_{n-1},y_{n})$ (which is the condition to be able to use an Auxiliary particle filter) and that $r_n$ exists.
\smallbreak
First, in the Bayesian HMM, the parameters are the emissions (which influence the observations and which are called $\theta_j$) and the transitions (which are called $\pi_j$ and represent the rows of the transition matrix of the Markov chain), so we have $\zeta=((\theta_j)_j,(\pi_j)_j)$. The transitions have a Dirichlet prior (with parameter $\alpha$) and as the hidden states are sampled from a Categorial distribution (with parameter the $\pi_j$), the prior is conjugate and we get the following posterior:
\[ \pi_j | x_{1:n} \sim Dir(\alpha_1+n_{j1},\cdots,\alpha_K + n_{jJ}) \text{ for } j=1,\cdots,J \]
with $n_{jl}=|\{i:x_{i-1}=j,x_i=l\}|$ (the number of transitions from state $j$ to state $l$) and $J$ the number of different states. This means that we have $p(\pi_j | x_{1:n})=p(\pi_j | (n_{jl})_{1\leq l\leq J})$, so $(n_{jl})_{1\leq j,l\leq J})$ are bayesian sufficient statistics for the transition distribution. In the same way, if we put a conjugate prior on the emissions, we will be able to find a bayesian sufficient statistic most of the time. In our disaggregation problem, we will suppose Normal distributed observations ($y_n \sim \mathcal{N}(\theta_{x_n},\sigma^2)$) and use a Normal prior ($\theta_j \sim \mathcal{N}(\mu_j, \sigma_j^2)$) on the mean. This way, we have:
\[ \theta_j | y_{1:n} \sim \mathcal{N}\left(\left(\dfrac{\mu_j}{\sigma_j^2}+\dfrac{\sum_{i=1}^{n}y_i}{\sigma^2}\right)\left(\dfrac{1}{\sigma_j^2}+\dfrac{n}{\sigma^2}\right)^{-1},\dfrac{1}{\sigma_j^2}+\dfrac{n}{\sigma^2}\right) \]
and so $\sum_{i=1}^{n}y_i$ is a bayesian sufficient statistics for the emissions distribution. We have $r_n = ((n_{jl})_{1\leq j,l\leq J}), \sum_{i=1}^{n}y_i)$ and we can easily update them sequentially: 
\[ \mathcal{R}(r_{n-1},x_{n-1:n},y_n)=(r_{n-1}[1]_{x_{n-1},x_n} \pluseq 1, r_{n-1}[2] \pluseq y_n) \]
for $n\geq 2$, where $r_n[i]$ is the $i^{th}$ element of $r_n$. For $n=1$, we only update the emissions statistic and we sample the transitions with their prior. The sampling of the parameters given their statistics ($p(\zeta_n | r_n)$) can be done with the two posterior distribution computed just above. Therefore, all that is left is the sampling of the particles and the importance weights. With the same method as seen in section~\ref{sec:SIS}, we can show that:
\[ p(x_n|x_{n-1},\zeta_{n-1},y_{n})= \dfrac{p(y_n|x_{n},\zeta_{n-1})p(x_n|x_{n-1},\zeta_{n-1})}{p(y_n|x_{n-1},\zeta_{n-1})}\]
with $p(y_n|x_{n-1},\zeta_{n-1})=\int p(y_n|x_{n},\zeta_{n-1})p(x_n|x_{n-1},\zeta_{n-1})dx_n$). We can compute this probability for each possible $x_n$ as the dimension of the hidden states is finite in this case and because we can compute $p(y_n|x_{n},\zeta_{n-1})$ which is the likelyhood of a normal distribution for parameter $\theta_{x_n}$ (with $\theta$ sampled at time step $n-1$) and $p(x_n|x_{n-1},\zeta_{n-1})$ which is the $x_n^{th}$ element of $\pi_{x_{n-1}}$ (with $\pi$ sampled at time step $n-1$). Then, we sample the particles according to Discrete distribution with the probabilities we've just computed. As the importance weights are defined as $\alpha_n(x_{1:n},r_{1:n},\zeta_{1:n})=p(y_n | x_{n-1},\zeta_{n-1})$, we can obtain them by doing the sum of the unormalized probabilities computed just before.

\section{Factorial bayesian particle filter}\label{FactorialPF}

The Bayesian Particle Filter that we derived last section allows us to do online learning for most of the models introduced in chapter~\ref{sec:Model}. However, for the disaggregation problem we are interested in, we presented a Factorial model in section~\ref{sec:FactorialHDPHSMM} and to use Particle Filters with this model, we will need to make some changes to the previous algorithm.
\smallbreak
First, let's recall our Factorial model. The idea was that each device would be represented by a HDP-HSMM, previous observations (the power consumption of each device) would now become hidden emissions $y_n^{(k)}$ and the new observation $\bar{y}_n$ would now be the aggregated power consumption (the sum of the emission from each device plus a noise). To add some structure and reduce computations, we also supposed that the hidden chains were independant:
\[ p(x_n | x_{n-1}) = \prod_{k=1}^{K}p(x_n^{(k)} | x_{n-1}^{(k)}) \]
where $K$ is the number of devices (hidden chains) and $x_n=(x_n^{(1)},\cdots,x_n^{(K)})$. We need the emissions $y_n^{(k)}$ to be able to sample sequentially all the components of each model but they are now hidden. One way around is to, in the same fashion as the last section, add them to the state space. This way, we are going to sample the hidden emissions given the aggregated observation and then sample the other components given these emissions. However, to do so, we need to add an other condition on our model to make the computations tractable. From now on, we will suppose that the hidden emissions and the aggregated observation are distributed as follows:
\begin{align*}
y_n^{(k)} | x_n^{(k)} &\overset{i.i.d}{\sim} \mathcal{N}(\theta_{x_n^{(k)}}^{(k)},\sigma_{(k)}^2) \text{ for } k=1,\cdots,K \label{eq:NormalEmissions}\tag{$\ast$}\\
\bar{y}_n=\sum_{k=1}^{K}y_n^{(k)} | x_n &\:\sim \mathcal{N}(\sum_{k=1}^{K}\theta_{x_n^{(k)}}^{(k)},\sum_{k=1}^{K}\sigma_{(k)}^2)
\end{align*}
From there, we can derive the target and the importance distribution for an Auxiliary Particle Filter. Let $x_{1:n}=(x_{1:n}^{(1)},\cdots,x_{1:n}^{(K)})$, $y_{1:n}=(y_{1:n}^{(1)},\cdots,y_{1:n}^{(K)})$, $r_{1:n}=(r_{1:n}^{(1)},\cdots,r_{1:n}^{(K)})$ and $\zeta_{1:n}=(\zeta_{1:n}^{(1)},\cdots,\zeta_{1:n}^{(K)})$. We have the following target distribution:
\begin{align*}
\pi_n(x_{1:n},y_{1:n},r_{1:n},\zeta_{1:n})&=\dfrac{\gamma_n(x_{1:n},y_{1:n},r_{1:n},\zeta_{1:n})}{Z_n} \\\gamma_n(x_{1:n},y_{1:n},r_{1:n},\zeta_{1:n})&=p(x_{1:n},y_{1:n},r_{1:n},\zeta_{1:n}, \bar{y}_{1:n}) \\
Z_n &=p(\bar{y}_{1:n})
\end{align*}
We then have the following importance distribution:
\[ q^{opt}_n(x_n, y_n, r_n, \zeta_n |x_{1:n-1},y_{1:n-1},r_{1:n-1},\zeta_{1:n-1})=p(x_n, y_n, r_n, \zeta_n |x_{1:n-1},y_{1:n-1},r_{1:n-1},\zeta_{1:n-1},\bar{y}_{1:n}) \]
that we can then develop this way:
\begin{align*}
p(x_n, y_n, r_n, \zeta_n |x_{1:n-1},y_{1:n-1},r_{1:n-1},\zeta_{1:n-1},\bar{y}_{1:n})&=p(x_n |x_{1:n-1},y_{1:n-1},r_{1:n-1},\zeta_{1:n-1},\bar{y}_{1:n})\\
&\quad \times p(y_n |x_{1:n},y_{1:n-1},r_{1:n-1},\zeta_{1:n-1},\bar{y}_{1:n})\\
&\quad \times \prod_{k=1}^{K}p(r_n^{(k)}|x_{1:n}^{(k)},y_{1:n}^{(k)},r_{1:n-1}^{(k)},\zeta_{1:n-1}^{(k)},\bar{y}_{1:n}) \\
&\quad \times \prod_{k=1}^{K}p(\zeta_n^{(k)} |x_{1:n}^{(k)},y_{1:n}^{(k)},r_{1:n}^{(k)},\zeta_{1:n-1}^{(k)},\bar{y}_{1:n}) \\
&=p(x_n |x_{n-1},\zeta_{n-1},\bar{y}_{n})\\
&\quad \times p(y_n |x_{n},\zeta_{n-1},\bar{y}_{n})\\
&\quad \times \prod_{k=1}^{K}p(r_n^{(k)}|x_{n}^{(k)},r_{n-1}^{(k)},y_{n}^{(k)}) \\
&\quad \times \prod_{k=1}^{K}p(\zeta_n^{(k)} |r_{n}^{(k)})
\end{align*}
We now have particles for each model and so there are hidden emissions, bayesian sufficient statistics and parameters attached to each particle from each model. The bayesian sufficient statistics and the parameters are computed in the same way as before (you just have to do it for each model now). For the hidden states, it is also similar:
\[p(x_n |x_{n-1},\zeta_{n-1},\bar{y}_{n})=\dfrac{p(\bar{y}_n | x_n, \zeta_{n-1})\prod_{k=1}^{K}p(x_n^{(k)} | x_{n-1}^{(k)},\zeta_{n-1}^{(k)})}{p(\bar{y}_n | x_{n-1},\zeta_{n-1})} \]
except that now the normal likelihood is evaluated for the parameters $\sum_{k=1}^{K}\theta_{x_n^{(k)}}^{(k)}$ and $\sum_{k=1}^{K}\sigma_{(k)}^2$ and that the number of possible states is $J^K$ (if we suppose that each model has the same number of possible states $J$).
\smallbreak
Finally, the hidden emissions can be easily sampled thanks to our Normal distributed condition. Indeed, from \eqref{eq:NormalEmissions} we get that $y_{n}=(y_{n}^{(1)},\cdots,y_{n}^{(K)})$ is distributed as a multivariate normal given $x_n$ (because the components are independant normals). So $(y_{n},\bar{y}_n)$ is also distributed as a multivariate normal given $x_n$ (by linear transform) and we can then apply the following result:
\begin{prop}
Let $X=(X_1,X_2)$ be a random vector distributed as a multivariate normal with parameters $\left( \begin{array}{c} \mu_1 \\ \mu_2 \end{array}\right)$ and $\left( \begin{array}{cc} \Sigma_{11} & \Sigma_{12} \\ \Sigma_{21} & \Sigma_{22} \end{array}\right)$. Then $X_1$ given $X_2$ is distributed as follows:
\begin{align*}
X_1 | X_2 \sim \mathcal{N}(\bar{\mu},\bar{\Sigma}) \text{ with } \bar{\mu}&=\mu_1 + \Sigma_{12}\Sigma_{22}^{-1}(X_2 - \mu_2) \\
\bar{\Sigma}&=\Sigma_{11} - \Sigma_{12}\Sigma_{22}^{-1}\Sigma_{21}
\end{align*}
\end{prop}
Using this proposition with $y_n$ being $X_1$ and $\bar{y}_n$ being $X_2$, we get that:
\begin{align*}
y_n |x_{n},\zeta_{n-1},\bar{y}_{n} \sim \mathcal{N}(\bar{\mu},\bar{\Sigma}) \text{ with } \bar{\mu}&=\left( \begin{array}{c} \theta_{x_n^{(1)}}^{(1)}+\sigma_{(1)}^2\dfrac{\bar{y}_n-\sum_{k=1}^{K}\theta_{x_n^{(k)}}^{(k)}}{\sum_{k=1}^{K}\sigma_{(k)}^2} \\ \vdots \\ \theta_{x_n^{(K)}}^{(K)}+\sigma_{(K)}^2\dfrac{\bar{y}_n-\sum_{k=1}^{K}\theta_{x_n^{(k)}}^{(k)}}{\sum_{k=1}^{K}\sigma_{(k)}^2} \end{array}\right) \\
\bar{\Sigma}&=(\bar{\Sigma}_{i,j})_{1\leq i,j \leq K}=\left\{ \begin{array}{cc} -\dfrac{\sigma_{(i)}^2\sigma_{(j)}^2}{\sum_{k=1}^{K}\sigma_{(k)}^2} & \text{ if } i\neq j \\ \sigma_{(i)}^2\dfrac{\sum_{k\neq i}\sigma_{(k)}^2}{\sum_{k=1}^{K}\sigma_{(k)}^2} & \text{ else } \end{array}\right.
\end{align*}
Therefore, we can easily sample from $p(y_n |x_{n},\zeta_{n-1},\bar{y}_{n})$ and we can present Algorithm~\ref{FactorialBayesianPF}

{\SetAlgoNoLine\begin{algorithm}
\caption{Factorial Bayesian Particle Filter}\label{FactorialBayesianPF}
At time $n=0$:

\Indp sample $\zeta_{0}$ according to the prior $p(\zeta)$.

\Indm At time $n=1$:

\Indp\For{$i\in\{1,\cdots,N\}$}{
	sample $X_{1}^i\sim q_1(x_1)$.
	
	sample $Y_{1}^i\sim p(y_1 |X_{1}^i,\zeta_{0},\bar{y}_{1})$.
	
\For{$k\in\{1,\cdots,K\}$}{	
	
	compute the bayesian sufficient statistics $r_1^{i(k)} = \mathcal{R}(X_1^{i(k)},Y_{1}^{i(k)})$.
	
	sample the parameters $\zeta_1^{i(k)} \sim p(\zeta_1 | r_1^{i(k)})$.
	
}
	
	compute the unormalized weights $w_1(X_{1}^i, Y_{1}^i, r_1^i,\zeta_1^i)$.
}
compute the normalized weights $W_1^1,\cdots,W_1^N$ with $W_1^i=\dfrac{w_1(X_{1}^i, Y_{1}^i,r_1^i,\zeta_1^i)}{\sum_{j=1}^N w_1(X_{1}^j, Y_{1}^j,r_1^j,\zeta_1^j)}$

\For{$i\in\{1,\cdots,N\}$}{
	resample $\overline{X}_{1}^i,\overline{Y}_{1}^i,\overline{r}_{1}^i,\overline{\zeta}_{1}^i\sim \hat{\pi}_1(x_1,y_1,r_1,\zeta_1)$.
	
	set uniform weights $W_1^i=\dfrac{1}{N}$
} 

\medbreak

\Indm At time $n\geq2$:

\Indp compute the unormalized weights $\alpha_{n-1}(X_{1:n-1}^1,Y_{1:n-1}^1,r_{1:n-1}^1,\zeta_{1:n-1}^1),\cdots,\alpha_{n-1}(X_{1:n-1}^N,Y_{1:n-1}^N,r_{1:n-1}^N,\zeta_{1:n-1}^N)$ 

compute the normalized weights $W_{n-1}^1,\cdots,W_{n-1}^N$ with $W_{n-1}^i=\dfrac{\alpha_{n-1}(X_{1:n-1}^i,Y_{1:n-1}^i,r_{1:n-1}^i,\zeta_{1:n-1}^i)}{\sum_{j=1}^N \alpha_{n-1}(X_{1:n-1}^j,Y_{1:n-1}^j,r_{1:n-1}^j,\zeta_{1:n-1}^j)}$

\For{$i\in\{1,\cdots,N\}$}{
	resample $\overline{X}_{1:n-1}^i,\overline{Y}_{1:n-1}^i,\overline{r}_{1:n-1}^i,\overline{\zeta}_{1:n-1}^i\sim \hat{\pi}_{n-1}(x_{1:n-1},y_{1:n-1},r_{1:n-1},\zeta_{1:n-1})$.
	
	set uniform weights $W_{n-1}^i=\dfrac{1}{N}$
}

\For{$i\in\{1,\cdots,N\}$}{
	sample $X_{n}^i\sim p(x_n|\overline{X}_{n-1}^i,\overline{\zeta}_{n-1}^i,\bar{y}_{n})$. 
	
	sample $Y_{n}^i\sim p(y_n|X_{n}^i,\overline{\zeta}_{n-1}^i,\bar{y}_{n})$.
	
\For{$k\in\{1,\cdots,K\}$}{	
	compute the bayesian sufficient statistics $r_n^{i(k)} = \mathcal{R}(r_{n-1}^{i(k)},X_n^{i(k)},Y_n^{i(k)})$.
	
	sample the parameters $\zeta_n^{i(k)} \sim p(\zeta_n | r_n^{i(k)})$.
}
	
	Put $X_{1:n}^i = (\overline{X}_{1:n-1}^i , X_{n}^i)$, $Y_{1:n}^i = (\overline{Y}_{1:n-1}^i , Y_{n}^i)$, $r_{1:n}^i = (\overline{r}_{1:n-1}^i , r_{n}^i)$ and $\zeta_{1:n}^i = (\overline{\zeta}_{1:n-1}^i , \zeta_{n}^i)$.

}
\end{algorithm}}

\section{Smoothing}

The Factorial Bayesian Particle Filter introduced last section works but can still be improve. It suffers from a major drawback as all the particle filters we presented before. Indeed, the marginals of our estimator $p(x_t | y_{1:n})$ doesn't approximate well the true marginals of the target distribution if too much time has passed $t<<n$. This is mainly due to resampling. If we have $t<<n$, then a lot of resampling steps has been done and there is a strong probability that there will be only one distinct particle at time $t$. Thus, we only use one particle to estimate the true marginal which makes it not a good approximation. We could do less often the resampling step but it would only push a bit further the time at which the estimation degenerates. We don't want to remove resampling neither because we've seen its advantages in the previous sections.
\smallbreak
Our particle filter estimate at time $t$ the target $p(x_t | y_{1:t})$. This means that it uses only the observations up until time $t$ to create particles. The resampling step after time $t$ makes the estimation of the marginal depends on the future observations too. However, it is done by removing particles. So we would like our algorithm to create particles for the estimation of the marginal (at time $t$) with the help of past observations $y_{1:t}$ and future observations $y_{t+1:n}$. A solution to this problem is to perform "Particle Smoothing". This method is presented in \cite{Doucet} and it has a few variants. We are mostly interested in the "Two Filter Formula" as it is the one that samples new particles. The others only compute new weights, for the particles generated by our particle filter, based on all observations (past and future). Thus, the problem is still the same: there will be not enough distinct particles to represent the true marginal.
\smallbreak
The idea behind the "Two Filter Formula" is that we run two particle filter. One "Forward" which is the one we already derived in the last section and one "Backward". The latter works in the same way as the former but it runs backward in time. This means that it estimates the value of the hidden states at time $t$ given all future observations $y_{t+1:n}$. Then, we merge the two filters to create a good estimator of each marginal that uses new particles sampled given all observations. The method is based upon the following result:
\[ p(x_t | y_{1:n}) = \dfrac{p(x_t | y_{1:t-1})p(y_{t:n} | x_t)}{p(y_{t:n} | y_{1:t-1})} \]
with $1\leq t\leq n$. Here, $p(x_t | y_{1:t-1})$ can be computed with the Forward filter: $p(x_t | y_{1:t-1})=\int p(x_{t} | x_{t-1})p(x_{t-1} | y_{1:t-1})dx_{t-1}$ and its estimation $\hat{p}(x_t | y_{1:t-1})=\sum_{i=1}^{N} p(x_{t} | X_t^i)W_t^i\delta_{X_t^i}(x_t)$. $p(y_{t:n} | x_t)$ is what is going to be estimated by the Backward filter. The latter can be computed sequentially:
\[ p(y_{t:n} | x_t) = p(y_t | x_t) \int p(x_{t+1}|x_t)p(y_{t+1:n} | x_{t+1})dx_{t+1}\]
However, it is not a density in $x_t$ and thus, we can't directly use a particle filter to estimate it. M. Briers and A.Doucet presented a solution to this problem in \cite{Briers} by introducing a distribution $\tilde{p}(x_t)$ and computing a particle filter for the following target:
\[ \tilde{p}_t(x_{t:n} | y_{t:n})\propto \tilde{p}(x_t) p(y_{t:n} | x_t) \]
If we put $\tilde{p}(x_t)=p(x_t)$ (the marginal of $p(x_{1:n})$), then we have $\tilde{p}_t(x_{t:n} | y_{t:n})=p(x_{t:n} | y_{t:n})$ and so we can have a good approximation of $p(y_{t:n} | x_t)$. $\tilde{p}(x_t)$ need to be computed exactly at each time step which makes it most of the time impossible to choose $p(x_t)$. Therefore, the authors explain that $\tilde{p}(x_t)$ should be close to $p(x_t)$ (a good approximation of it) and should verify this equation:
\[ \tilde{p}(x_t)=\int p(x_t |x_{t-1})\tilde{p}(x_{t-1})dx_{t-1}\]
To compute $\tilde{p}(x_t)$, we can put $\tilde{p}(x_1)=p(x_1)$ and then use a Monte Carlo estimator of $p(x_t)$ for each time $t$: $\tilde{p}(x_t)=\sum_{i=1}^{M}p(x_t |X_{t-1}^i)$ where $X_{1:n}^1,\cdots,X_{1:n}^M$ are samples from the prior (the hidden Markov chain) $p(x_{1:n})$.
\smallbreak
Now that we can compute $\tilde{p}(x_t)$, we should be able to derive a backward particle filter for the target $\tilde{p}_t(x_{t:n} | y_{t:n})$. First, let us define the importance distribution. Recall that we are using the factorial bayesian model and thus, we want to sample the parameters and particles for each chain. Using result from the last section, we could define the importance distribution like this:
\begin{align*}
q^{opt}_n(x_t, y_t, r_t, \zeta_t |x_{t+1:n},y_{t+1:n},r_{t+1:n},\zeta_{t+1:n})&=\tilde{p}(x_t, y_t, r_t, \zeta_t |x_{t+1:n},y_{t+1:n},r_{t+1:n},\zeta_{t+1:n},\bar{y}_{t:n}) \\
&=\tilde{p}(x_t |x_{t+1:n},y_{t+1:n},r_{t+1:n},\zeta_{t+1:n},\bar{y}_{t:n})\\
&\quad \times p(y_t |x_{t:n},y_{t+1:n},r_{t+1:n},\zeta_{t+1:n},\bar{y}_{t:n})\\
&\quad \times \prod_{k=1}^{K}p(r_t^{(k)}|x_{t:n}^{(k)},y_{t:n}^{(k)},r_{t+1:n}^{(k)},\zeta_{t+1:n}^{(k)},\bar{y}_{t:n}) \\
&\quad \times \prod_{k=1}^{K}p(\zeta_t^{(k)} |x_{t:n}^{(k)},y_{t:n}^{(k)},r_{t:n}^{(k)},\zeta_{t+1:n}^{(k)},\bar{y}_{t:n}) \\
&=\tilde{p}(x_t |x_{t+1},\zeta_{t+1},\bar{y}_{t})\\
&\quad \times p(y_t |x_{t},\zeta_{t+1},\bar{y}_{t})\\
&\quad \times \prod_{k=1}^{K}p(r_t^{(k)}|x_{t}^{(k)},r_{t+1}^{(k)},y_{t}^{(k)}) \\
&\quad \times \prod_{k=1}^{K}p(\zeta_t^{(k)} |r_{t}^{(k)})
\end{align*}
We suppose that the backward filter is independant of the forward, this means that at time $n$ we don't have information from time $1$ to $n$. So the initial step (time $n$) uses the same prior as the initial step from the forward filter (time $1$). This assumption let us sample easily $y_t$, $r_t$ and $\zeta_t$: we use the same method as last section. However, a difficulty appears in the sampling of $x_t$. In \cite{Briers}, the authors use this relation to sample $x_t$:
\[ \tilde{p}(x_t |x_{t+1},y_{t})\propto p(y_t | x_t)\tilde{p}(x_t)p(x_{t+1} |x_{t})\]
However in our context, this relation is:
\[ \tilde{p}(x_t |x_{t+1},\zeta_{t+1},\bar{y}_{t})\propto p(\bar{y}_t | x_t,\zeta_{t+1})\tilde{p}(x_t,\zeta_{t+1})p(x_{t+1} |x_{t},\zeta_{t+1})\]
but we can not compute $\tilde{p}(x_t,\zeta_{t+1})$ easily, we can only compute $\tilde{p}(x_t,y_t,r_t,\zeta_{t})$ for each time $t$. An other approach is to not decompose the importance distribution as we did just before but to jointly sample everything at the same time:
\begin{align*}
\tilde{p}(x_t, y_t, r_t, \zeta_t |x_{t+1:n},y_{t+1:n},r_{t+1:n},\zeta_{t+1:n},\bar{y}_{t:n})&\propto p(\bar{y}_t | x_t, y_t, r_t, \zeta_{t})\tilde{p}(x_t,y_t,r_t,\zeta_{t}) \\
&\quad \times p(x_{t+1},y_{t+1},r_{t+1},\zeta_{t+1} |x_{t},y_{t},r_{t},\zeta_{t})
\end{align*}
However, there is also a problem because $p(\bar{y}_t | x_t, y_t, r_t, \zeta_{t})=\delta_{\sum_{k=1}^{K}y_t^{(k)}}(\bar{y}_t)$. So, our Monte Carlo estimator $\tilde{p}(x_t,y_t,r_t,\zeta_{t})$ will almost surely be equal to zero (as $y_t$ is an absolute continuous random vector and so the event $\sum_{k=1}^{K}y_t^{(k)}=\bar{y}_t$ has probability zero).
\smallbreak
For now, we don't know how to derive an importance distribution for the backward filter.

%% file: Control.tex
\chapter{Control}\label{chap:control}

The power network is like a market, it needs balance between supply and demand. Usually, the demand is predicted by different models and the power producer will produce according to this prediction. However, sometimes the predictor makes some errors or unpredicted events happen on the grid which creates a need to balance supply and demand. The resources required to maintain this balance are called "ancillary services". Most of the time, these ancillary services are provided by supply. They switch off or on means of production or use batteries.  However, this type of balancing is not very flexible because it takes too long to react to sudden changes. Moreover, the increase of renewable energy makes it even harder because we cannot control these resources.
\smallbreak
Therefore, the idea is that demand could also provide ancillary services. One way to do so is "demand response" where the customers (essentially retail customers or industrial plants) are requested to reduce their power consumption to help balancing. Another way is "Demand Dispatch" which was introduced by A. Busic in \cite{Busic}. This method does not require any actions from the customers. It uses distributed controlled algorithms which increase or decrease the power consumption of customers' devices based on information sent from the Balancing Authority (BA). The customers do not see the difference because the algorithms maintain for each device quality of service and over time, its total energy deviation is zero. So, this method exploits the inherent flexibility in power consumption of devices to create virtual energy storage which are then used to help balancing supply and demand.
\smallbreak
In \cite{Busic}, the authors developed the control architecture to perform "Demand Dispatch". They introduced several designs for local randomized controllers, showed performance through the bode plot of a linearized mean-field model and applied it to pool pumps and thermostatically controlled loads. During this chapter, we will present this architecture and how we can leverage our work on disaggregation.

\section{Local control design}

We suppose that we have for each load a Markov chain, with transition matrix $P_0$ and with state-space $X=\{x^1,\cdots,x^d\}$, which models its normal operating behavior. The control architecture revolves around using a distributed randomized controller for each load. A controller transform the transition matrix of the load in $P_{\zeta}$ at each time step $t$ based on signal $\zeta_t$ which comes from the BA (this signal is computed using another controller, see section~\ref{sec:PIcontrol}). This way, the controller influences the load to switch on or off based on what is happening on the grid.
\smallbreak
Now, we have to define the family of transition matrix $P_{\zeta}$ that the controller will use. In many cases, the behavior of load depends on elements that we cannot control (for example the temperature inside the tank of a water heater). To take into account these uncontrollable dynamics, we suppose that the state-space of the Markov chain is the cartesian product of two finite state-space $X=X_u \times X_n$ where $X_u$ is the set of states which can be controlled and $X_n$ is the set of states which cannot. Moreover, we suppose that for a new state $x'=(x_u ',x_n ')$, $x_u '$ and $x_n '$ are independant given the previous state $x$. This means that the family of transition matrix $P_{\zeta}$ have the following structure:
\[\forall\: x\in X, x'=(x_u ',x_n ')\in X\quad P_{\zeta}(x,x') = R_{\zeta}(x,x_u ') Q_0 (x,x_n ') \]
$R_{\zeta}$ gives the transition probabilities to the controlled part of the new state, so $0\leq R_{\zeta}(x,x_u ')\leq 1$ $\forall\: x\in X, x_u ' \in X_u$  and $\sum_{x_u ' \in X_u} R_{\zeta}(x,x_u ')=1$. $Q_0$ gives the transition probabilities to the uncontrolled part of the new state, so $0\leq Q_0 (x,x_n ')\leq 1$ $\forall\: x\in X, x_n ' \in X_n$  and $\sum_{x_n ' \in X_n} Q_0 (x,x_n ')=1$. In \cite{Busic}, it is assumed that $R_{\zeta}$ is of the form:
\[ R_{\zeta}(x,x_u ')=R_{0}(x,x_u ')exp\left(h_{\zeta}(x,x_u ') -\Lambda_{h_{\zeta}}(x)\right)\]
where $h_{\zeta}$ is continuously differentiable in $\zeta$ and $\Lambda_{h_{\zeta}}$ is the normalizing constant:
\[ \Lambda_{h_{\zeta}}(x)=log\left(\vphantom{\sum R_{0}}\right. \sum_{x_u ' \in X_u}R_{0}(x,x_u ')exp\left(h_{\zeta}(x,x_u ')\right)\left.\vphantom{\sum R_{0}}\right)\]
The design of the local controller comes down to choosing a specific $h_{\zeta}$. Several designs are presented in \cite{Busic}, such as the Individual Perspective Design (IPD), the System Perspective Design (SPD) or the exponential family. We choose to use the myopic design (which is a special case of the exponential family) where $h_{\zeta}(x,x_u '):=\zeta\, \mathcal{U}(x_u ')$. $\mathcal{U}(x_u )$ is defined as the power consumption of the load when it is in the controlled state $x_u$. For example, we could have $X_u = \{\oplus,\ominus\}$ where $\oplus$ means that the load is ON and $\ominus$ means that the load is OFF. Then, $\mathcal{U}(\oplus )$ is the power consumption of the load when it is ON and $\mathcal{U}(\ominus )$ when it is OFF.
\smallbreak
To combine this control architecture with our disaggregation algorithm, we replaced the current state and the power consumptions by estimations from the Factorial bayesian particle filter (seen in section~\ref{FactorialPF}). In our application, we only estimate the controlled part $x_u$ of the state but we could also consider the estimation of $x_n$. However, the latter requires a different model for each type of device, which were not developped in this paper. Each local controller receives at time step $t$ the aggregated power consumption (of the place they are in charge). They use the disaggregation algorithm to obtain samples for each device (what we called particles in the previous chapter) of the hidden states $x_{1:t}^1,\cdots,x_{1:t}^N$ ($N$ is the number of particles) and all the parameters ($\theta_{j}^1,\cdots,\theta_{j}^N$ for the observations parameters) from the posterior joint distribution. Then, they compute an estimator $\hat{x}_u$ of the current state of each of their devices and an estimator $\hat{\mathcal{U}}$ of the power consumption of these states:
\begin{align*} 
&\hat{x}_u=\underset{x_u\in X_u}{argmax}\sum_{i=1}^N \mathds{1}_{x_t^i =x_u} &\text{ MAP (Maximum A Posteriori) estimator}\\
\forall\, x_u\in X_u \quad &\hat{\mathcal{U}}(x_u)=\frac1N \sum_{i=1}^N \theta_{x_u}^i &\text{ Empirical posterior mean estimator}
\end{align*} 
Finally, they change the state of each device at time $t+1$ using the following transitions probabilities:
\[ P_{\zeta}(\hat{x},x') = Q_0 (\hat{x},x_n ')R_{0}(\hat{x},x_u ')exp\left(\zeta\,\hat{\mathcal{U}}(x_u ') -\Lambda_{\zeta\,\hat{\mathcal{U}}}(\hat{x})\right) \text{  with } \hat{x}=(\hat{x}_u,x_n) \]
Another important part of the control architecture is the Quality of Service (QoS) of a load (i.e how well a load fulfill its purpose). Indeed, the goal of "demand dispatch" is to use the inherent flexibility of the loads without decreasing the QoS. This means that the QoS should be the same with or without control. In our application, we only looked at Thermostatically Controlled Loads (TCLs), so we maintained QoS (which for this type of load translates into keeping the temperature within specific bounds) through the design of the nominal model $P_0$ (see section ??). See \cite{QoS} for more details on the control of QoS.

\section{Mean-field model}

\begin{figure}[h]
\caption*{Magnitude plot}
\hspace*{-1cm}\includegraphics[scale=0.5]{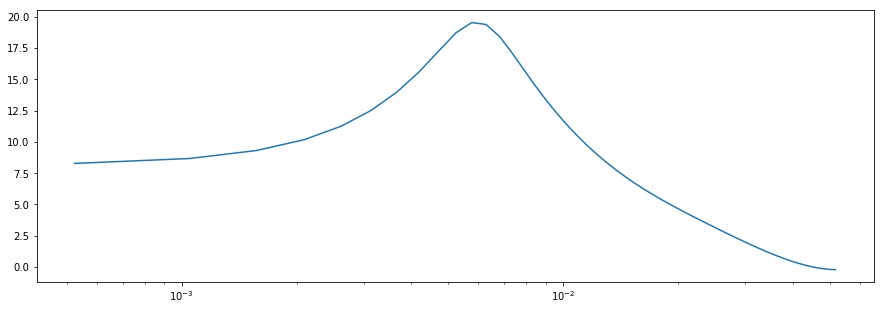}
\caption*{Phase plot}
\hspace*{-1cm}\includegraphics[scale=0.5]{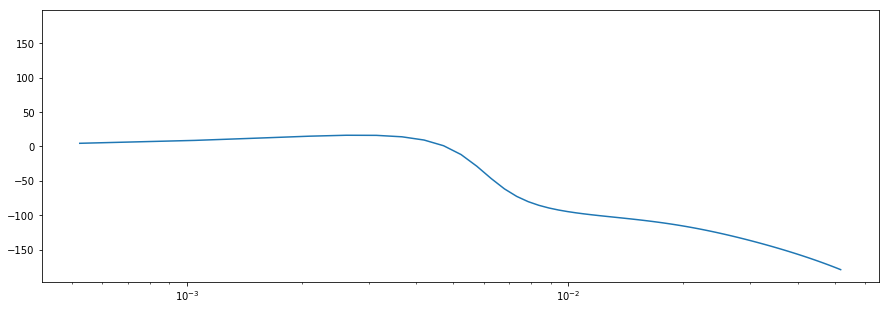}
\caption{\label{bodeLocal} Bode plot of the mean-field model linearized in $\zeta=0$}
\end{figure}

In \cite{Busic}, the authors introduced a mean-field model which serves two purposes. The first one is to evaluate the performance of the control architecture and the second one is to be able to define a PI controller (see section~\ref{sec:PIcontrol}) at the BA level. In this section, we will present the model and how it is fulfilling the first purpose. A mean-field model can be defined as a simple model which approximates a more complex stochastic model (composed of several small components). For our application, the complex stochastic model is the control architecture (that we defined in the previous section) and the small components are the local controllers. The interaction we are looking at is how the signal $\zeta$ (which is broadcast from the BA to every local controller) changes the total power consumption (because we want to control the total power consumption to help balance supply and demand).
\smallbreak
Let $X_t^i$ be the state of the $i^{th}$ load at time $t$ and $N$ the number of loads. If we assume that we are only looking at the same type of loads and that each load of a same type follows the same nominal model. Then, we can define the following empirical probability mass function:
\[\forall\, x\in X,\quad \mu_t^N (x) = \frac1N \sum_{i=1}^N \mathds{1}_{X_t^i=x}\]
Under general conditions on the model, $\mu_t^N$ tends to $\mu_t$ (when $N$ tends to infinity) where $\mu_t$ is defined by $\mu_0$ and the following dynamics:
\begin{equation}\label{eq:mu} \mu_{t+1} = \mu_t P_{\zeta_t}\quad \forall\, t\geq 0 \end{equation}
Moreover, we have a similar result for the average power consumption:
\[ y_t^N = \frac1N \sum_{i=1}^N \mathcal{U}(X_t^i)\]
and $y_t^N$ tends to $y_t$ (when $N$ tends to infinity) where $y_t$ is defined as:
\begin{equation}\label{eq:power} y_{t} = \sum_{x\in X}\mu_t (x) \mathcal{U}(x)\quad \forall\, t\geq 0 \end{equation}
Therefore, we choose our mean-field model as the deterministic system defined by \eqref{eq:mu} and \eqref{eq:power}. This simple model approximates (assuming $N$ is large enough) the interaction between $\zeta$ and the total power consumption in our control architecture (this interaction is described through $\mu_t^N$ and $y_t^N$).
\smallbreak
We will use a bode plot of the whole system to evaluate the performance of our control architecture. However, to plot the bode plot we need a linear system and the system defined by \eqref{eq:mu} and \eqref{eq:power} is not linear in $\zeta$. Thus, we need to linearized it in $\zeta$. Using proposition 3.1 of \cite{Busic}, we have that:
\[ \forall\, z\in\mathbb{C},\quad G_{\zeta}(z)=C(I_{|X|}z - A)^{-1}B \] 
is the transfer function of the linearization of the system defined by \eqref{eq:mu} and \eqref{eq:power} at a particular value $\zeta$, with $A=P_{\zeta}^T$, $C_i = \tilde{\mathcal{U}}(x^i)$ and $B_i = \sum_{x\in X}\pi_{\zeta} (x)\mathcal{E}_{\zeta} (x,x^i)$ for all $i=1,\cdots,d$. $\pi_{\zeta}$ is the invariant probability mass function of $P_{\zeta}$ (assuming $P_{\zeta}$ is irreductible), $\tilde{\mathcal{U}}(x)=\mathcal{U}(x)-\sum_{x'\in X}\pi_{\zeta} (x')\mathcal{U}(x')$ and $\mathcal{E}_{\zeta}=\dfrac{d}{d\zeta}P_{\zeta}$. The transfer function of a system is a function which describes the behavior of the output given a specific input, it is equal to the Laplace transform of the output divided by the Laplace transform of the input. In order to better see the properties of this function we can look at a graphical representation called the bode plot. This representation consists in two plots. The first one looks at the magnitude $20\, log_{10}(|G_{\zeta}(z)|)$ (in $dB$) of the transfer function depending on frequencies ($z=w\di$) on a log scale. The second looks at the phase shift $arg(G_{\zeta}(z))$ (in degrees) of the transfer function depending on frequencies ($z=w\di$) on a log scale. Figure~\ref{bodeLocal} is the bode plot of the linearized mean-field model in our application.
From this plot, we can see that the magnitude is constant around $10^{-3}$ frequency (which is a period of about $30$ minutes). This means that the control architecture should work for signals with a frequency around $10^{-3}$.

\section{Feedback loop}\label{sec:PIcontrol}

\begin{figure}[h]

\hspace*{-1cm}\includegraphics[scale=0.5]{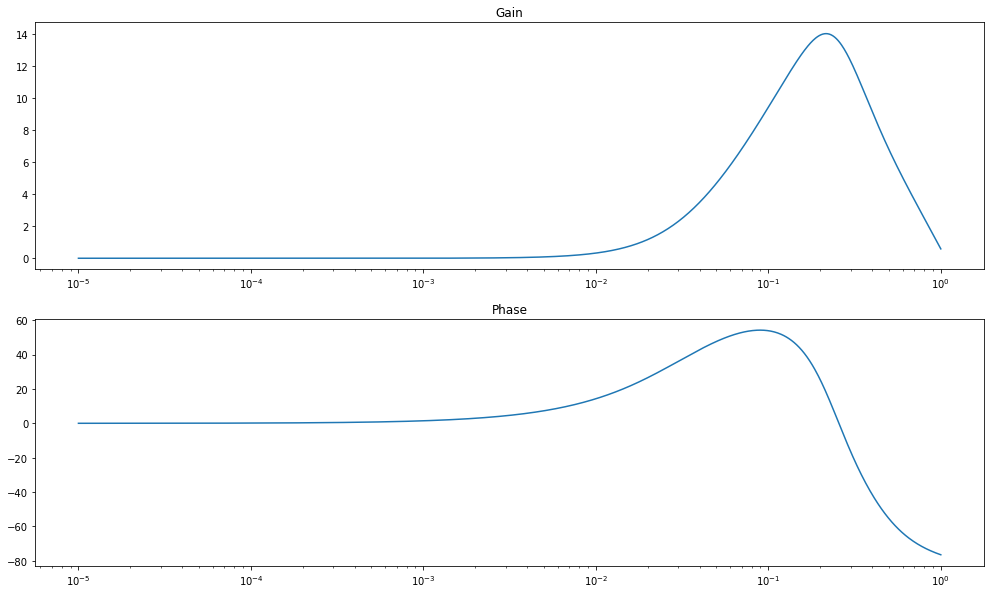}
\caption{\label{bodeGlobal} Bode plot of the feedback loop}
\end{figure}

In \cite{Busic}, the authors tackle balancing supply and demand of power by first decomposing the power production curve into several signals with different frequencies, see Figure 2 and 3 in \cite{Busic} for more details. Then, they suppose that the components with low frequency should be handled by supply and that the components with high frequency could be handled by demand. Therefore, the goal of the whole control architecture is to increase or decrease the total power consumption in order to track a reference signal $r_t$ which is the result of a high pass filter on the power production curve. To do so, the authors used a feedback loop: at each time step $t$, the BA mesures the total power consumption $y_t$, computes the power deviation $\tilde{y}_t = y_t - \bar{y}_t$ (where $\bar{y}_t$ is the nominal power consumption without control, $\tilde{y}_t$ represents the output of the mean-field model) which should be equal to the reference signal. So then, the BA computes the error $e_t = r_t - \tilde{y}_t$, sends it to a controller which computes the signal $\zeta_t$ that is broadcasted to all of the local controllers.
\smallbreak
In our application, we chose a Proportional Integral (PI) controller for the controller at the BA. It computes $\zeta_t$ given $e_t$ with the following formula:
\[ \zeta_t = K_P\, e_t + K_I \sum_{l=0}^t e_l\]
where $K_P$ and $K_I$ are two scalar parameters that need to be fitted so that the output of the mean-field model $\tilde{y}_t$ can properly track the reference signal $r_t$. We fitted these parameters by hand: we looked at the magnitude $m$ of the transfer function for frequencies where it is constant (see Figure~\ref{bodeLocal}), we set $K_P = \frac{m}{20}$ and $K_I = 60 \frac{w_c}{5}K_P$ with $w_c$ the cutoff frequency (the frequency just before the phase shifts too much, $3\times 10^{-3}$ in our application) and $60$ because we looked at $1-min$ sampled data and the frequency in the bode plot is in rad/s. Figure~\ref{bodeGlobal} is the bode plot of the transfer function $H(z)=\dfrac{K(z)G_0(z)}{1+K(z)G_0(z)}$ of the feedback loop, with $K(z)= K_P + \dfrac{K_I}{z}$ the transfer function of the PI controller.

From this plot, we can see that the whole control architecture should be able to track a reference signal with frequency around $10^{-3}$ because the magnitude/gain is $0$ (so the output $\tilde{y}_t$ and the input $r_t$ have the same amplitude) and the phase is $0$ (so the two signals evolves at the same time).

%% file: Applications.tex
\chapter{Application}\label{sec:Applications}

\begin{figure}[h]
\hspace*{-3cm}\includegraphics[width=0.97\paperwidth,height=9cm]{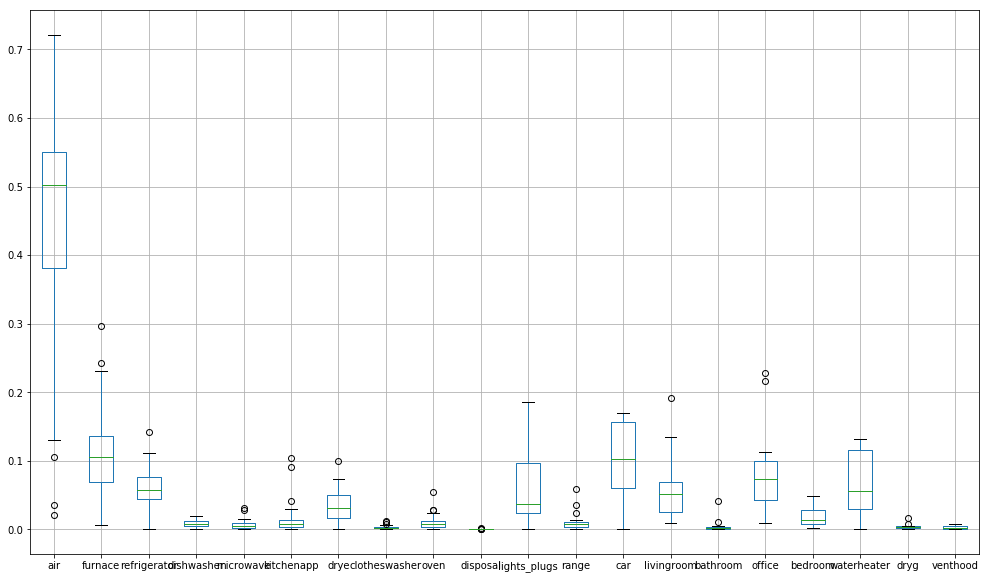}
\caption{\label{devices} Devices usage}
\end{figure}

\begin{figure}[h]
\hspace*{-3cm}\includegraphics[width=0.97\paperwidth]{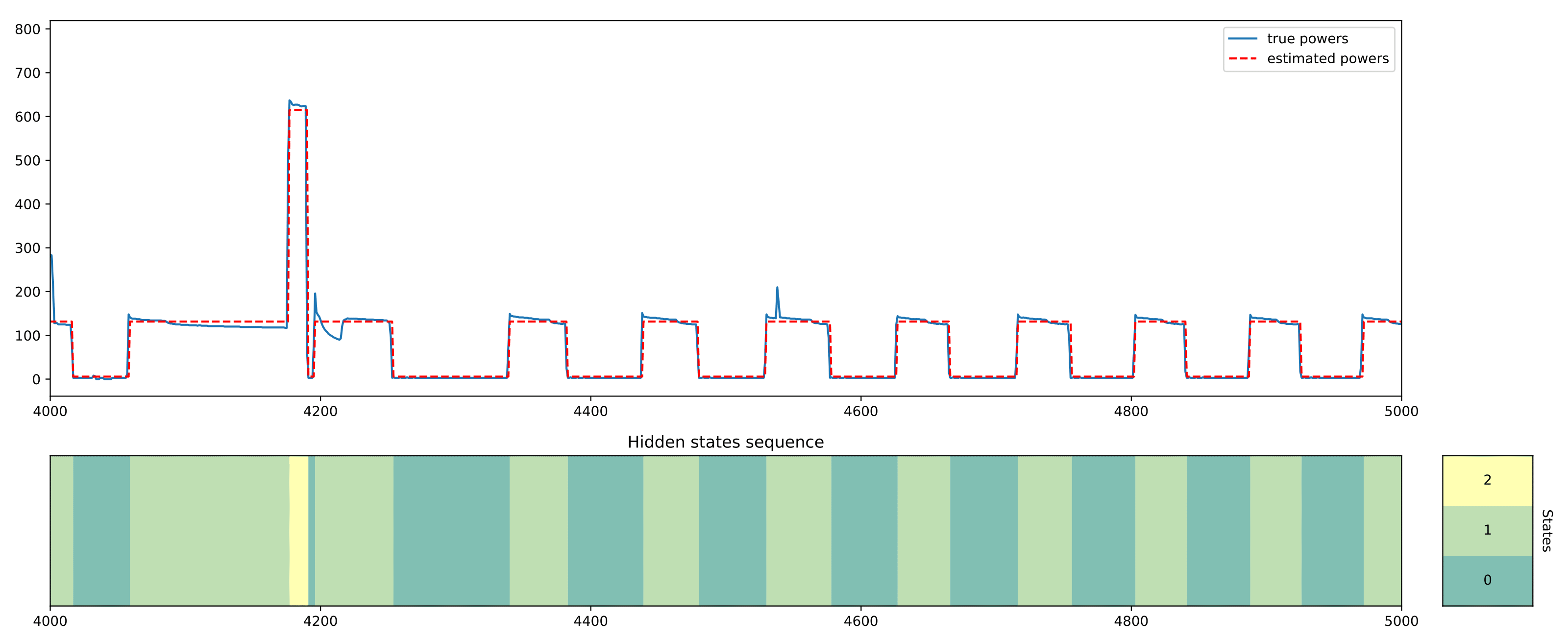}
\caption{\label{stateEstimation} Hidden states estimation on the refrigerator of house 189}
\end{figure}

\begin{figure}[h]
\hspace*{-3cm}\includegraphics[width=0.97\paperwidth]{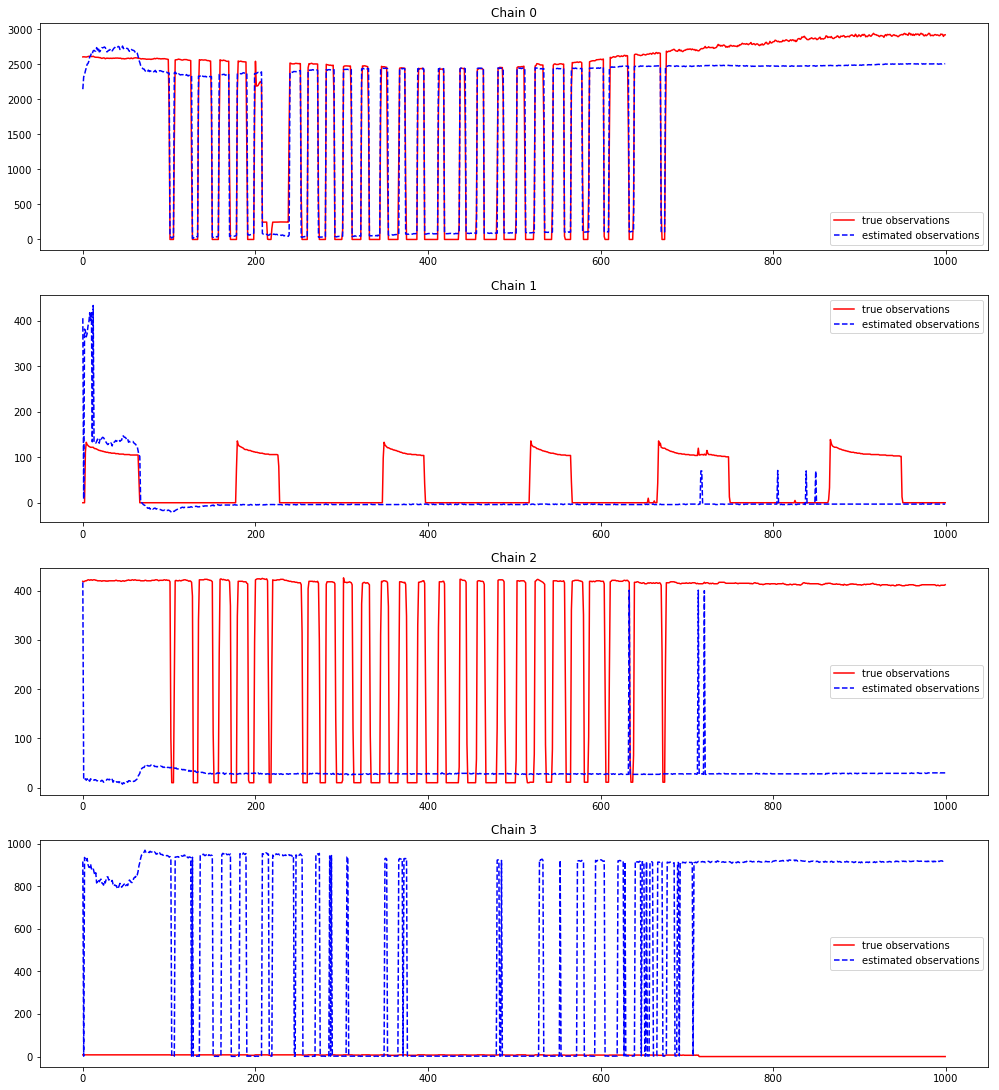}
\caption{\label{disagg} Disaggregation of house 1830}
\end{figure}

We applied the disaggregation algorithm and the control techniques to data from \href{http://www.pecanstreet.org/}{Pecan Street} using Python. We collected, from their \href{https://dataport.cloud/}{Dataport}, the power consumption of about a hundred houses (id 26 to 2401) over one month: august, 2016. The measurements were taken every minute and give the power consumption of several devices and the total power consumption of the house. We stored the data in a sqlite3 database to be able to ask queries about the usage of the different devices in those houses. 
\smallbreak
One drawback with the model we chose is that you have to choose the devices (i.e the components in the factorial model) which should explain the total power consumption (i.e the aggregated observations). However, each house has different devices and if we choose a model with a device that is not present in the house, it will still try to force it to explain the part that is not explained by the other components or noise. Therefore, we decided to carefully choose the devices of our model. We asked queries to our database in order to find devices that are often used and that explain an important part of the total power consumption. Figure~\ref{devices} is one result of these queries. 
\smallbreak
This figure shows a boxplot for each of the 20 most used devices (we consider that a house uses a device if the latter has at least once a power consumption strictly greater than zero during the observed period). The devices are ordered from the most used one (left) to the least used one (right). Each boxplot is constructed with the part of the device power consumption in the total power consumption for all houses. In our application we chose the following devices to construct our model: the air compressor, the furnace, the refrigerator, the dishwasher.
\smallbreak
In order to apply our disaggregation algorithm, we had to select the parameters of our priors (which we call hyper-parameters). Usually, these hyper-parameters are given by experts of the field who know which parameter would best fit. Because we had not any knowledge about which parameter to choose, we decided to estimate them with Pecan Street data. First, we selected some houses where the four devices represent the most part of the total consumption and some houses where they represent only a small part. This way, we can test the disaggregation algorithm in different situations. We also separated the houses in a training set and a test set. The training set is there to compute estimators of the priors hyper-parameters and is the test set is there to test the algorithm on houses where we did not learn. Then, for each device of each house from the training set, we estimated hidden states with a HDP-HSMM (using the \href{https://github.com/mattjj/pyhsmm}{pyhsmm} package from \cite{HDPHMM}) and separated the observations given the hidden states because the hyper-parameters depends on the hidden states. See Figure~\ref{stateEstimation} for a result of the hidden states estimation on the refrigerator of house 189. Finally we computed estimators of the hyper-parameters using the method of moments (we used an EM algorithm for the durations hyper-parameters because it is a mixture distribution) and plug-in techniques (when we did not have access to a value but we could compute an estimator).
\smallbreak
We applied the factorial Bayesian particle filter (from section~\ref{FactorialPF}) to the test houses using the computed priors. The algorithm performed well for the devices with the highest power consumption, which is the air compressor in our application. Figure~\ref{disagg} is the result of the algorithm for house 1830. The estimation of the refrigerator, the furnace and the dishwasher are not good enough but the method estimated well the hidden states of the air compressor and thus, we tried to apply control for this device.
\smallbreak

\begin{figure}[h]
\hspace*{-3cm}\includegraphics[width=0.97\paperwidth]{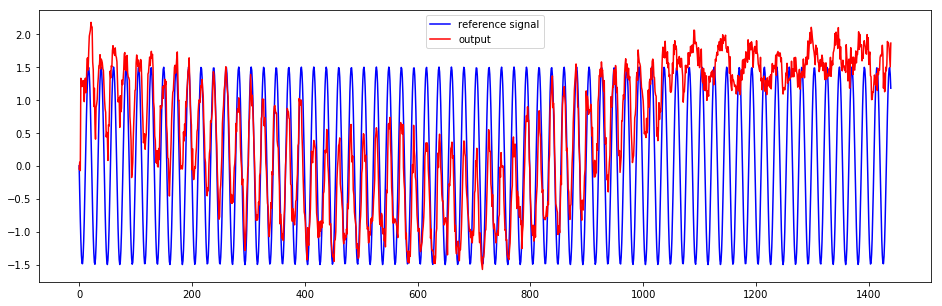}
\caption{\label{control} }
\end{figure}

In order to apply the control architecture (seen in chapter~\ref{chap:control}), we have to specify the nominal model $P_0$ for an air compressor. To do so, we used the model for thermostatically controlled loads presented in \cite{Busic}. We simulated the power consumption of the four devices, using the priors computed previously, for 100 houses. At each time step $t$, each local controller measures the aggregated power, disaggregates it and uses the air compressor state and power consumption estimation to sample the new state with $P_{\zeta_t}$. The BA measures the new total power consumption $y_{t+1}$ and uses the PI controller (seen in section~\ref{sec:PIcontrol}) given the reference signal $r_{t+1}$. We did the experiment with a simple reference signal (a sinusoid) and the result is shown in figure~\ref{control}. We were able to track the reference signal except for the hotter part of the day (from observation 1100 to the end) but this could be a consequence of poorly chosen parameters for the ODE of the inside temperature. We did not have time to experiment with a reference signal obtained through real data but this should be done to confirm these results. All the code in Python for this application can be found \href{https://github.com/ArnaudCadas/Internship-report}{here}.

%% file: report_appendix.tex
\begin{appendices}

\chapter{Preliminary knowledge}

\section{Graphical models}\label{graphmodel}

In this section, we will very briefly introduce graphical models as they are quite useful to represent the structure and the dynamics in the models of chapter~\ref{sec:Model}. We also review two important results about the graphical models that greatly helped us for the inference part.

\begin{definition}{Graphical Model}

A graphical model is a probabilistic model in which a graph represents the dependancy structure of a set of random variables. The graph can be directed or not.

When the graph is directed and acyclic (DAG), we can construct the moral graph which is the undiricted graph obtained by linking the parents of a same node (for each node) and by removing the direction of the edges.
\end{definition}

\begin{example}
\smallbreak

\begin{tikzpicture}[->,>=stealth',shorten >=1pt,auto,node distance=2cm,
                    semithick,{baseline=(current bounding box.north)}]
	
	\tikzstyle{every state}=[fill=white,draw=none,text=black]
	
	\node[state]         (X2)                                 {$X_{2}$};
	\node[state]         (X1) [above left=0.5cm and 0.5cm of X2]  {$X_{1}$};
	\node[state]         (X3) [below left=0.5cm and 0.5cm of X2]  {$X_{3}$};
	\node[state]         (X4) [above right=0.5cm and 0.5cm of X2] {$X_{4}$};
	\node[state]         (X5) [below right=0.5cm and 0.5cm of X2] {$X_{5}$};
	\node[state]         (X6) [right=1cm of X5]             {$X_{6}$};
	\node[text width=3cm, align=center]         (Xlegend) [below=1cm of X2] {Graphical model};
	
	\node[state]         (M3) [right=0.8cm of X6] {$X_{3}$};
	\node[state]         (M2) [above right=0.5cm and 0.5cm of M3]  {$X_{2}$};
	\node[state]         (M1) [above left=0.5cm and 0.5cm of M2]  {$X_{1}$};
	\node[state]         (M5) [below right=0.5cm and 0.5cm of M2]  {$X_{5}$};
	\node[state]         (M4) [above right=0.5cm and 0.5cm of M2] {$X_{4}$};
	\node[state]         (M6) [right=1cm of M5]             {$X_{6}$};
	\node[text width=3cm, align=center]         (Mlegend) [below=1cm of M2] {Moral graph};

  \path 
	      (X1) edge             node    {}         (X2)
				(X3) edge             node    {}         (X2)
				(X2) edge             node    {}         (X4)
				(X2) edge             node    {}         (X5)
				(X5) edge             node    {}         (X6)
		    (M1) edge [-]             node    {}         (M2)
				(M3) edge [-]            node    {}         (M2)
				(M2) edge [-]            node    {}         (M4)
				(M2) edge [-]            node    {}         (M5)
				(M1) edge [-]            node    {}         (M3)
				(M5) edge [-]            node    {}         (M6);
\end{tikzpicture}

\end{example}

In chapter~\ref{sec:Inference}, we computed challenging distributions. Using the graphical models that are presented in chapter~\ref{sec:Model}, the next two propositions allowed us to ease these derivations.

\begin{prop}{Factorisation of the joint distribution}

If the graph is a DAG, the joint probability density function can be factorized given the parents of each random variable:
\[ p(\{X_i\}_{i\in V})=\prod_{i\in V} p(X_i | parents(X_i)) \]
where $parents(X_i)$ is the set of parents of the node $X_i$.

\end{prop}

\begin{example}

Using the factorisation of the joint distribution with the previous graphical model, we obtain:
\[ p(X_1,\cdots,X_6)=p(X_1)p(X_3)p(X_2 | X_1, X_3)p(X_4 | X_2)p(X_5 | X_2)p(X_6 | X_5) \]

\end{example}

\begin{prop}{Markov property}

Let $\mathcal{G}(V,E)$ a DAG, $I$,$J$,$K$ disjoint subsets of $V$. Then, in the moral graph associated to $\mathcal{G}$, if all paths from $I$ to $J$ go through $K$, we have $\{X_i\}_{i\in I} \indep \{X_j\}_{j\in J} | \{X_k\}_{k\in K}$

\end{prop}

\begin{example}

Using the Markov property with the previous graphical model, we have that:
\begin{itemize}
\item $X_3$ is independent of $X_4$ given $X_2$.
\item $X_1$ and $X_3$ are not independent.
\item $\{X_5,X_6\}$ is independent of $\{X_1, X_3, X_4\}$ given $X_2$.
\end{itemize}

\end{example}

\section{Bayesian statistics}\label{BayesStats}

In inferential statistics, we suppose that we observe a realization of independent and identically distributed (i.i.d) random variables. The objective is to infer some properties about the distribution of these variables. Bayesian statistics tackle the same problem but with a different perspective. In the latter, we suppose that we have some information about the data (through what is called the "prior") and then with the realizations we update this information (through what is called the "posterior"). Finally, we infer some properties on the distribution with the updated information.

To make this more formal, let us introduce some notations. Let $\textbf{X}:\Omega\rightarrow E$ and $\boldsymbol{\theta}:\Omega\rightarrow \Theta$ be two random variables, $\nu$ a measure on $E$ and $\lambda$ a measure on $\Theta$. We note $\pi$ the density of $\boldsymbol{\theta}$ (also called the prior) with respect to $\lambda$, $f(\cdot|\theta)$ the conditional density of $X$ given $\boldsymbol{\theta}=\theta$ (also called the likelihood) with respect to $\nu$ and $\pi(\cdot | x)$ the conditional density of $\boldsymbol{\theta}$ given $\textbf{X}=x$ (also called the posterior) with respect to $\lambda$. Usually, the prior is given by the knowledge we have or some expert has on the data we are looking at and the likelihood is given by the statistical model you choose to represent your data. However, the posterior depends on the realizations and has to be computed. To do so, we often use a well known theorem:

\begin{Thm}{Bayes' theorem}
\[ \pi(\theta | x)=\frac{f(x|\theta)\pi(\theta)}{f(x)}=\frac{f(x|\theta)\pi(\theta)}{\int_{\Theta}f(x|t)\pi(t)\lambda(dt)}\propto f(x|\theta)\pi(\theta) \]
\end{Thm}

Here the symbol $\propto$ means "proportional to", so $\pi(\theta | x)$ equals $f(x|\theta)\pi(\theta)$ up to a normalization constant (everything that doesn't depend on $\theta$). This symbol is often used in Bayesian statistics as we do not need to compute the normalization constant if we recognize the distribution of $f(x|\theta)\pi(\theta)$ or if we only want to sample from the posterior (like for Monte Carlo methods). As with inferential statistics, we can define estimators and their expected loss:

\begin{definition}{Posterior expected loss}

Let $\hat{\theta}=\hat{\theta}(\textbf{X})$ be a statistic to estimate $\theta$ and $l: \Theta \times \Theta \rightarrow \mathbb{R}^{+}$ a loss function. We define the posterior expected loss as:
\[ R_B ( \pi, \hat{\theta})=\mathbb{E}[l(\boldsymbol{\theta},\hat{\theta}(\textbf{X}))]=\int_{\Theta} \left( \int_E l(\theta,\hat{\theta}(x))f(x|\theta)\nu(dx)\right)\pi(\theta)\lambda(d\theta)\]
\end{definition}

\begin{definition}{Bayesian statistic}

Let $\tilde{\theta}=\tilde{\theta}(\textbf{X})$ a statistic to estimate $\theta$ and $l: \Theta \times \Theta \rightarrow \mathbb{R}^{+}$ a loss function. $\tilde{\theta}$ is a Bayesian statistic for the loss $l$ if:
\[ \tilde{\theta} = \underset{\bar{\theta}\in \mathcal{S}}{argmin}\: R_B ( \pi, \bar{\theta})\]
where $\mathcal{S}=\{\phi:E\rightarrow\Theta\}$ is the set of estimators of $\theta$. 
\end{definition}

We give the examples of the Bayesian estimators for the most used loss functions: 

\begin{example}

\begin{enumerate}[-]
 \item $\tilde{\theta}=\mathbb{E}[\boldsymbol{\theta}|\textbf{X}]$ for the loss $l(\theta,t)=(\theta - t)^2$
 \item $\tilde{\theta}=$ The median of $\boldsymbol{\theta}$ given $\textbf{X}$ for the loss $l(\theta, t)=|\theta - t|$
 \item $\tilde{\theta}=\underset{\theta\in\Theta}{argmax}\: \pi(\theta | x)$ for the loss $l(\theta, t)=\mathbbm{1}_{\theta\neq t}$
\end{enumerate}
\end{example}

In the following example, we are given a statistical model and a prior on the parameter of the model. We compute the posterior distribution and then a Bayesian estimator of the parameter for a specific loss:

\begin{example}

Let $\textbf{X}=(X_i)_{i\in \{1,\cdots,n\} }$ with $X_i \overset{i.i.d}{\sim}Binomial(N,\theta)$ and $\boldsymbol{\theta}\sim Beta(\alpha,\beta)$. So we have the following likelihood 
\[f(x | \theta)=\theta^{\sum_{i=1}^{n}x_i}(1-\theta)^{nN-\sum_{i=1}^{n}x_i}\prod_{i=1}^{n}\binom{N}{x_i}\] 
and the following prior 
\[\pi(\theta)=\frac{\Gamma(\alpha +\beta)}{\Gamma(\alpha)\Gamma(\beta)}\theta^{\alpha -1}(1-\theta)^{\beta -1}\mathbbm{1}_{[0,1]}(\theta)\]
We compute the posterior by using Bayes' theorem:
\begin{align*}
\pi(\theta | x)&\propto f(x|\theta)\pi(\theta) \\
&\propto \theta^{\alpha +\sum_{i=1}^{n}x_i -1}(1-\theta)^{\beta + nN - \sum_{i=1}^{n}x_i -1}\mathbbm{1}_{[0,1]}(\theta)
\end{align*}
We recognize the beta distribution and so we have $\boldsymbol{\theta}|x\sim Beta(\alpha +\sum_{i=1}^{n}x_i,\beta + nN - \sum_{i=1}^{n}x_i)$. We can then infer a bayesian estimator for the loss $l(\theta,t)=(\theta - t)^2$. As the mean of a Beta-distributed random variable is equal to $\dfrac{\alpha}{\alpha+\beta}$, we have:
\[ \tilde{\theta}(\textbf{X})=\mathbb{E}[\boldsymbol{\theta}|\textbf{X}]=\dfrac{\alpha +\sum_{i=1}^{n}x_i}{\alpha +\beta + nN}\] 

\end{example}

The models that are presented in chapter~\ref{sec:Model} use the Bayesian framework as they have priors on the parameters of the models. The inference of these models that is made in chapter~\ref{sec:Inference} follows the same idea as the previous example, we compute the posterior distribution of these parameters and then derive Bayesian estimators for them (using loss functions adapted to the space on which they are defined).

\section{Dirichlet distribution}\label{sec:Dirichlet}

This section presents some properties about the Dirichlet distribution and their proof. The motivation behind this section is that we believe these results will help in the understanding of the different models in chapter~\ref{sec:Model} and of section~\ref{sec:NonParamBayes}. First, let us recall the definition of the Dirichlet distribution introduced in section~\ref{sec:BayesianHMM}:

\begin{definition}{Dirichlet distribution}

Let $X=(X_1,\cdots,X_K)$ be a random vector with $K\in\mathbb{N}^{*}$. We say that $X$ is distributed as a Dirichlet of parameter $\alpha=(\alpha_1,\cdots,\alpha_K)\in\mathbb{R}_{+}^K$  (noted $X\sim Dir(\alpha_1,\cdots,\alpha_K)$) if for every $x\in\Delta_{K-1}=\{(t_1,\cdots,t_K): t_i \geq 0, \sum_{i=1}^K t_i =1\}$, its density (with respect to the Lebesgue measure on $\mathbb{R}^{K-1}$) is:
\begin{equation}\label{dirdef1} f(x)=\frac{\Gamma(\sum_{k=1}^K \alpha_k)}{\prod_{k=1}^K \Gamma(\alpha_k)}\prod_{k=1}^K x_k^{\alpha_k -1} \text{ with } \Gamma(y)=\int_0^{+\infty} t^{y-1}e^{-t}dt 
\end{equation}
If we have $\alpha_i = 0$ ($i\in \{1,\cdots,K\}$), we say that $X_i$ is degenerate and we put $X_i=0$.
\end{definition}

The Lebesgue measure is on $\mathbb{R}^{K-1}$ because the simplex $\Delta_{K-1}$ is of dimension $K-1$: if we know the value of $K-1$ coordinates, then the last one is $1$ minus the sum of these coordinates. Because of this, the density should only have $K-1$ variables and a more suited definition should be with the following density:
\begin{equation}\label{dirdef2}f(x)=\frac{\Gamma(\sum_{k=1}^K \alpha_k)}{\prod_{k=1}^K \Gamma(\alpha_k)}\left( \prod_{k=1}^{K-1} x_k^{\alpha_k -1} \right) \left( 1-\sum_{k=1}^{K-1}x_k \right)^{\alpha_K -1}\quad \forall\:x\in S_{K-1} \end{equation}
where $S_{K-1}=\{(t_1,\cdots,t_{K-1}): t_i \geq 0, \sum_{i=1}^{K-1} t_i \leq 1\}$. So, in \eqref{dirdef1}, $x_K$ is just a notation and we have $x_K=1-\sum_{k=1}^{K-1}x_k$ (we have chosen $x_K$ but we could have chosen any $i\in\{1,\cdots,K\}$) which turns $\Delta_{K-1}$ into $S_{K-1}$ because $\sum_{i=1}^K x_i = \sum_{i=1}^{K-1} x_i + 1 - \sum_{i=1}^{K-1} x_i = 1$ is always verified and $x_K \geq 0$ is equivalent to $ \sum_{i=1}^{K-1} x_i \leq 1$. We kept \eqref{dirdef1} throughout the paper because it is more clear, it eases the notations and we do not go into any integral computation in most of the paper. So, if no precision is given, \eqref{dirdef1} is assumed. However, \eqref{dirdef2} is much more useful to prove properties about the Dirichlet distribution. So, we will use \eqref{dirdef2} for the next two results in this appendix.

\begin{prop}\label{prop:DirMean}
Let $X=(X_1,\cdots,X_K)$ be a random vector distributed as a Dirichlet of parameter $\alpha=(\alpha_1,\cdots,\alpha_K)$. We have the following result:
\[ \mathbb{E}[X_k]=\dfrac{\alpha_k}{\sum_{i=1}^{K}\alpha_i} \text{   for }k\in\{1,\cdots,K\}\] 
\end{prop}

\begin{proof}
Let $X=(X_1,\cdots,X_K)$ with $(X_1,\cdots,X_{K-1})\sim Dir(\alpha_1,\cdots,\alpha_K)$ and $X_K=1-\sum_{k=1}^{K-1}X_k$. For any $k\in\{1,\cdots,K-1\}$:
\begin{align*}
\mathbb{E}[X_k]&=\mathbb{E}[g_k(X_1,\cdots,X_{K-1})] \text{ with } g_k:y=(y_1,\cdots,y_{K-1})\mapsto y_k \\
&=\underset{S_{K-1}}{\int \cdots \int} x_k f(x_1,\cdots,x_{K-1})dx_1 \cdots dx_{K-1} \text{ with }f\text{ as \eqref{dirdef2}} \\
&=\underset{S_{K-1}}{\int \cdots \int} \dfrac{\Gamma(\sum_{i=1}^{K}\alpha_i)}{\prod_{i=1}^{K}\Gamma(\alpha_i)}x_k^{\alpha_k +1 -1}\left( \prod_{i\neq k}x_i^{\alpha_i -1}\right) \left( 1-\sum_{i=1}^{K-1}x_i \right)^{\alpha_K -1} dx_1 \cdots dx_{K-1} \\
&=\dfrac{\Gamma(\alpha_k +1)}{\Gamma(\alpha_k)}\dfrac{\Gamma(\sum_{i=1}^{K}\alpha_i)}{\Gamma(1+\sum_{i=1}^{K}\alpha_i)} \\
 &\:\times \underset{S_{K-1}}{\int \cdots \int}\underbrace{ \dfrac{\Gamma(1+\sum_{i=1}^{K}\alpha_i)}{\Gamma(\alpha_k +1)\prod_{i\neq k}\Gamma(\alpha_i)}x_k^{\alpha_k +1 -1}\left( \prod_{i\neq k}x_i^{\alpha_i -1}\right) \left( 1-\sum_{i=1}^{K-1}x_i \right)^{\alpha_K -1}}_{\text{density of a $Dir(\alpha_1,\cdots,\alpha_k +1,\cdots,\alpha_K)$}}dx_1 \cdots dx_{K-1} \\
&=\dfrac{\alpha_k}{\sum_{i=1}^{K}\alpha_i}
\end{align*}
For $k=K$, we have $\mathbb{E}[X_K]=1-\sum_{j=1}^{K-1}\mathbb{E}[X_j]=1-\sum_{j=1}^{K-1}\dfrac{\alpha_j}{\sum_{i=1}^{K}\alpha_i}=\dfrac{\alpha_K}{\sum_{i=1}^{K}\alpha_i}$.
\end{proof}

\begin{prop}\label{prop:DirNeutrality}
Let $X=(X_1,\cdots,X_K)$ be a random vector distributed as a Dirichlet of parameter $\alpha=(\alpha_1,\cdots,\alpha_K)$. For any $k\in\{1,\cdots,K\}$, define $Y=(Y_{-k},Y_k)$ where $Y_{-k}=(Y_1,\cdots,Y_{k-1},Y_{k+1},\cdots,Y_K)$, $Y_i = \dfrac{X_i}{1-X_k}$ for $i\in\{1,\cdots,K\}\backslash\{k\}$ and $Y_k=X_k$. Then, we have:
\begin{itemize}
\item $Y_{-k}$ is distributed as a Dirichlet of parameter $\alpha_{-k}$.
\item $Y_k$ is distributed as a Beta of parameter $\alpha_k$ and $\sum_{i\neq k} \alpha_i$.
\item $Y_{-k}$ and $Y_{k}$ are independent.
\end{itemize}
Because $Y_{-k}$ and $Y_{k}$ are independent for any $k\in\{1,\cdots,K\}$, we also say that $X$ is neutral.
\end{prop}

\begin{proof}
Let $X=(X_1,\cdots,X_K)$ be a random vector distributed as a Dirichlet of parameter $\alpha$. Without loss of generality, we prove this proposition for $k=1$. We choose $X_K$ to be our "dummy variable" (its subscript must be different from $k$), so we have $(X_1,\cdots,X_{K-1})\sim Dir(\alpha_1,\cdots,\alpha_K)$ and $X_K=1-\sum_{k=1}^{K-1}X_k$. We define $Y=(Y_{-1},Y_1)$ where $Y_{-1}=(Y_2,\cdots,Y_K)=(\dfrac{X_2}{1-X_1},\cdots,\dfrac{X_K}{1-X_1})$ and $Y_1=X_1$. For any measurable function $\phi :\mathbb{R}^{K-1}\longrightarrow\mathbb{R}_{+}^{*}$, we have:
\begin{align*}
\mathbb{E}[\phi(& Y_1,Y_2,\cdots,Y_{K-1})] \\
&=\mathbb{E}[\phi(X_1,\dfrac{X_2}{1-X_1},\cdots,\dfrac{X_{K-1}}{1-X_1})] \\
&=\underset{S_{K-1}}{\int \cdots \int}\phi(x_1,\dfrac{x_2}{1-x_1},\cdots,\dfrac{x_{K-1}}{1-x_1}) \dfrac{\Gamma(\sum_{i=1}^{K}\alpha_i)}{\prod_{i=1}^{K}\Gamma(\alpha_i)}\left( \prod_{i=1}^{K-1}x_i^{\alpha_i -1}\right) \left( 1-\sum_{i=1}^{K-1}x_i \right)^{\alpha_K -1} dx_1 \cdots dx_{K-1}
\end{align*}
we use the transformation $(y_1,y_2,\cdots,y_{K-1})=(x_1,\dfrac{x_2}{1-x_1},\cdots,\dfrac{x_{K-1}}{1-x_1})$ which has the following Jacobian:
\[ J = 
 \begin{pmatrix}
  1        &         &         &  (O) \\
  -y_2     & (1-y_1) &         &  \\
  \vdots   &         & \ddots  &   \\
  -y_{K-1} & (O)     &         & (1-y_1)
 \end{pmatrix} \text{ and } |det(J)|=(1-y_1)^{K-2}\]

\begin{align*}
&=\underset{[0,1]\times S_{K-2}}{\int \cdots \int}\phi(y_1,y_2,\cdots,y_{K-1}) \dfrac{\Gamma(\sum_{i=1}^{K}\alpha_i)}{\prod_{i=1}^{K}\Gamma(\alpha_i)}y_1^{\alpha_1 -1} \left( \prod_{i=2}^{K-1}(y_i(1-y_1))^{\alpha_i -1}\right) \quad\quad\quad\quad\quad\quad\quad \\
&\quad \times \left( 1-y_1-\sum_{i=2}^{K-1}y_i(1-y_1) \right)^{\alpha_K -1}(1-y_1)^{K-2} dy_1 \cdots dy_{K-1} \\
&=\underset{[0,1]\times S_{K-2}}{\int \cdots \int}\phi(y_1,y_2,\cdots,y_{K-1}) \underbrace{\dfrac{\Gamma(\sum_{i=2}^{K}\alpha_i)}{\prod_{i=2}^{K}\Gamma(\alpha_i)}\left( \prod_{i=2}^{K-1}y_i^{\alpha_i -1}\right)\left( 1-\sum_{i=2}^{K-1}y_i \right)^{\alpha_K -1}}_{\text{density of a $Dir(\alpha_{-1})$}} \\
&\quad \times \underbrace{\dfrac{\Gamma(\sum_{i=1}^{K}\alpha_i)}{\Gamma(\alpha_1)\Gamma(\sum_{i=2}^{K}\alpha_i)}y_1^{\alpha_1 -1} (1-y_1)^{\sum_{i=2}^{K-1}\alpha_i -1}}_{\text{density of a $Beta(\alpha_1,\sum_{i=2}^{K}\alpha_i)$}} dy_1 \cdots dy_{K-1} 
\end{align*}
\end{proof}

\section{Nonparametric Bayesian statistics}\label{sec:NonParamBayes}

The HDP-HSMM that we presented in section~\ref{sec:HDPHSMM} is based on the idea that we do not want to specify the number of hidden states but we want the model to infer it. This means that we can have, in theory, an infinite number of hidden states and therefore, a transition matrix of infinite dimension. In addition, we want to choose a prior on this matrix. To do so, we need some models from the nonparametric Bayesian statistics that we are going to study in this section. First we will present the Dirichlet process and two of its representations to better understand it. Then we will introduce the Hierarchical Dirichlet Process and the extension of the two previous representations. This section is merely a preamble of these two objects, see \cite{TehHDP} for more details.

\subsection{Dirichlet Process}

The Dirichlet process (DP) is stochastic process which was introduced by Thomas Ferguson in 1973. It has been used a lot in machine learning, genetics, information retrieval and speaker diarization problems. We will first present it with a formal definition given by Y. W. Teh in \cite{TehHDP}:

\begin{definition}{Dirichlet process}

Let $(E,\mathcal{E})$ be a measurable space, $H$ a base probability measure on that space and $\gamma$ a positive real number. A Dirichlet process, noted $DP(\gamma,H)$, is defined as the distribution of a random probability measure G over $(E,\mathcal{E})$ such that for any partition $(A_1,\cdots,A_r)$ of $E$, the random vector $(G(A_1),\cdots,G(A_r))$ is distributed as a finite-dimensional Dirichlet distribution with parameters $(\gamma H(A_1),\cdots,\gamma H(A_r))$:
\[ (G(A_1),\cdots,G(A_r))\sim Dir(\gamma H(A_1),\cdots,\gamma H(A_r))\]
\end{definition}

Ferguson has proven that draws from $G$ are almost surely discrete probability measures. 
\smallbreak
We are going to see briefly what is the influence of the two parameters $H$ and $\gamma$ in the draws we can get from $G$. First, let us talk about $H$:

\begin{prop}\label{prop:DPmean}
Let $G\sim DP(\gamma,H)$. For any subset $A$ of $E$, we have:
\[ \mathbb{E}[G(A)]=H(A)\]
\end{prop}

\begin{proof}
Let $G\sim DP(\gamma,H)$. For any subset $A$ of $E$, $\{A,E\backslash A\}$ is a partition of $E$. So, by definition, $(G(A),G(E\backslash A))\sim Dir(\gamma H(A),\gamma H(E\backslash A))$. Using proposition \ref{prop:DirMean}, we have that:
\[ \mathbb{E}[G(A)]=\dfrac{\gamma H(A)}{\gamma (H(A)+H(E\backslash A))}=H(A)\]
because $H$ is a probability measure on $E$ and $\{A,E\backslash A\}$ is a partition of $E$.
\end{proof}

This proposition shows that the base probability measure $H$ is the expected value of the process. Therefore, a realization of $G$ will be "near" this base probability measure. For example, if $H$ is a normal distribution, then a realization of $G$ would "look" like a discretization of this normal distribution.
\smallbreak
The parameter $\gamma$ acts as a concentration parameter. When $\gamma$ is close to $0$, the mass (of a realization of $G$) is spread among very few values. However, when $\gamma$ tends to infinity, the realization tends to a continuous probability measure. To justify this result, we will now present another representation of the Dirichlet process which uses the stick-breaking process.

\subsubsection{Stick-breaking representation of the DP}\label{sec:stickDP}

The stick-breaking representation is a more constructive view of the Dirichlet process. The idea is that because draws from $G$ are discrete probability measure, they can be decomposed in two parts: their support and the probability they attach to each of its elements. Therefore, we could construct the random support, the random probabilities and combine them to get $G$.
\smallbreak
First, let us note that the support of $G$ is contained in the support of $H$ because if we have $H(A)=0$ (for a subset $A$ of $E$) then by definition of the Dirichlet distribution $G(A)$ is degenerate and equals zero. So, taking proposition \ref{prop:DPmean} into account, it seems natural to construct the random support by sampling from $H$ so that the values stay within the support of $H$ and are more likely to be in the regions of high density.
\smallbreak
Then, to construct the random probabilities, the idea is to generalize the Dirichlet distribution. We want to be able to sample an infinite number of positive values that all sum to $1$. This is where the stick-breaking name takes its full meaning, it is an image to represent the process of sampling these probabilities. We start with a stick of length $1$, we choose a random point on the stick by sampling from a Beta distribution with parameters $1$ and $\gamma$ and we break the stick, at that point, into two parts. The length of the first part (the value of our sample) is our first probability. Then, we consider the second part to be our new stick of length $1$ and we repeat the process (an infinite number of times).
\smallbreak
Finally, to combine both parts, we just have to attach each random probability to an element of the random support. The whole construction can be sum up as:
\begin{align}\label{stick}
\beta_k^{\prime} &\:\sim Beta(1,\gamma)& \nonumber \\
\beta_k &\:= \beta_k^{\prime} \prod_{l=0}^{k-1}(1-\beta_l^{\prime})& \text{ with }\beta_0 = \beta_0^{\prime}\nonumber \\
\mu_k &\overset{i.i.d}{\sim} H \quad &\text{ for } k=0,1,2,\cdots \nonumber \\
G&\:=\sum_{k=0}^{\infty}\beta_k \delta_{\mu_k}& 
\end{align}
If $G$ is a random probability measure constructed as \eqref{stick}, then Sethuraman showed in \cite{Sethuraman}, that $G$ is distributed as a Dirichlet process with parameters $\gamma$ and $H$. We can also consider only $\beta$ which is in this case a random probability measure on $\mathbb{N}$ (so a generalization of the categorical distribution) and we note $\beta \sim GEM(\gamma)$ (GEM stands for Griffiths, Engen and McCloskey).
\smallbreak
In this representation of the DP, we can see more clearly why $\alpha$ is a concentration parameter. If $\gamma$ is close to $0$, then the realizations of the $\beta_k^{\prime}$ have a high probability to be close to $1$. This means that the first few $\beta_k$ will get all the mass and the probability measure will be highly concentrated on the first few $\mu_k$. If $\gamma$ is very high, then the realizations of the $\beta_k^{\prime}$ have a high probability to be close to $0$. This means that the mass will be spread among a lot of the $\beta_k$ and the probability measure will be spread among a lot of the $\mu_k$.

\subsubsection{Chinese Restaurant Process}\label{sec:CRP}

Another representation of the Dirichlet process is the Chinese Restaurant Process (CRP) which uses a metaphor to express the probability of $\theta_i$ given $ \theta_{i-1},\cdots,\theta_{1},\gamma,H$ where $\theta_{1},\cdots,\theta_{i}$ are random variables distributed according to $G$ with $G\sim DP(\gamma,H)$. In this representation, we do not explicitly define $G$ but we can observe its properties and create a generative model with the help of realisations from $G$. 
\smallbreak
The metaphor of the CRP is as follows: a client $\theta_i$ enters the restaurant where $K$ tables are indexed by distinct values $(\phi_k)_{1\leq k \leq K}$. He sits at a table indexed by $\phi_k$ with probability proportional to number of clients $n_k$ already seated there (we put $\theta_i=\phi_k$ and $n_k=n_k +1$), and sits at a new table with probability proportional to $\gamma$ (we put $K=K+1$, we draw $\phi_K\sim H$ and $\theta_i=\phi_K$). Therefore, the probability of $\theta_i$ given $ \theta_{i-1},\cdots,\theta_{1} $ is defined by this mixture:
\[ \theta_i|\theta_{i-1},\cdots,\theta_{1} \sim \sum_{k=1}^{K}\frac{n_k}{i-1 + \gamma}\delta_{\phi_k}+\frac{\gamma}{i-1  + \gamma}H\]
This representation clearly shows that draws from the DP have a clustering property ("Rich gets richer").

\subsection{Hierarchical Dirichlet Process}\label{sec:HDP}

A Hierarchical Dirichlet Process (HDP) defines a set of random probability mesures $G_j$ and a global random probability measure $G_0$. All the random measures $G_j$ are conditionally independent given $G_0$ and distributed as a Dirichlet Process with concentration parameter $ \alpha $ and base probability measure $G_0$. $G_0$ is distributed as a Dirichlet Process with concentration parameter $ \gamma $ and base probability measure $H$:
\begin{align*}
G_0 &\sim DP(\gamma,H) \\
G_j|G_0 &\sim DP(\alpha,G_0)
\end{align*}
As we introduced the Stick-Breaking process to better understand what is a Dirichlet Process, we can do the same for the HDP. 

\subsubsection{Stick-breaking representation of the HDP}

First, as $G_0$ is distributed as a $DP(\gamma,H)$ we can use its stick-breaking representation:
\[G_0=\sum_{k=0}^{\infty}\beta_k \delta_{\theta_k}\]
with $ \beta \sim GEM(\gamma) $ and $ \theta_k \overset{i.i.d}{\sim} H  $. Then, all the $G_j$ are distributed as a $ DP(\alpha,G_0) $, so we can also use their stick-breaking representation:
\[G_j=\sum_{k=0}^{\infty}\pi_{jk} \delta_{\theta_k}\]
Note that we used the same atoms $\theta_k$ as with $G_0$. This is because the base probability measure of the $G_j$ is $G_0$, so when we will draw atoms for the $G_j$, we will draw them from $G_0$ and therefore, $G_j$ and $G_0$ will have the same support. We can also note that the new probabilities $\pi_{j}=(\pi_{jk})_{k=0}^{\infty}$ are independent given $ \beta $ because the $G_j$ are independent given $G_0$. 
\smallbreak
If we suppose that $H$ is a non-atomic probability measure (which will be the case in our next model), we can show the link between the $\pi_j$ and $\beta$. Let $(A_1,\cdots,A_r)$ be a partition of $E$ and let $K_l=\{k:\theta_k \in A_l \}$ for $l=1,\cdots,r$. Then, $(K_1,\cdots,K_r)$ is a partition of $\mathbb{N}$ and by the stick-breaking representation, we have $G_j(A_l)=\sum_{k\in K_l}\pi_{jk}$ (for each $j$) and $G_0(A_l)=\sum_{k\in K_l}\beta_{k}$ for $l=1,\cdots,r$. Because all the $G_j$ are distributed as a $DP(\alpha,G_0)$, by definition, we have for each $j$:
\begin{equation}\label{HDPstick} (\sum_{k\in K_1}\pi_{jk},\cdots,\sum_{k\in K_r}\pi_{jk})\sim Dir(\alpha \sum_{k\in K_1}\beta_{k},\cdots,\alpha\sum_{k\in K_r}\beta_{k}) 
\end{equation}
With the assumption that $H$ is non-atomic, all the $\theta_k$ are almost surely distinct. This means that for any partition  $(K_1,\cdots,K_r)$ of $\mathbb{N}$, there is a partition of $E$ ($A_l=\{\theta_k : k \in K_l \}$ for $l=1,\cdots,r$) which verifies \eqref{HDPstick}. So, by definition, we have $\pi_j \sim DP(\alpha,\beta)$ for each $j$. 
\smallbreak
Furthermore, we can also specify a stick-breaking construction for the $\pi_j$. Let us use \eqref{HDPstick} with the partition $(\{0,\cdots,k-1\},\{k\},\{k+1,k+2,\cdots\})$:
\[ (\sum_{l=0}^{k-1}\pi_{jl},\pi_{jk},\sum_{l=k+1}^{\infty}\pi_{jl})\sim Dir(\alpha \sum_{l=0}^{k-1}\beta_{l},\alpha \beta_k,\alpha\sum_{l=k+1}^{\infty}\beta_{l}) \]
Using the proposition~\ref{prop:DirNeutrality}, we have:
\[ \dfrac{1}{1-\sum_{l=0}^{k-1}\pi_{jl}}(\pi_{jk},\sum_{l=k+1}^{\infty}\pi_{jl})\sim Dir(\alpha \beta_k,\alpha\sum_{l=k+1}^{\infty}\beta_{l}) \]
If we define $\pi_{jk}^{\prime}:=\dfrac{\pi_{jk}}{1-\sum_{l=0}^{k-1}\pi_{jl}}$ with $\pi_{j0}^{\prime}:=\pi_{j0}$, we get:
\begin{align*}
\pi_{jk}^{\prime} &\sim Beta(\alpha \beta_k,\alpha(1-\sum_{l=0}^{k-1}\beta_{l}))\\
\pi_{jk}&=\pi_{jk}^{\prime}\prod_{l=0}^{k-1}(1 - \pi_{jl}^{\prime})\text{ by induction}
\end{align*}
because a Dirichlet distribution of dimension 2 is a Beta distribution and $\sum_{l=k+1}^{\infty}\beta_{l}=1-\sum_{l=0}^{k-1}\beta_{l}$. This result shows that given $\beta$, we can construct the $\pi_j$ in a stick-breaking way.

\subsubsection{Chinese Restaurant Franchise}\label{sec:CRF}

To extend the CRP as the HDP extended the DP, we are going to see what is the Chinese Restaurant Franchise (CRF). In the same way of the CRP, the CRF uses a metaphor to express the probability of random variables (given the previous draws) that are distributed according to a set of random probability measures which are themselves distributed according to a HDP. In this representation, each $ G_j $ represents a restaurant and all of them are tied together by $G_0$ which represents the dishes served at each restaurant. 
\smallbreak
The metaphor is as follows: a client $\theta_{ji}$ enters the restaurant $j$ where there are $m_{j \cdot}$ tables (with $m_{j \cdot}=\sum_{k}m_{j k}$ and $m_{j k}$ the number of tables in restaurant $j$ where dish $k$ is served). He sits at a table (indexed by $t_{ji}$) with probability proportional to number of clients $n_{jt}$ that already seated there (we put $t_{ji}=t$), and sits at a new table with probability proportional to $ \alpha $ (we put $m_{j \cdot} = m_{j \cdot} +1$ and $t_{ji}=m_{j \cdot}$). If the client is the first to sit at that table (which means it is a new table), he orders a dish $\psi_{jt}$ that all the clients that will sit there afterwards will share. To choose the dish, the client will look at the menu $H$ and the popularity of already $K$ unique served dishes across all restaurants ($\phi_1,\cdots,\phi_K$), he will choose a dish already served (we put $\psi_{jt}=\phi_k$ and $k_{jt}=k$) with probability proportional to $m_{\cdot k}$ (with $m_{\cdot k}=\sum_{j}m_{j k}$), and will chose a new dish from the menu with probability proportional to $ \gamma $ (we put $K=K+1$, we draw $\phi_K\sim H$, we put $\psi_{jt}=\phi_K$ and $k_{jt}=K$). If the client sits at a table where a dish is already served, he will share this dish with the people already present at that table, so every table serves only one dish. In the end, we have $\theta_{ji}=\psi_{jt_{ji}}=\phi_{k_{jt_{ji}}}$. 
\smallbreak
The order of arrivals from the client is not really defined in \cite{TehHDP} but to link this representation to our next model, we suppose that there exists a bijection $f: n\mapsto j,i$ and that if client $\theta_{f(n)}$ arrives at a restaurant $j$ and sits at a table with a dish $k$ served, then the next client $\theta_{f(n+1)}$ will arrive at restaurant $k$. To sum it all up, we can define the probabilities of the clients and the dishes (given the previous arrivals) by:
\begin{align*}
\theta_{ji}|\theta_{ji-1},\cdots,\theta_{j1}, G_0 &\sim \sum_{t=1}^{m_{j \cdot}}\frac{n_{jt}}{i-1 + \alpha}\delta_{\psi_{jt}}+\frac{\alpha}{i-1  + \alpha}G_0 \\
\psi_{jt}|(\bar{\psi}_{l})_l &\sim \sum_{k=1}^{K}\frac{m_{\cdot k}}{m_{\cdot \cdot} + \gamma}\delta_{\phi_k}+\frac{\gamma}{m_{\cdot \cdot}  + \gamma}H
\end{align*}
with $\bar{\psi}_{l}=(\psi_{l1},\cdots,\psi_{lm_{l \cdot}})$ and $m_{\cdot \cdot}=\sum_{k=1}^{K}m_{\cdot k}$.

\section{Gibbs Sampling}\label{sec:GibbsSampling}

Gibbs sampling is a Markov chain Monte Carlo (MCMC) method to obtain a sequence of samples which approximates samples from a target probability distribution. This method is used when direct sampling from the target is too difficult.
\smallbreak
Let $p(x)=p(x_1,\cdots,x_d)$ be a density, with respect to a reference measure $\nu$ on $\mathcal{X}^d$, that we want to sample from. We note $x_{-l}=(x_1,\cdots,x_{l-1},x_{l+1},\cdots,x_d)$. Suppose that we know how to sample from the conditional density $p(\cdot|x_{-l})$ for all $l\in\{1,\cdots,d\}$ and all $x_{-l}\in\mathcal{X}^{d-1}$. A Gibbs sampler can be derived using Algorithm~\ref{RGibbs}.

\begin{algorithm}
\caption{Random Gibbs sampler}\label{RGibbs}
Sample $x^1$ with an initial distribution.

\For{$n\in\{1,\cdots,N\}$}{
    Draw uniformely a coordinate $l$ between $1$ and $d$.
    
	Sample $x_l '$ following $p(\cdot|x_{-l}^{n})$.
	
	Set $x^{n+1}=(x_l ',x_{-l}^{n})$.
}
\end{algorithm}

The deterministic variant of the Gibbs sampler makes the same assumptions as the original one but uses a deterministic order for the sampling step: Algorithm~\ref{DGibbs}.

\begin{algorithm}
\caption{Deterministic Gibbs sampler}\label{DGibbs}
Sample $x^1$ with an initial distribution.

\For{$n\in\{1,\cdots,N\}$}{
    \For{$l\in\{1,\cdots,d\}$}{
    
	     Sample $x_l '$ following $p(\cdot|x_1 ',\cdots,x_{l-1}',x_{l+1}^{n},\cdots,x_{d}^{n})$.
}

    Set $x^{n+1}=(x_1 ',\cdots,x_d ')$.
    
}
\end{algorithm}

The initial distribution for both algorithms can be anything, for example we could draw $x^1$ uniformly on $\mathcal{X}^d$. However, choosing an initial distribution close to the target one would generate better results. In order to understand why these two algorithms can produce samples from the target density, we can look at the following result. 
\smallbreak
Suppose that we are given a sample $x$ and that we choose coordinate $l$ (either it was drawn uniformely or it is the $l^{th}$ step for the deterministic sampler). Then, the next sample $x'$ is obtained following the density $p(x_l '|x_{-l})$ if $x_{-l}'=x_{-l}$. Let us define the following transition kernel for any $l\in\{1,\cdots,d\}$:
\[ P_l(x,x')=\left\{
  \begin{array}{ccc}
    p(x_l '|x_{-l}) & if &x_{-l}'=x_{-l} \\
    0 & else & \\
  \end{array}
\right.\]
The target density is reversible for this transition kernel because
\[ p(x)P_l(x,x')=p(x_l,x_{-l})\dfrac{p(x_l ',x_{-l})}{p(x_{-l})}=p(x_l,x_{-l}')\dfrac{p(x_l ',x_{-l}')}{p(x_{-l}')}=p(x')P_l(x',x)\]
if $x_{-l}'=x_{-l}$ and $p(x)P_l(x,x')=p(x')P_l(x',x)=0$ if $x_{-l}'\neq x_{-l}$. Therefore, the target density is invariant for $P_l$: 
\[ p(x') = \int_{\mathcal{X}^d}p(x)P_l(x,x')\nu(dx) \]
We can view the samples of a Gibbs sampler as a realisation of a Markov chain with a specific transition kernel. For the random sampler, we have the following kernel:
\[ P_R = \dfrac{1}{d}(P_1 + \cdots + P_d)\]
because we choose a coordinate $l$ with probability $\dfrac{1}{d}$ and then use the kernel $P_l$. For the deterministic sampler, we have the following kernel:
\[ P_D = P_d \circ \cdots \circ P_1 \]
because we use successively each kernel $P_l$. We can show that the target density is invariant for both transition kernel $P_R$ and $P_D$:
\[ \int_{\mathcal{X}^d}p(x)P_R(x,x')\nu(dx)= \dfrac{1}{d}\left(\int_{\mathcal{X}^d}p(x)P_1(x,x')\nu(dx) + \cdots + \int_{\mathcal{X}^d}p(x)P_d(x,x')\nu(dx)\right)=p(x') \]

\begin{align*}
 \int_{\mathcal{X}^d}p(x)P_D(x,x')\nu(dx)&=\int_{\mathcal{X}^d} \cdots \int_{\mathcal{X}^d}p(x)P_1(x,x^1)\nu(dx)\cdots P_d(x^d,x')\nu(dx^d) \\
&=\int_{\mathcal{X}^d} \cdots \int_{\mathcal{X}^d}p(x^1)P_2(x^1,x^2)\nu(dx^1)\cdots P_d(x^d,x')\nu(dx^d) \\
& \ \; \vdots \\
& =p(x')
\end{align*}
Under some conditions (irreductibility if $|\mathcal{X}|<\infty$ or positive Harris recurrent if $\mathcal{X}$ is more general) on the transition kernel, we can use the Law of Large Numbers (LNN) for Markov chains. Because the target density is invariant for both transition kernel $P_R$ and $P_D$, the LNN shows that with enough sampling steps, we can use the samples to create Monte-Carlo estimators of $\int \phi(x)p(x)\nu(dx)$ for any $\phi$. Moreover, if the transition kernel is aperiodic,  with enough sampling steps, the samples are drawn following the target density. 

\end{appendices}